\documentclass[aps,prd,twocolumn,showpacs,superscriptaddress,groupedaddress,nofootinbib]{revtex4}  
\usepackage{graphicx}  
\usepackage{dcolumn}   
\usepackage{placeins}
\usepackage{tikz}
\usepackage{soul}
\usepackage{physics}
\usepackage{siunitx}
\usepackage[justification=centering]{caption}

\usepackage{graphicx}
\usepackage{amsmath}
\usepackage{amsfonts}
\usepackage{epsfig}
\usepackage{slashed}
\usepackage{braket}
\usepackage{subfig}
\usepackage{comment}
\usepackage{cases}
\usepackage{appendix}
\usepackage{color}
\usepackage{booktabs}
\usepackage{lipsum}
\definecolor{mygreen}{rgb}{0,0.5,0}

\def\bec{\begin{center}}
\def\eec{\end{center}}
\def\beq{\begin{equation}}
\def\eeq{\end{equation}}
\def\bea{\begin{eqnarray}}
\def\eea{\end{eqnarray}}

\usepackage{lineno}
\begin{document}
\title{Quantum Ising model on two dimensional anti-de Sitter space}

\author{Muhammad Asaduzzaman}
\affiliation{Department of Physics and Astronomy, University of Iowa, Iowa City, IA 52242, USA}
\author{Simon Catterall}
\affiliation{Department of Physics, Syracuse University, Syracuse, NY 13244, USA }
\author{Yannick Meurice}
\affiliation{Department of Physics and Astronomy, University of Iowa, Iowa City, IA 52242, USA}
\author{Goksu Can Toga}
\affiliation{Department of Physics, Syracuse University, Syracuse, NY 13244, USA }

\date{\today}

\begin{abstract}
This paper investigates the transverse Ising model on a discretization of two-dimensional anti-de Sitter space. We use classical and quantum algorithms to simulate real-time evolution
and measure out-of-time-ordered correlators (OTOC). The latter can probe thermalization and
scrambling of quantum information under time evolution. We compared tensor network-based methods both with simulation on gate-based superconducting quantum devices and analog
quantum simulation using Rydberg arrays.
While studying this system's thermalization properties, we observed different regimes depending on the radius of curvature of the space. 
In particular, we find a region of parameter space where the thermalization time depends only logarithmically 
on the number of degrees of freedom.
\end{abstract}

\pacs{}
\maketitle

\section{\label{intro}Introduction}

One of the most fruitful ideas in theoretical physics developed over the last twenty-five years has been the concept of holographic duality -- that the physical content of a gravitational theory in anti-de Sitter space can be captured by a non-gravitational conformal
field theory (CFT) living on the boundary of that space. Since the duality maps strong coupling to 
weak coupling, it has frequently 
been used to probe the strong coupling dynamics of a CFT living at the boundary by 
solving a classical gravity problem in the bulk \cite{Beisert:2010jr,Hubeny:2014bla}. To gain insight into
quantum gravity, one would like to invert the direction of this logic and use the non-perturbative quantum dynamics of the CFT to infer aspects of bulk quantum gravity.

As a first step in this direction,
one performs a Wick rotation on anti-de Sitter space to obtain 
hyperbolic space, followed by a discretization of the latter to obtain a lattice theory.

There have been recent efforts to perform classical simulations 
of such theories using Monte Carlo methods \cite{Asaduzzaman:2020hjl,Brower:2019kyh,Asaduzzaman:2021bcw,Brower:2022atv}, tensor network methods \cite{swingle2012entanglement,swingle2012constructing,steinberg2022conformal, jahn2022tensor} and other numerical techniques \cite{basteiro2023breitenlohner}. 
However such studies
cannot probe the real-time dynamics of such systems, and in this manuscript, we return to a simple toy model that can be {\it quantum} simulated directly in anti-de Sitter space -- the
transverse Ising model formulated in two-dimensional anti-de Sitter space ($\rm AdS_2$).

This paper will study this model using exact diagonalization, tensor network methods, noiseless quantum simulators, and simulation on superconducting quantum devices. Since the boundary theory
is conformal quantum mechanics, a prime focus of our work will be time-dependent correlation
functions and, in particular, so-called ``out-of-time-ordered" correlators (OTOCs). These provide information on how fast quantum information can propagate through the lattice and how long
thermalization takes in such an interacting quantum system.

Contrary to naive expectation it is possible for a quantum mechanical system to undergo thermalization {\it locally} \cite{Rigol_2008,PhysRevE.50.888}. Indeed such
thermalization has also
been observed experimentally \cite{Kaufman_2016}.

The key idea is that one needs to focus on a subset $A$ of the composite system comprising
$A$ and its environment $B$. 
If $A$ is entangled
with $B$ then one naturally obtains a 
density matrix for $A$
by tracing out the degrees of freedom in the Hilbert space of $B$. If $\ket{\psi}\bra{\psi}$ denotes a pure state of the combined
system, the density matrix of $A$ is given by
\begin{equation}
	\rho_A=\Tr_{{\cal H}_B} \ket{\psi}\bra{\psi}.
\end{equation}
This density matrix corresponds to a mixed state if there is entanglement between $A$ and $B$, and this is manifested by a non-zero entanglement entropy given by the von Neumann formula:
\begin{equation}
	S=-\Tr_{{\cal H}_A}\rho_A\ln{\rho_A}\label{eqn_vne}.
\end{equation}

In this paper, we are particularly interested in mixed states corresponding to thermal systems.
One simple way to construct a thermal density matrix for $A$ is to start from a composite system comprising two identical
copies of $A$
\begin{equation}
	\ket{\Psi}=\frac{1}{Z^{\frac{1}{2}}}\sum_n e^{-\frac{\beta}{2} E_n}\ket{n_A}\ket{n_B}.
\end{equation}
In this case, tracing out $B$ yields 
\begin{equation}
	\rho_A=\frac{1}{Z}\sum_n e^{-\beta E_n}\ket{n}\bra{n},
\end{equation}

in the case where the quantum mechanical system corresponds to a conformal
field theory there is a holographic interpretation of the density 
matrix as describing a black hole
in a dual geometry which contains the CFT on its boundary. Indeed the entanglement entropy in this
case can then be shown to correspond
to the Bekenstein-Hawking entropy associated with the area of the event horizon of the black hole \cite{Bekenstein:1973ur,Bekenstein:2003dt,Eisert:2008ur,Aharony:1999ti}. 

The next most obvious question that arises is how long it takes to realize this density matrix 
under Hamiltonian evolution starting from some pure non-generic state $\ket{\psi}$. In general, this process resembles classical chaotic dynamics with initial states 
that differ only by small perturbations
yielding radically different states at large times. This
thermalization process is called scrambling and has been the focus of many
previous studies \cite{xu_accessing_2020,kusuki_entanglement_2019,yuan_quantum_2022,bhattacharyya_quantum_2022,xu_scrambling_2022,Tsuji:2017fxs,Bentsen:2018uph,Campisi:2016qlj,PhysRevB.98.134303,Bohrdt:2016vhv,smith2019logarithmic}. The scrambling time $\tau_S$ is determined by the speed at which information can propagate across the system under time evolution and is related to the dimensionality of the system and the locality of the Hamiltonian. There are theoretical bounds on the scrambling time  $\tau_S$ which  is bounded
from below by
\begin{equation*}
	\tau_S \sim \beta \ln{V},
\end{equation*}
where, $V$ counts the number of microscopic degrees of freedom. Attaining
this bound depends on an exponentially fast spread of information through the system \cite{lieb2004finite,Shenker:2013pqa,Shenker:2014cwa,Maldacena:2015waa,Aleiner:2016eni}. 

It has been conjectured that CFTs with black hole duals provide one example of a system capable of
such ``fast scrambling" \cite{sekino_fast_2008,lashkari_towards_2013}. Systems that show fast scrambling typically
involve non-local Hamiltonians and all-to-all interactions such as the SYK model \cite{sachdev1993gapless,noauthor_alexei_nodate,noauthor_alexei_nodate2,PhysRevD.94.106002,Polchinski:2016xgd}. In this paper we will show that in certain regions of the parameter space the
transverse quantum Ising model with nearest neighbor
interactions living on a discretization of two dimensional
anti-de Sitter space appears to exhibit similar behavior. However one should be careful
with this interpretation -- the spatial boundary of our system is just points and our quantum
spins populate the bulk space as well as the boundary. So we are primarily looking at information
spread in the bulk. To understand the thermalization
properties better one would need to extend the model to three dimensional anti-de Sitter space
which possesses a non-trivial spatial boundary.

	We have performed both classical and quantum simulations of this system.
	In Sec.~\ref{Ham}, we find the ground state of this model using the density matrix renormalization (DMRG) algorithm \cite{white2004real,PhysRevB.48.10345,RevModPhys.77.259} and time-evolve it  
	with the time evolving block decimation (TEBD) algorithm using the ITensor library \cite{PhysRevLett.91.147902,verstraete2004matrix,PhysRevLett.93.040502,Fishman:2020gel}. In Sec.~\ref{qm_sim}, real time evolution of the magnetization is discussed and implemented for a thirteen qubit system and compared to the tensor method results. We discuss the information propagation in this model in Sec.~\ref{scrambling}. To study the
	scrambling properties of the model we have used matrix product operator (MPO) methods to calculate the OTOCs \cite{Huang:2016knw,Chen:2016qpx} in Sec.~\ref{classical_OTOC}. In the next subsection~\ref{qm_OTOC}, the computation of OTOCS using a protocol developed by Vermersch et al. \cite{vermersch_probing_2019} is discussed and implemented for a model with seven qubits. Successful implementation of the model on quantum devices required applying some additional
	error mitigation techniques. We discuss the influence of the mitigation techniques on the results and other numerical aspects of the digital quantum simulation in Appendix~\ref{quantum_ap}.
	We also sketch out how to implement 
	this Hamiltonian via analog quantum devices like Rydberg arrays and perform simulations of the system on the Bloqade simulator developed by QuEra in Appendix~\ref{ryd_app}. In Appendix~\ref{app_otoc_protocol}, we include some details of the protocol used for the computation of the OTOC using a quantum computer. 

\section{\label{Ham} Transverse Ising Model on a Hyperbolic Space}
In this section, we describe the Transverse Field Ising (TFI) model formulated
on a one dimensional hyperbolic space. The model is an analogue of the classical Ising model on a two dimensional tessellation of hyperbolic space \cite{asaduzzaman2022holography,brower2021lattice}.
The Hamiltonian that describes this Ising chain can be represented as a sum of local terms \cite{ueda_hyperbolic_2011,ueda_transverse_2010,ueda_scaling_2010} 

\begin{align}
	\hat{H} &= \frac{-J}{4}\sum_{<ij>}\frac{\cosh(l_i)+\cosh(l_j)}{2}\sigma_i^z\sigma_j^z +\frac{h}{2}\sum_i \cosh(l_i)\sigma_i^x \nonumber\\
	&+\frac{m}{2}\sum_i \cosh(l_i)\sigma_i^z.  \label{hamiltonian}
\end{align} 
Here, $\sigma^p_i$ is a local Pauli operator at site $i$ with $p=\{x,y,z\}$. The first term corresponds to a nearest neighbour interaction term coupling site $i$ and site  $j=i+1$.
The deformation factors $\cosh{l_i}$  arise from the metric of Euclidean $\mathrm{AdS}_2$ given in
Eq. (\ref{ads2}) and 
give rise to a site-dependent coupling for the Ising chain 
\begin{equation}
	ds^2= \ell^2(\cosh^2(\rho)dt^2+d\rho^2)\label{ads2}.
\end{equation}
For an $N$ site lattice the site-dependent deformation factor $l_i$ is given by
\begin{equation}
	l_i=-l_{\rm max}+i\frac{2l_{\rm max}}{N-1},
\end{equation}
where $l_{\rm max}$ denotes a length scale that determines the degree of deformation. In the limit of $l_{\rm max} \to 0$, the usual transverse Ising model is recovered.

\begin{figure}[!h]
	\centering
	\includegraphics[width=0.50\textwidth,height=0.25\textwidth]{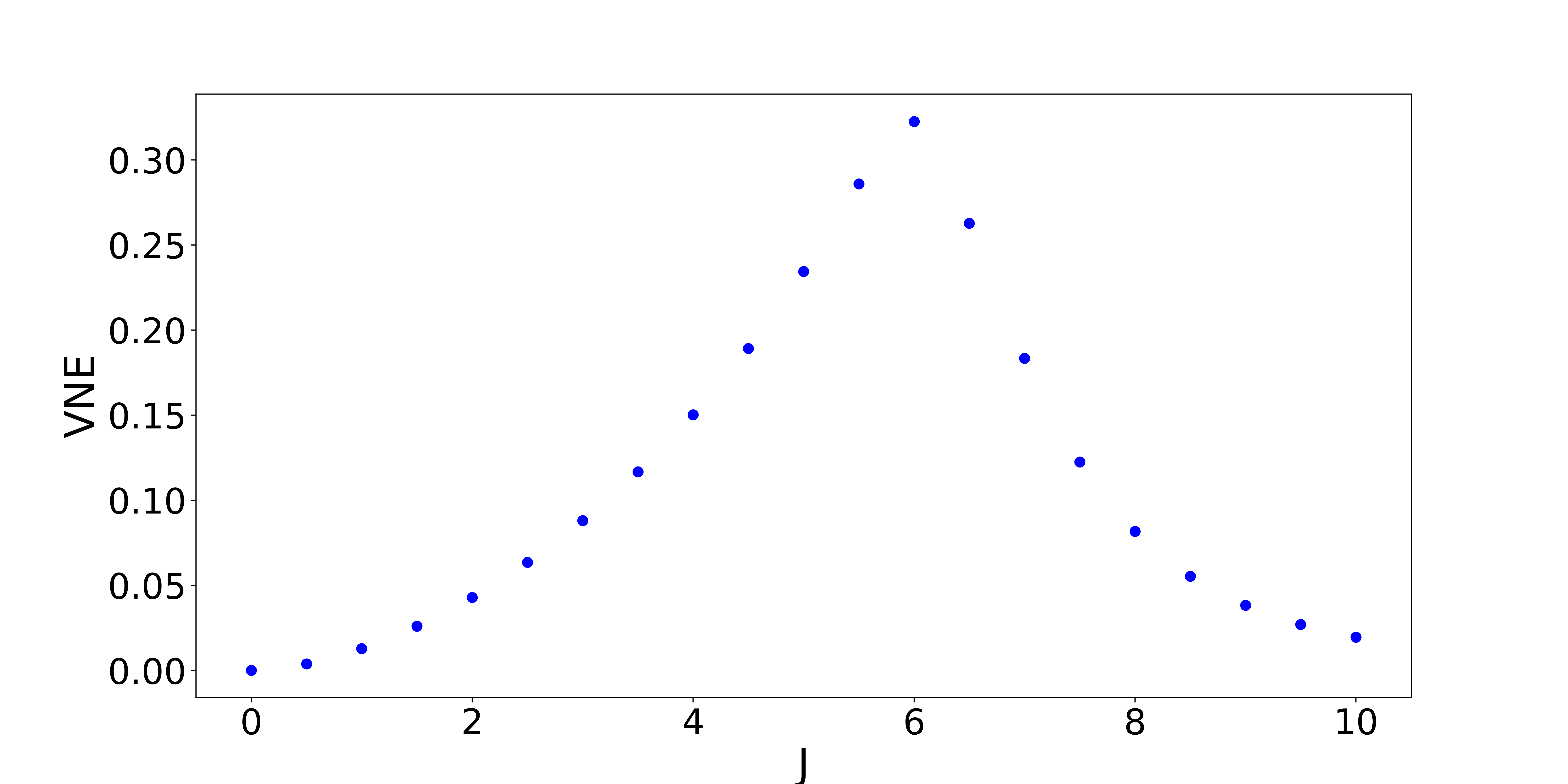}
	\caption{von Neumann Entropy versus $J$ for $N=37$, $l_{\rm max}=3.0, h=3.0, m=0.25$.}
	\label{fig:vne_N37}
\end{figure}

\begin{figure}[!h]
	\centering
	\includegraphics[width=0.50\textwidth,height=0.25\textwidth]{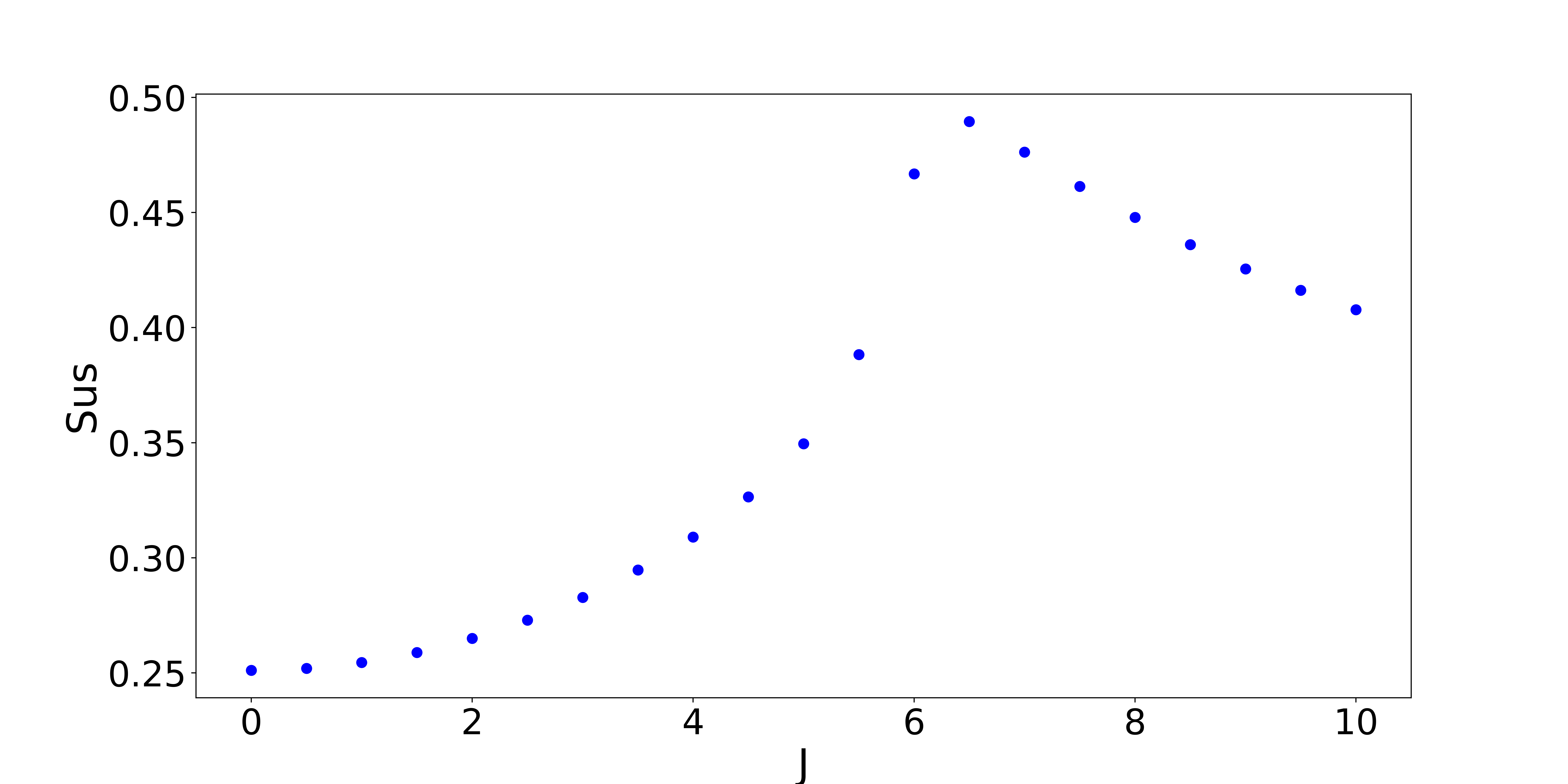}
	\caption{Magnetic susceptibility versus $J$ for $N=37$, $l_{\rm max}=3.0, h=3.0, m=0.25$.}
	\label{fig:sus_N37}
\end{figure}

We start the discussion of our numerical results 
with the von Neumann entropy Eq.~(\ref{eqn_vne}) 
calculated from the reduced density matrix obtained by tracing out one half of
the spins. Fig.~\ref{fig:vne_N37}  shows a plot of the half chain entropy and Fig.~\ref{fig:sus_N37} shows  the magnetic susceptibility at $l_{\rm max}=3.0$ $h=3.0$ and $m=0.25$ using $N=37$ spins as a function of $J$. 

For our DMRG calculation we used $50$ sweeps of the chain with a cutoff of order $\epsilon=10^{-12}$ which resulted in a bond dimension of order $\chi=10$ on average.  We see that there are  peaks in the entropy and the susceptibility signaling a possible phase transition in the model. In our later work on OTOCs we will
always tune our couplings to be close to their critical values. 

\section{\label{qm_sim}Time evolution of the magnetization}

\begin{figure}[!htb]
	\centering
	\includegraphics[scale=.24]{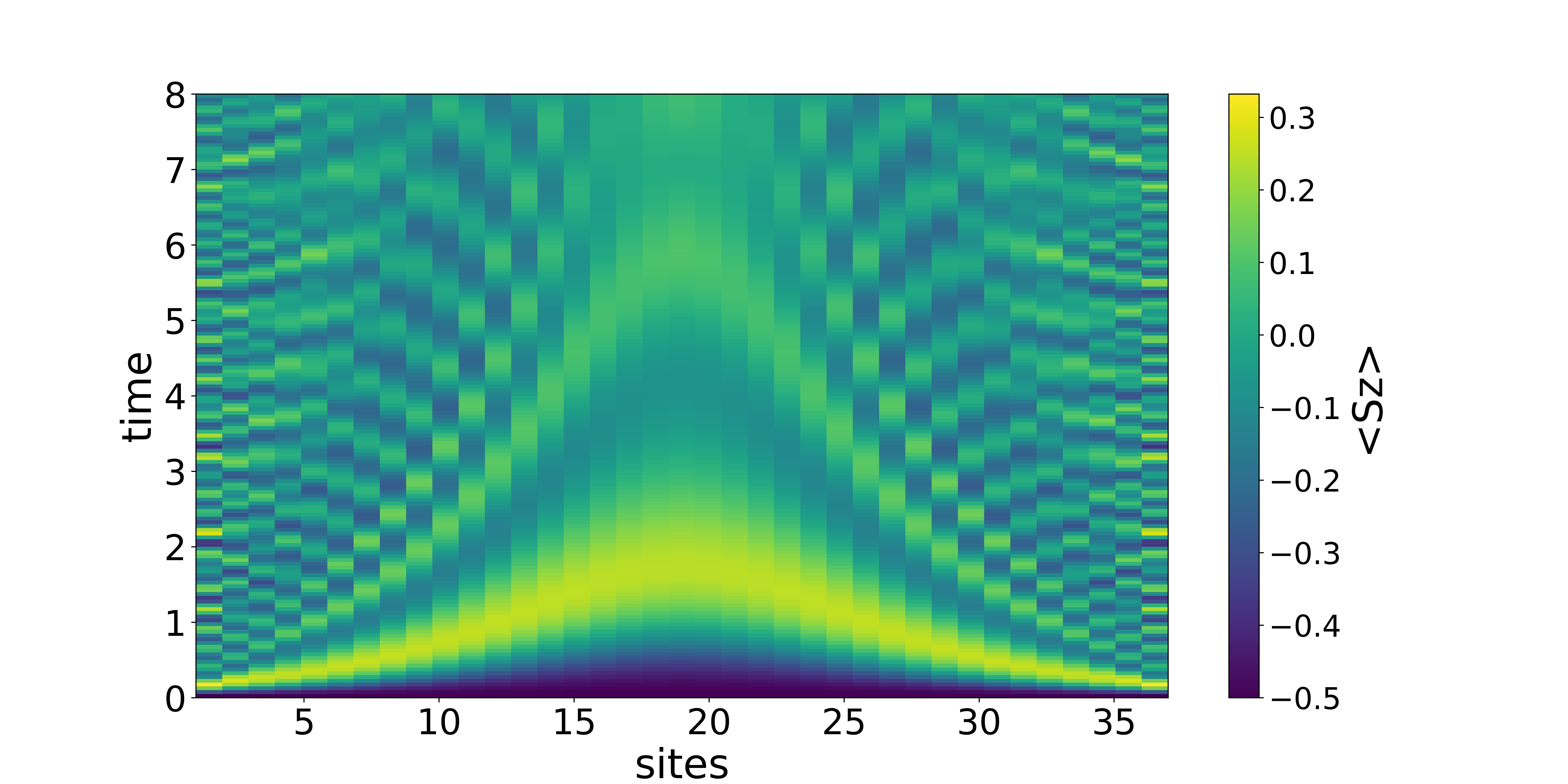}
	\caption{$\langle S^z \rangle$ for a lattice with $N=37$ spins and  parameters set at $J=2.0, h=2.0, m=0.25, l_{\rm max}=3.0$. }
	\label{fig:tebd_sz}
\end{figure}

\begin{figure}[!htb]
	\centering
	\includegraphics[width=0.50\textwidth]{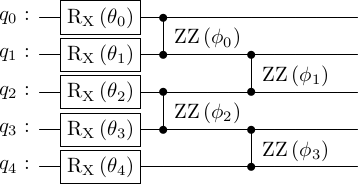}
	\caption{Trotter evolution circuit for 5-Qubit deformed hyperbolic spin chain. Here, $\theta_i$  parameters are symmetric to the center of the lattice chain, for example, in the above circuit $\theta_0=\theta_4$.}
	\label{digital_ckt}
\end{figure}

\begin{figure*}
	\centering
	{
		\includegraphics[width=\textwidth]{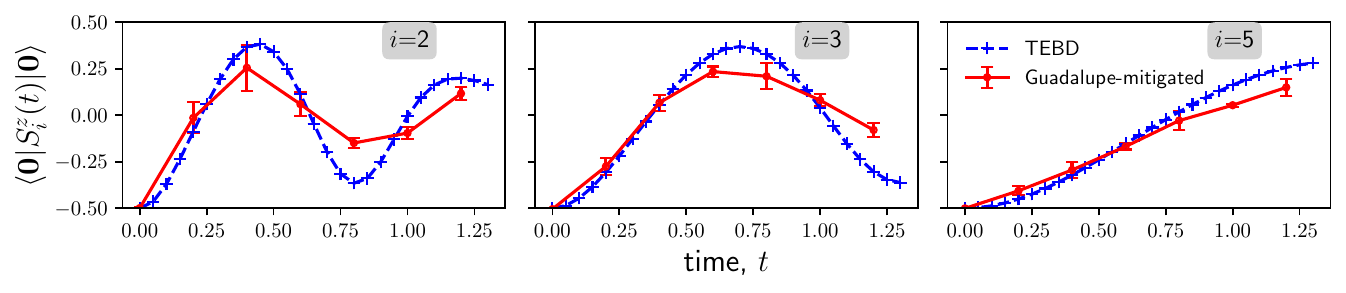}
		\caption{Local magnetization $\langle S_z^i(t)\rangle$ at site $i$ for TFI model on hyperbolic lattice chain with 13 lattice sites. Parameters: $J=2.0$, $h=1.05$, $l_{\rm max}=3.0$. }\label{fig_local_m}
	}
\end{figure*}

\begin{figure}
	\centering
	{
		\includegraphics[scale=.4]{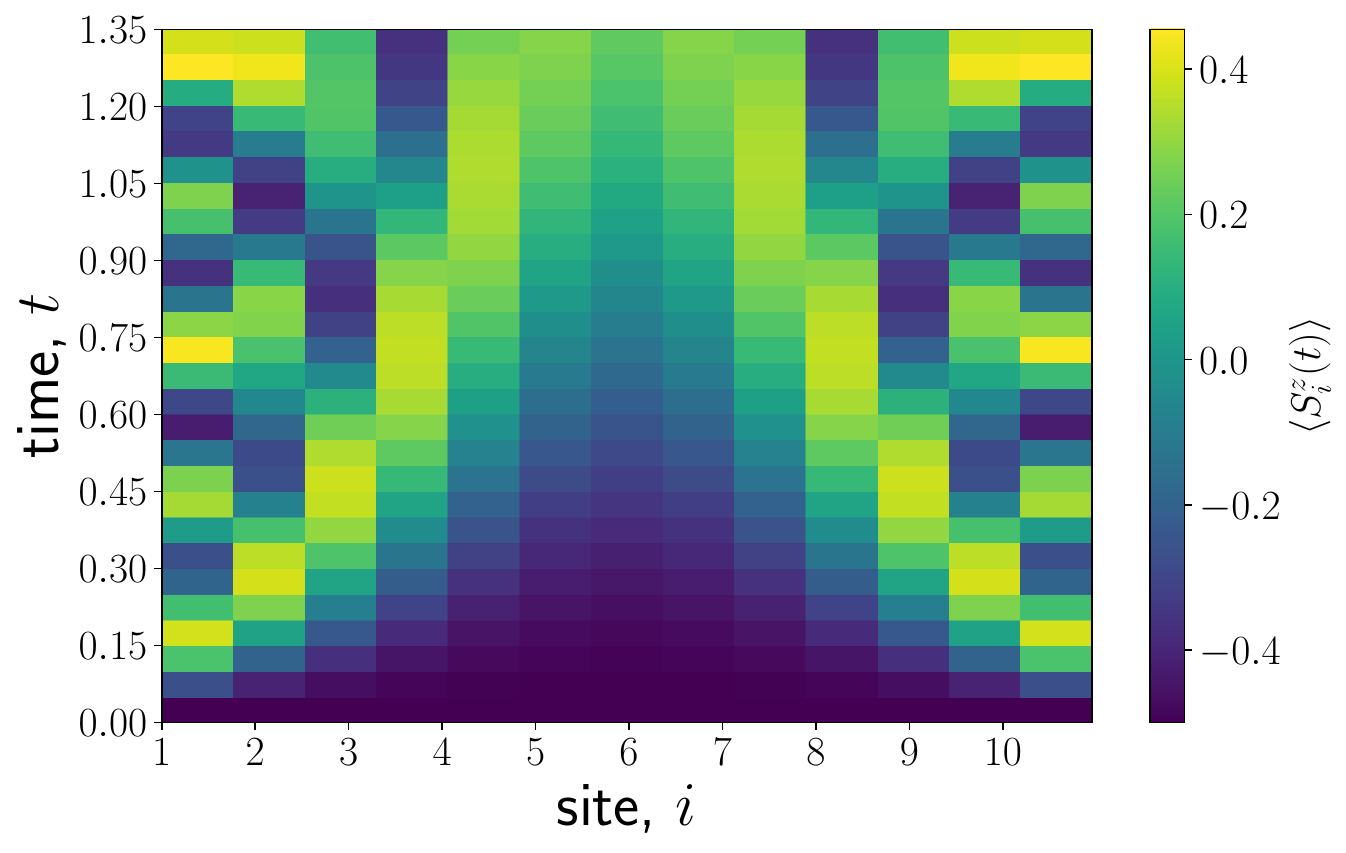}
		\caption{Trotter evolution of local magnetization $\langle S^z_i(t)\rangle$ with exact diagonalization. Parameters: $N=13$, $J=2.0$, $h=1.05$, $l_{\rm max}=3.0$. }\label{ideal}
	}
\end{figure}

\begin{figure}
	{
		\includegraphics[scale=.4]{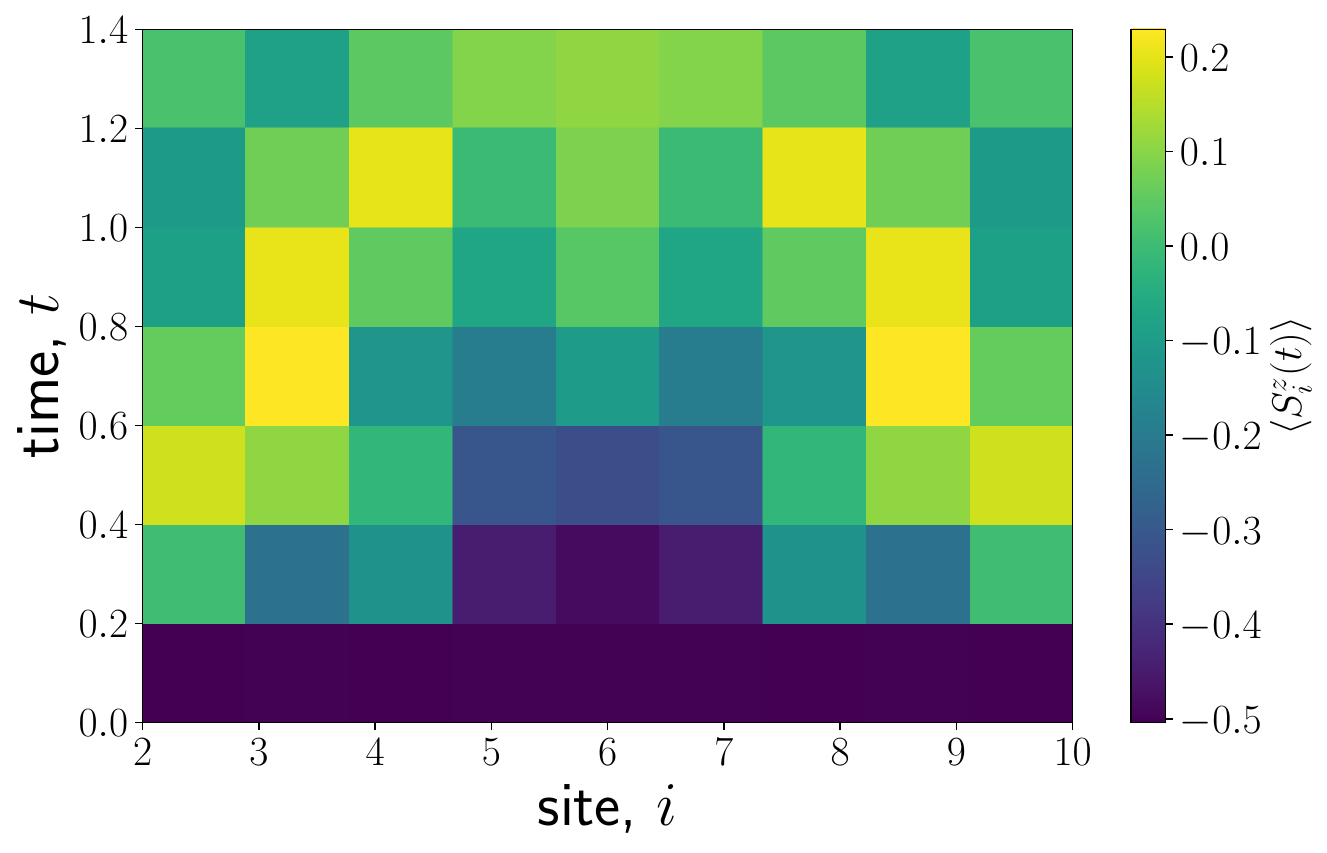}
		
	}
	\caption{ Trotter evolution of local magnetization $\langle S^z_i(t)\rangle$ computed using guadalupe Quantum Processing Unit (QPU). Parameters: $N=13$, $J=2.0$, $h=1.05$, $l_{\rm max}=3.0$. Magnetization data on the edges of the lattice chain are omitted due to the large Trotter-error associated with it. Note that the deformation strength is stronger on the edges. }\label{noisy_mitigated}
	\label{szq_2d}
\end{figure}

In this section, we show results on the time evolution of the magnetization $\langle S^z\rangle =\frac{1}{2}\langle \sigma^z\rangle$ computed using tensor methods compared with simulation on quantum devices. 

We start by time evolving the system using the Time Evolving Blocked Decimation (TEBD) algorithm \cite{suzuki1976generalized}. Historically, TEBD was adapted 
from the Suzuki-Trotter approximation for the Matrix Product State (MPS) \cite{vidal2003efficient}. In Fig.~\ref{fig:tebd_sz}, the Trotter evolution of the magnetization $\langle S^z_i (t) \rangle$ is plotted at each lattice site $i$ for a lattice chain with $N=37$ sites, and $l_{\mathrm{max}}=3.0$, $h=2.0$, $J=2.0$, and $m=0.25$ starting with all spins in the down state. Clearly,
the dynamics of the magnetization
shows warping effects in the bulk due to the curved background. One can think of
this warping effect as due to time dilation effects in the bulk. 

Next, we attempt to investigate the model using a quantum platform -- namely
the IBM Guadalupe machine. Currently, quantum devices experience both
large coherent and incoherent noise in any given computation. Thus, we 
have attempted to investigate a system with modest system size of $N=13$ spins where
there is limited device noise and the warping effects can be observed. Using the quantum circuit representation of operations given in Fig.~\ref{digital_ckt},  we have computed the time-dependent expectation value of the magnetization  $\langle S^z_i (t)\rangle$ at site $i$ using a first order Trotter approximation.  Different orderings of the operators can be used for this approximation, see the discussion in the Appendix~\ref{ordering}. In this section, all the results presented use what we denote as `odd-even' ordering in the Appendix.  

Local magnetization results are shown for three different sites in the Fig.~\ref{fig_local_m} and compared against classical simulation results obtained from TEBD. The parameters used were $J=2.0$, $h=1.05$ and $l_{\rm max}=3.0$. The gate cost of such a circuit is similar to that of the Ising spin chain on a flat lattice \cite{gustafson2021indexed}. The difference in our Trotter evolution of the deformed Hamiltonian lies in the site dependent phase factors of the rotation and entangling gates. This brings an inherent complication to the problem of selecting the optimal Trotter step $\delta t$. Previous studies have shown that theoretical bounds of the first-order Trotter approximation can be relaxed for observing time evolution with current NISQ-era machines \cite{gustafson2021indexed,meurice2021quantum,asaduzzaman2022quantum}. The phases ($\theta_i$, $\phi_i$) of the rotation and entangling gates are of the form $ C_i \times \delta t$ and the optimal choice for the Trotter step is different for local operators $\langle S_z^i \rangle$  at different sites. Thus, one constraint for choosing the optimal Trotter step $(\delta t)_{\rm optimal}$ comes from the local couplings $C_i$. In NISQ-era devices, the other constraint comes from the maximum possible circuit depth $d_{\rm max}= n\times \delta t$ that can be simulated before the noise swamps the signal. The accumulation of gate errors effectively limits our ability to go beyond a certain number of Trotter steps. We found that a compromise value of $(\delta t)_{\rm optimal} \sim 0.2$ and $t_{\rm max}\sim 1.2$ is a good choice for time evolution of the system. For the computation of the local magnetization, the number of shots used is $N_{\rm shots}=1000$. See Appendix.~\ref{shot_error} for a discussion of the statistical noise associated with different  $N_{\rm shots}$.

In Fig.~\ref{fig_local_m}, classical simulation results of the local magnetization with the TEBD algorithm are compared with the mitigated results obtained from the Guadalupe machine. The error-bars in the figures represent statistical errors associated with six different measurements. The measurements were performed on different days to demonstrate reliable systematic error on the current devices. Various error mitigation techniques were applied to obtain the results. 
Dynamical Decoupling (DD) \cite{viola1998dynamical,charles2023simulating} was applied to reduce the coherent noise and the Mthree (M3) method \cite{nation2021scalable} was used to reduce readout errors. We also created noise-scaled circuits with three-fold and five-fold amplification of
the noise in comparison to the original circuit and applied the Zero Noise Extrapolation (ZNE) \cite{temme2017error,PhysRevX.7.021050} mitigation technique to reduce the incoherent noise. We used the built-in features of the IBM runtime system
to apply DD and M3 while noise-scaled circuits were created by inserting an appropriate number of identity operators for each CNOT gate. This choice is justified for current IBM devices, where two-qubit gates have significantly larger errors than single-qubit rotation gates\footnote{For the Guadalupe machine, the ratio of the median-errors for the two-qubit and single-qubit gates is $\frac{\epsilon_{\mathrm{CNOT}}}{\epsilon{1qubit-gate}} \sim 25$.}.  See Appendix.~\ref{mitigation} for the discussion of how different error mitigation techniques improved our results. 

After post-processing the data with different error mitigation techniques, we found that the magnetization results obtained from the Guadalupe machine Fig.~\ref{szq_2d} show the warping effects. For comparison purposes, the TEBD results are plotted in Fig.~\ref{ideal}.
The CNOT gate cost for computing time-evolution with first order Trotter approximation of a $N$-qubit quantum spin chain is $2(N-1)$ per Trotter-step and the circuit depth at Trotter step $n=6$ is $d=48$.
The results from the QPU track the peak of the local magnetization quite well. The QPU results also demonstrate that the initial state with all-down spins is disrupted by the boundary at a slower rate as we move from the edge to the center of the lattice chain. While the quantum simulation results align qualitatively with tensor methods, it is clear that larger numbers of qubits
would be needed to identify the warping effects in a greater detail. We 
have also explored a possible implementation of the real-time magnetization evolution on QuEra's analog quantum computers based on Rydberg arrays. See Appendix.~\ref{ryd_app} for the discussion of the analog computation of the local magnetization.

\section{\label{scrambling}Out-of-time-ordered Correlators}

We now turn to the question of how information spreads
in the model.  To answer that, we computed an out-of-time-ordered-correlator (OTOC). 
This observable is known to capture information spread and scrambling in quantum systems \cite{Swingle:2018ekw,PhysRevB.97.144304,Garcia-Mata:2022voo} and can be thought of as a quantum mechanical 
counterpart of the Loschmidt echo \cite{goussev2012loschmidt}. To construct the OTOC, we
use two operators $W_i(t)$ and $V_j$ where  $W(t)=\exp^{iHt}W(0)\exp^{-iHt}$.
From these we construct the commutator of these operators
\begin{equation}
	C(t) = \langle ||[W_i(t),V_j]||^2 \rangle=2(1-\mathrm{Re}[F_{ij}(t)]), \label{double_comm}
\end{equation}
where
$F_{ij}(t)$ is the the required out of time ordered correlator (OTOC)
\begin{equation}
	F_{ij}(t) = \langle W_i(t)^{\dagger}V_j(0)^{\dagger} W_i(t). V_j(0) \rangle.\label{F_def}
\end{equation}  
This equality is obtained under the assumption that $W$ and $V$ are unitary and that terms that correspond to local observables thermalize to a constant after a short time and hence
can be omitted.  
The connection between $F_{ij}(t)$ and the information spread can be 
made clear by considering $W$ as a simple local perturbation. Under time evolution this 
Heisenberg operator becomes more and more non-local. The growth of these non-local effects can be captured by calculating the commutator of $W(t)$ with another local operator $V$.
When the operators commute, $C$ vanishes and $F$ is one. So by measuring the double commutator or the OTOC we can track the propagation of $W(t)$ along the system.

The relationship between the double commutator and operator growth can be made clear by considering a simpler setup. Let's start by representing a unitary time evolution operator out of local two qubit unitaries. Using this representation we can obtain the Heisenberg time evolution for a local operator $A(t)=U^\dagger A U$. 

\begin{figure}[!ht]
	\centering
	\includegraphics[width=0.50\textwidth]{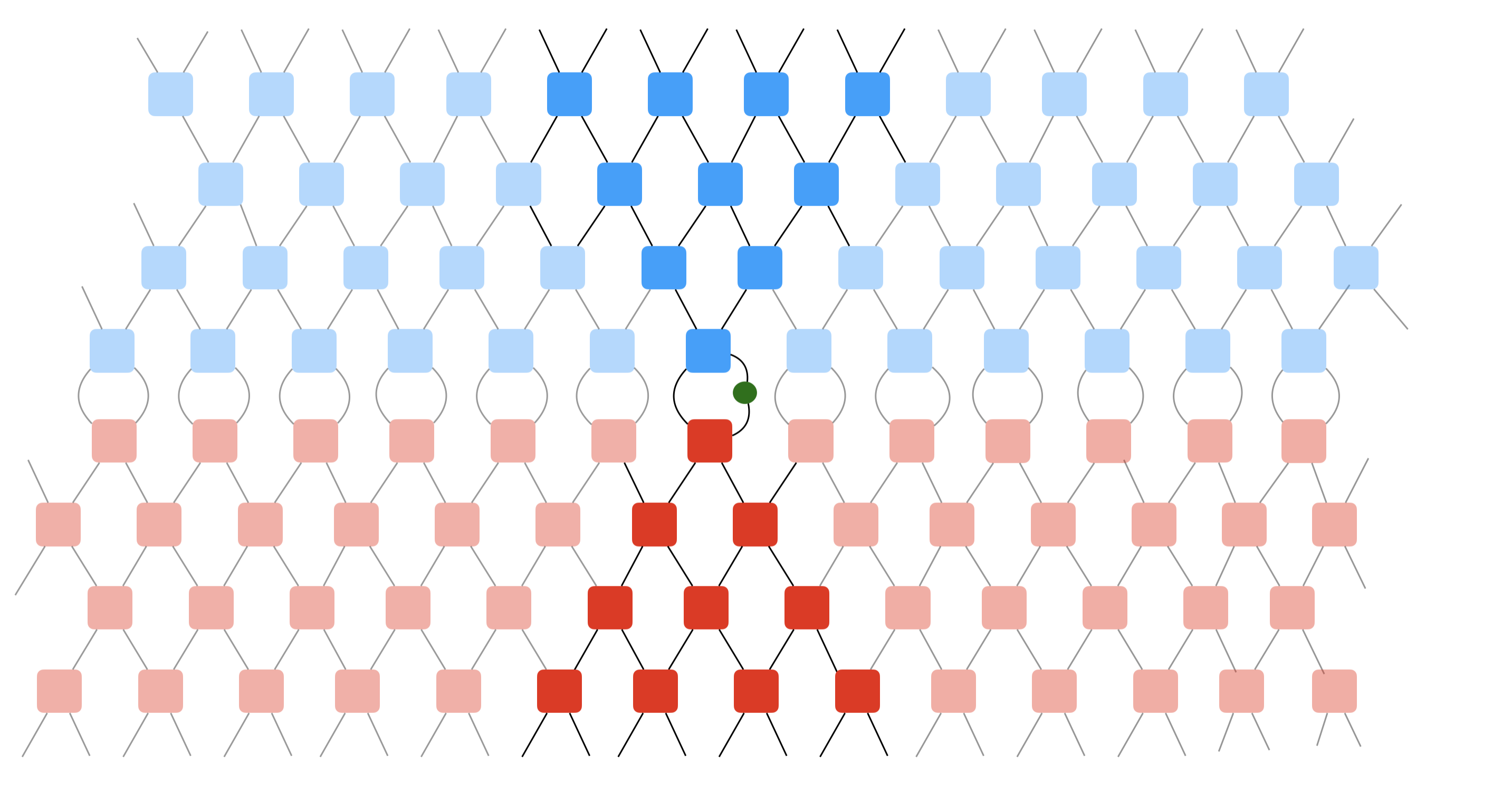}
	\caption{Heisenberg time evolution for a local operator.}
	\label{fig:lightcone_A}
\end{figure}
Where in Fig.~\ref{fig:lightcone_A} blue and red boxes represent $U^\dagger$ and $U$ while the green circle represents the operator $A$. One can clearly see from the above figure that any contraction that doesn't involve the operator $A$ will be the identity so we can ignore those and focus on the contractions that involve the operator. This clearly shows us the lightcone for the operator growth in the Heisenberg picture and demonstrates that the OTOCs capture the characteristics of the operator spread in the system. 

However, this general form of the OTOC is not the easiest to deal with in our simulations. 
Instead, we choose the following form for the OTOC  operator 
which can be seen from Eq.~\ref{otoc1} \cite{vermersch_probing_2019,joshi_probing_2022} 
\begin{equation}
	O_i(t) = \frac{{\rm Tr}(\rho W(t)_{\frac{N+1}{2}} V_i^\dagger 
		W(t)_{\frac{N+1}{2}} V_i)}{{\rm Tr}(\rho W(t)^2 V^\dagger V)}. \label{otoc1}
\end{equation}

In our calculations, we take $W(t)=\sigma^z(t)$, $V=\sigma^z$ 
and fix the position of $W(t)$ operator at the center of the lattice chain. To see the effect of the interaction of two local operators, we then place the 
operator $V$ at different lattice sites $i$. We have focused on the infinite temperature limit which corresponds to taking a density matrix $\rho \sim \mathbf{I}$ in Eq.~\ref{otoc1}. Infinite-temperature OTOCs bear the signature of entanglement growth after a quench is applied to an energy eigenstate \cite{fan2017out} and are easier to compute.
Furthermore, many of the protocols used in finite-temperature-OTOCs can be developed from the
corresponding protocols used in the infinite temperature case \cite{sundar2022proposal,vermersch_probing_2019}. Additionally, the exponents computed from the infinite-temperature OTOCs are insensitive to slightly different OTOC definitions that exist in the literature, see the appendix in \cite{sundar2022proposal}.

\subsection{\label{classical_OTOC}Classical Simulations of OTOCs}
\begin{figure}[!ht]
	\centering
	\includegraphics[width=0.50\textwidth]{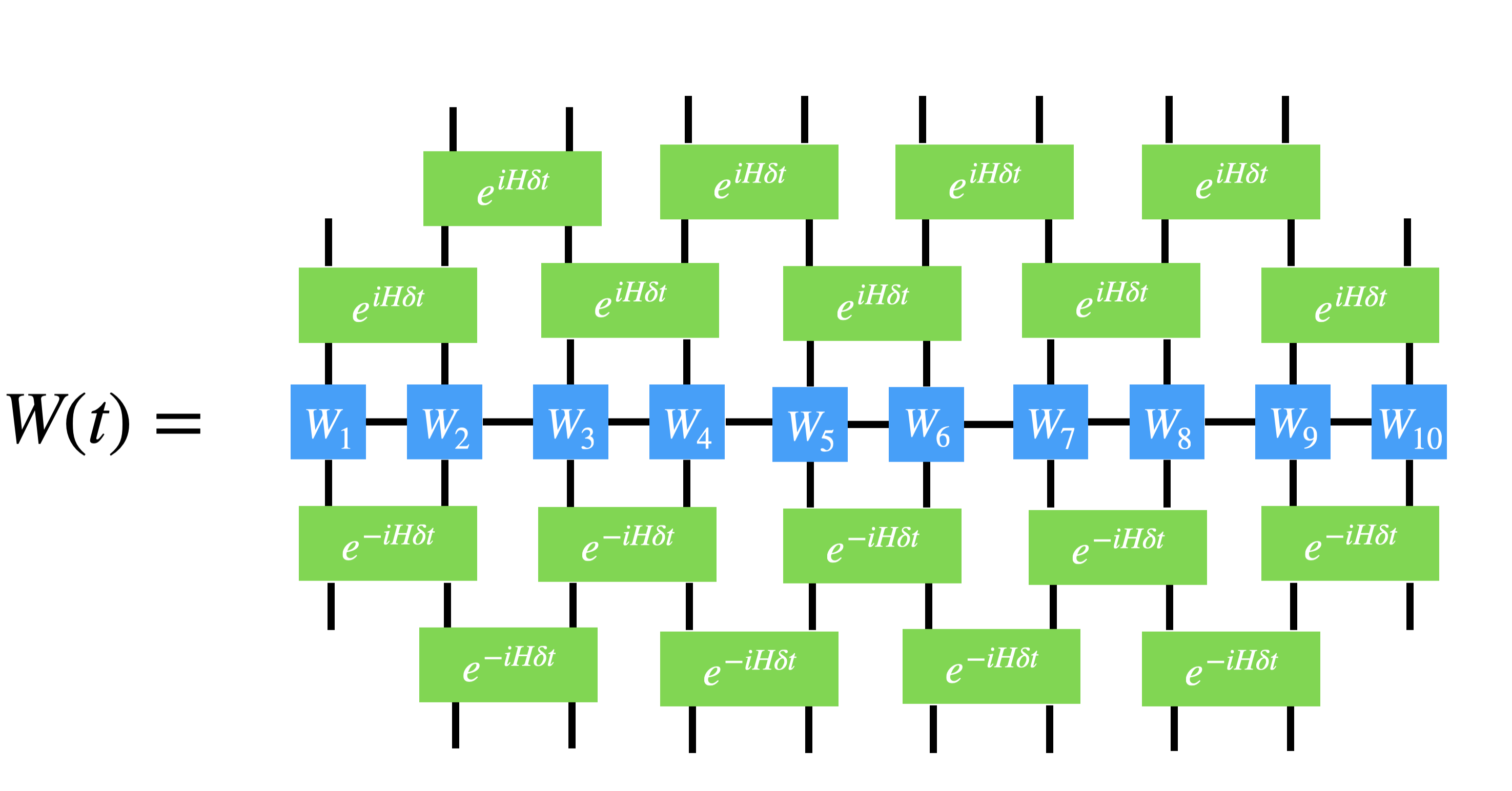}
	\caption{Heisenberg time evolution for W(t)}
	\label{fig:mpo_heisenberg}
\end{figure}

For the classical computation of the OTOC
the time evolution for $W(t)$ is obtained by expressing $W$ and $V$ as
matrix product operators (MPO) and applying 
Heisenberg time evolution to $W$ via the TEBD algorithm. 
In Fig.~\ref{fig:mpo_heisenberg}, we show how to apply Heisenberg 
time evolution to a generic operator $W$ for one Trotter step. 

Then the resulting time evolved operator $W(t)$ can be plugged into the OTOC calculation.

Initially we show  results for the flat space transverse Ising model and
as expected  we see a linear light cone.

\begin{figure}[!ht]
	\centering
		\includegraphics[scale=.25]{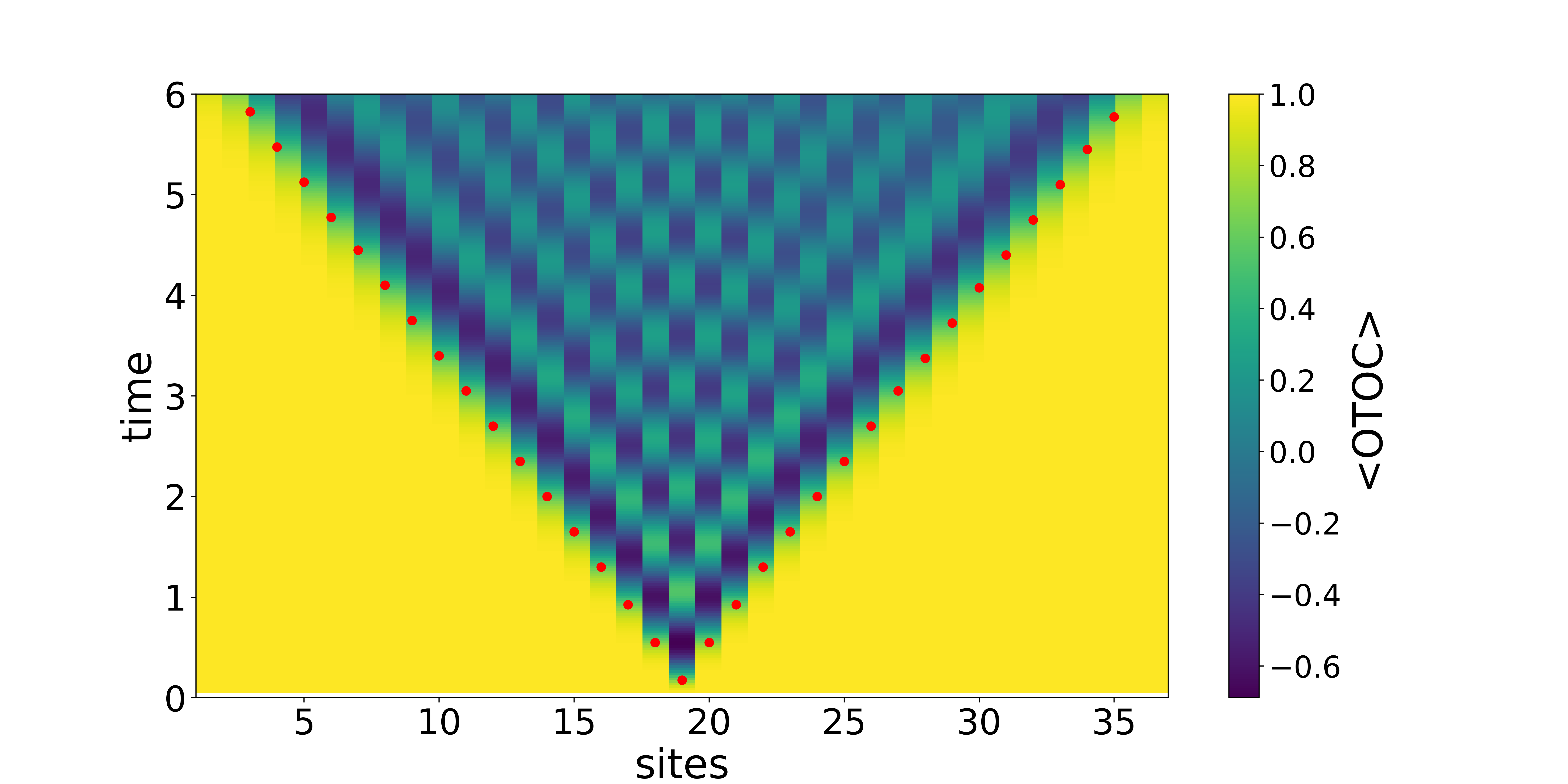}
	\caption{$O_i$ for $l_{\rm max }=0.0,J=6.0$}
	\label{fig:otoc_in_l0}
\end{figure}

\begin{figure}[!htb]
	\centering
	\includegraphics[scale=.25]{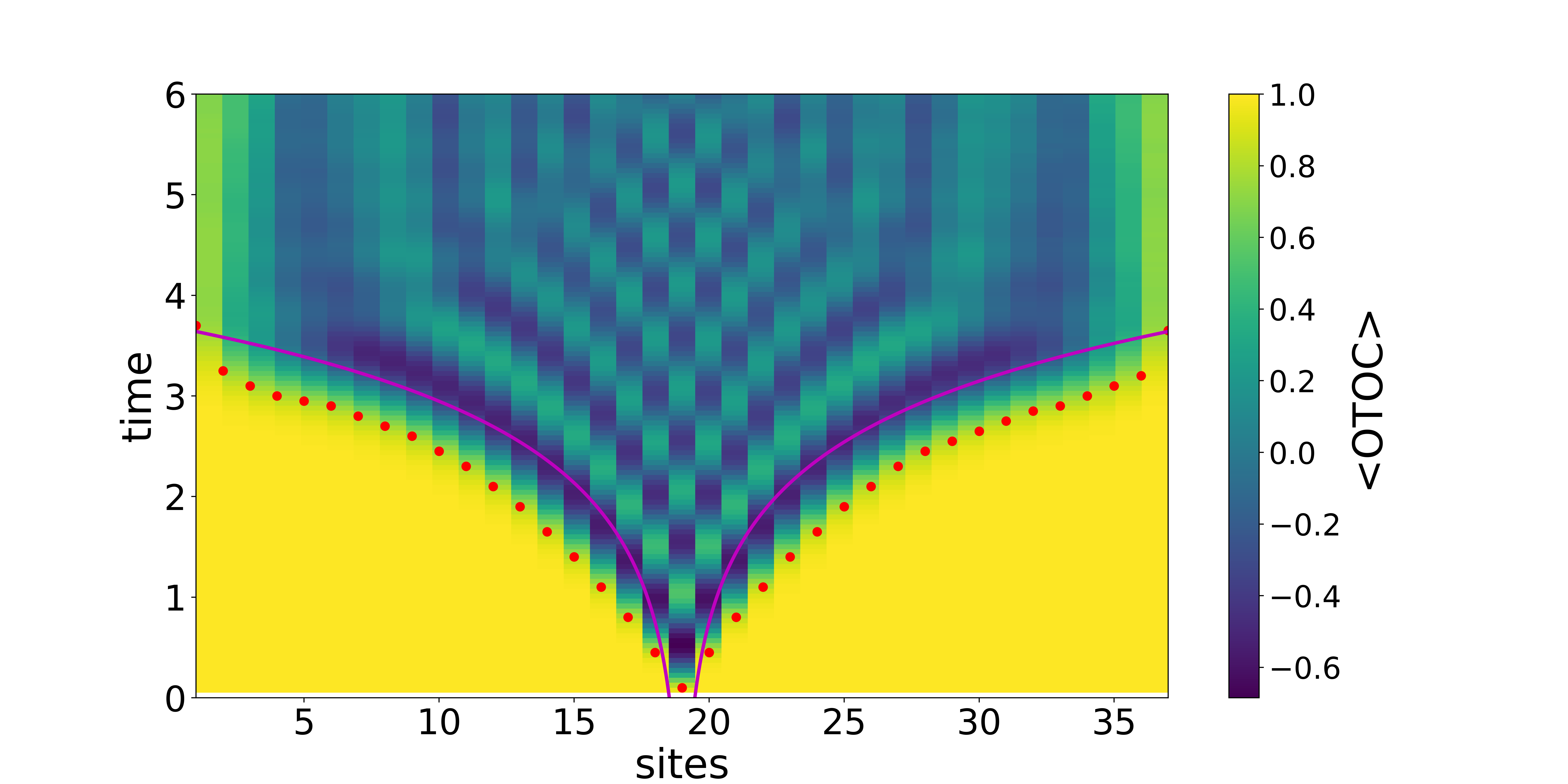}
	\caption{$O_i$ for $l_{\rm max}=3.0,J=6.0$}
	\label{fig:otoc_in_l3}
\end{figure}

If we turn on the hyperbolic deformation and set $l_{\rm max}=3.0$ we see that the system develops a warped lightcone and faster than linear propagation of information. 
In Fig.~\ref{fig:otoc_in_l0}, Fig.~\ref{fig:otoc_in_l3} and the rest of the plots of OTOCs, the
red dots represent the times where the OTOC at that lattice site first deviates 
from $1.0$ by some amount $\epsilon$. Here $\epsilon=0.25$. 
These resultant points trace out the lightcone shown in the plot.
The purple line which is shown to guide the eye corresponds to a  curve of the form  
\begin{equation*}
	t= \log \Big|x-\frac{N+1}{2}\Big|+B,
\end{equation*}
where, $B$ is a constant.  

\begin{figure}[!htb]
	\centering
	\includegraphics[width=0.50\textwidth,height=0.25\textwidth]{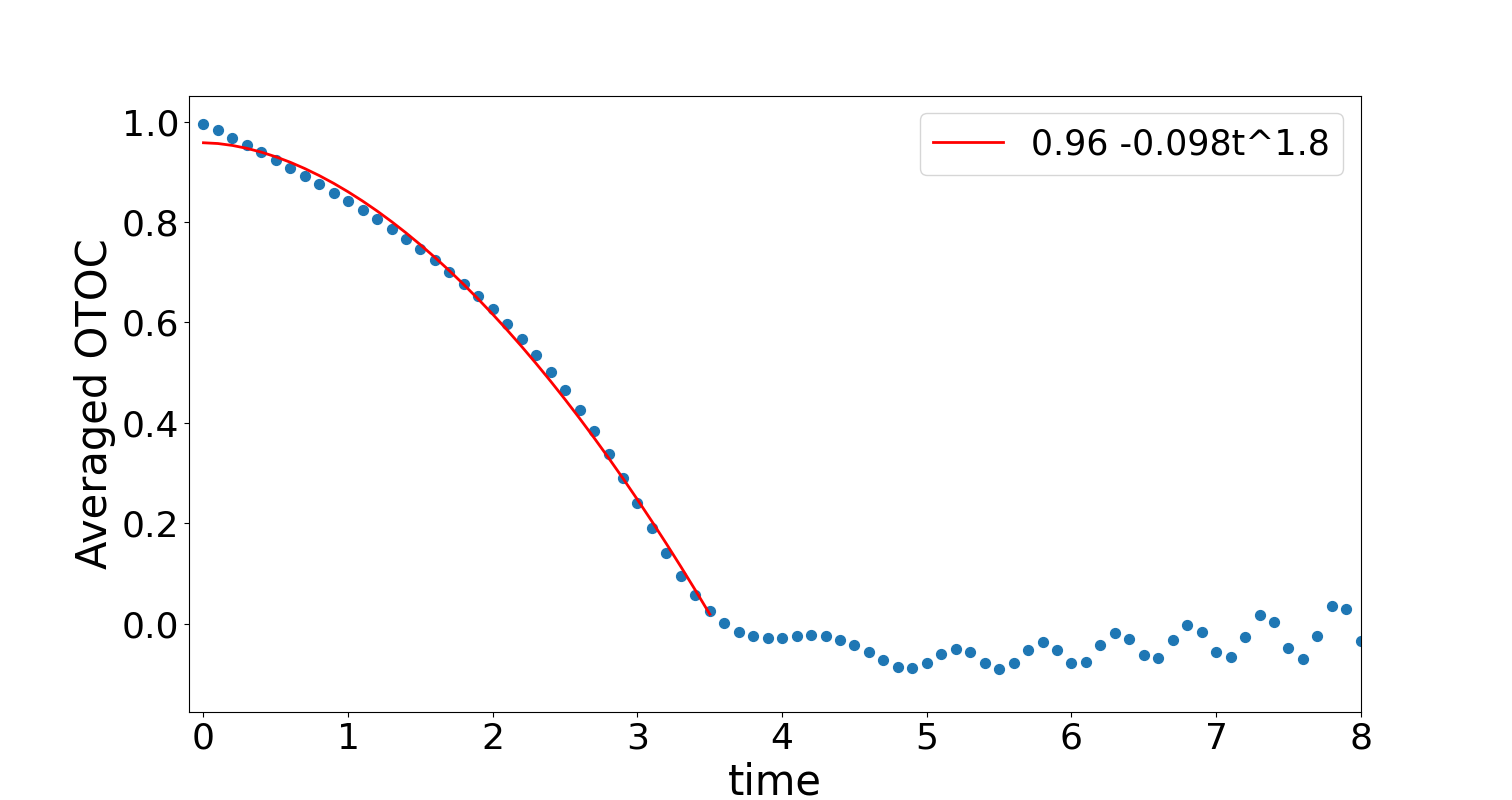}
	\caption{Site averaged OTOC for $J=6, h=3.05, m=0.25, l_{\rm max}=3.0$}
	\label{fig:av_otoc}
\end{figure}

\begin{figure}[!htb]
	\centering
	\includegraphics[scale=.3]{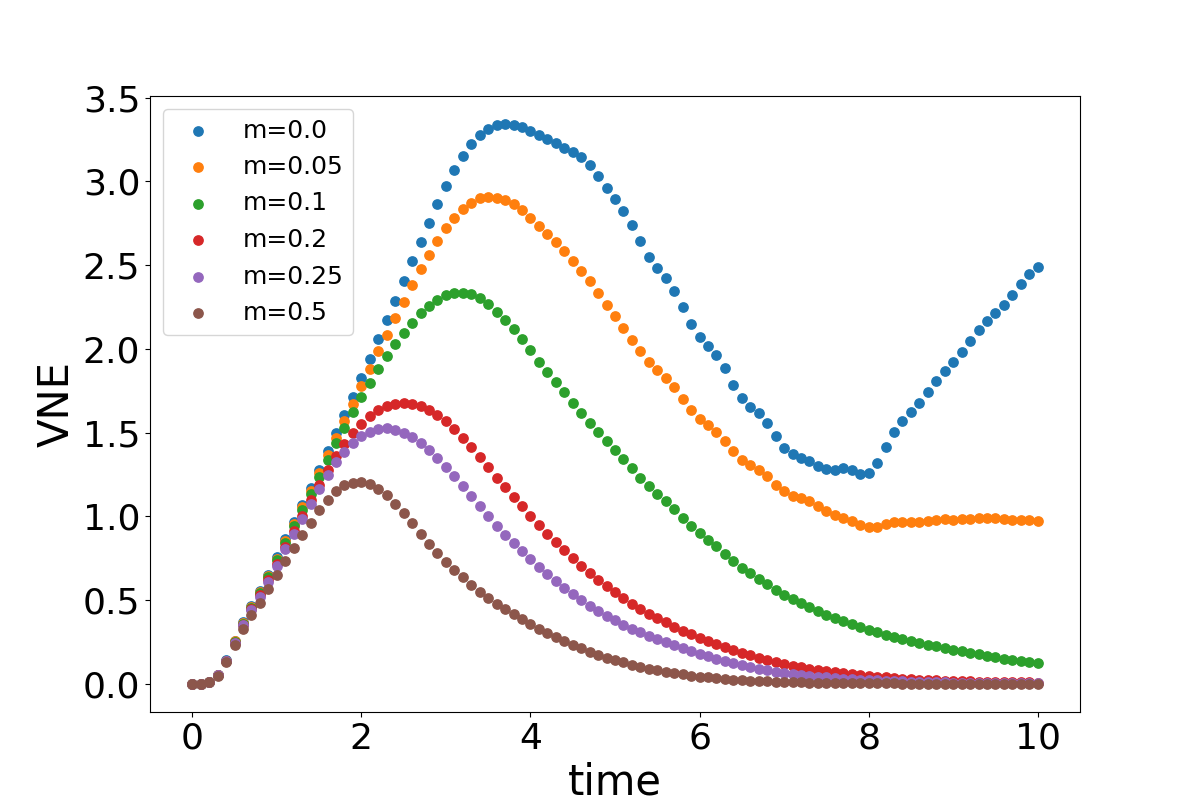}
	\caption{von Neumann entropy for $J=6, h=3.05, l_{\rm max}=3.0$}
	\label{fig:ent_mps}
\end{figure}

We found that to access the logarithmic regime of the 
model the physical couplings $J$ and $h$ need to be tuned to be close to their critical values. 
The remaining physical coupling $m$ then controls the thermalization dynamics. In Fig.~\ref{fig:ent_mps} we plot the time evolution of the half-chain von Neumann entropy which shows how $m$ controls the thermalization. We can also look at the site-averaged OTOCs which 
are plotted in Fig.~\ref{fig:av_otoc}. This clearly show a power-law dependence on $t$ as the system thermalizes. 

\begin{figure}[!htb]
	\centering
	\includegraphics[width=0.50\textwidth]{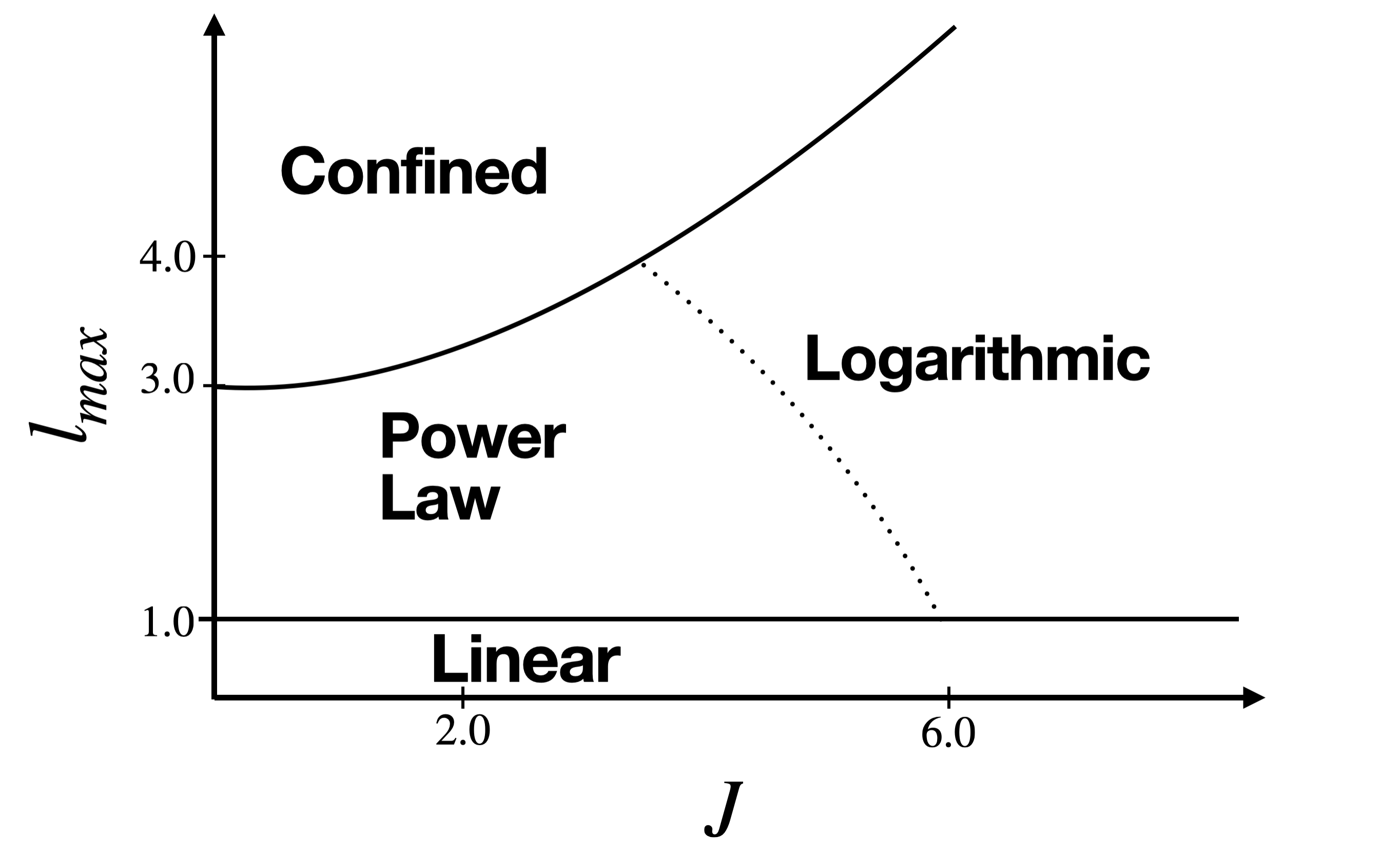}
	\caption{OTOC Phase Diagram for $N=37$}
	\label{fig:phase_space}
\end{figure}

Note that the 
value of $m$ doesn't affect the structure of the light cone 
and only controls the thermalization time. 
In fact the shape of the lightcone is 
determined by the value of $l_{\rm max}$. For $N=37$ we found four distinct
behaviors for the lightcone. For $0.0<l_{\rm max}<1.0$ we find a linear lightcone. 
Then for $1.0<l_{\rm max}<2.0$ we see a power-law behavior while
for $2.0<l_{\rm max}\leq 3.0$ the light cone
takes on a logarithmic behavior. Finally for $l_{\rm max}>4.0$ the system 
confines and a deformation that's been initialized in the bulk 
never reaches the  boundaries of the chain. 
We summarize this structure in  the following cartoon of the OTOC phase diagram of the model for $N=37$ Fig.\ref{fig:phase_space}.

\begin{figure}[!h]
	\centering
	\includegraphics[scale=.3]{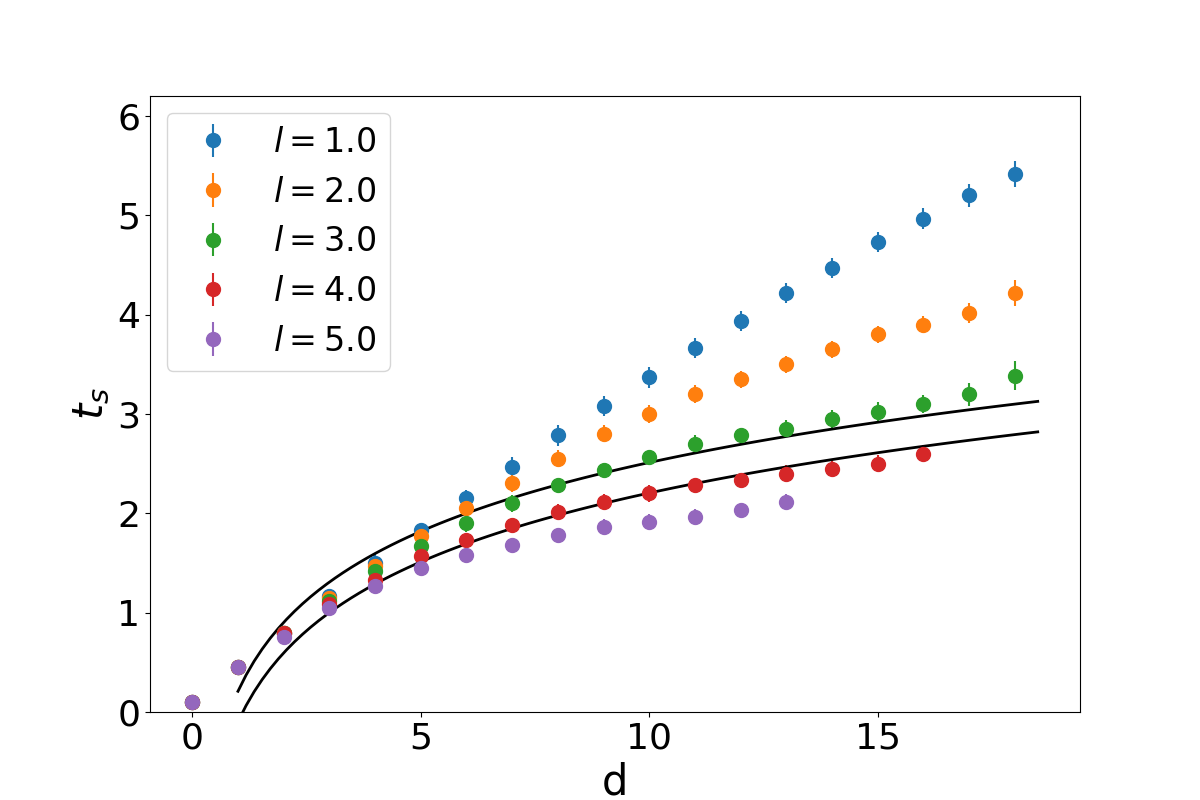}
	\caption{Curvature dependence of the propagation behavior of OTOC for $N=37,J=6.0, h=3.05, m=0.25$}
	\label{fig:scram_l}
\end{figure}

The dependence on $l_{\rm{max}}$ can be clearly seen in Fig.\ref{fig:scram_l} where 
we plot the local light-cone time obtained from OTOC calculations vs the lattice site,
starting from the middle of the chain and ending at the first site. 
The black curves show the logarithmic fits for $l\geq 3.0$. 
Error bars on the points are obtained by taking multiple cutoff values and averaging over them. 

\begin{figure}[!h]
	\centering
	\includegraphics[scale=.25]{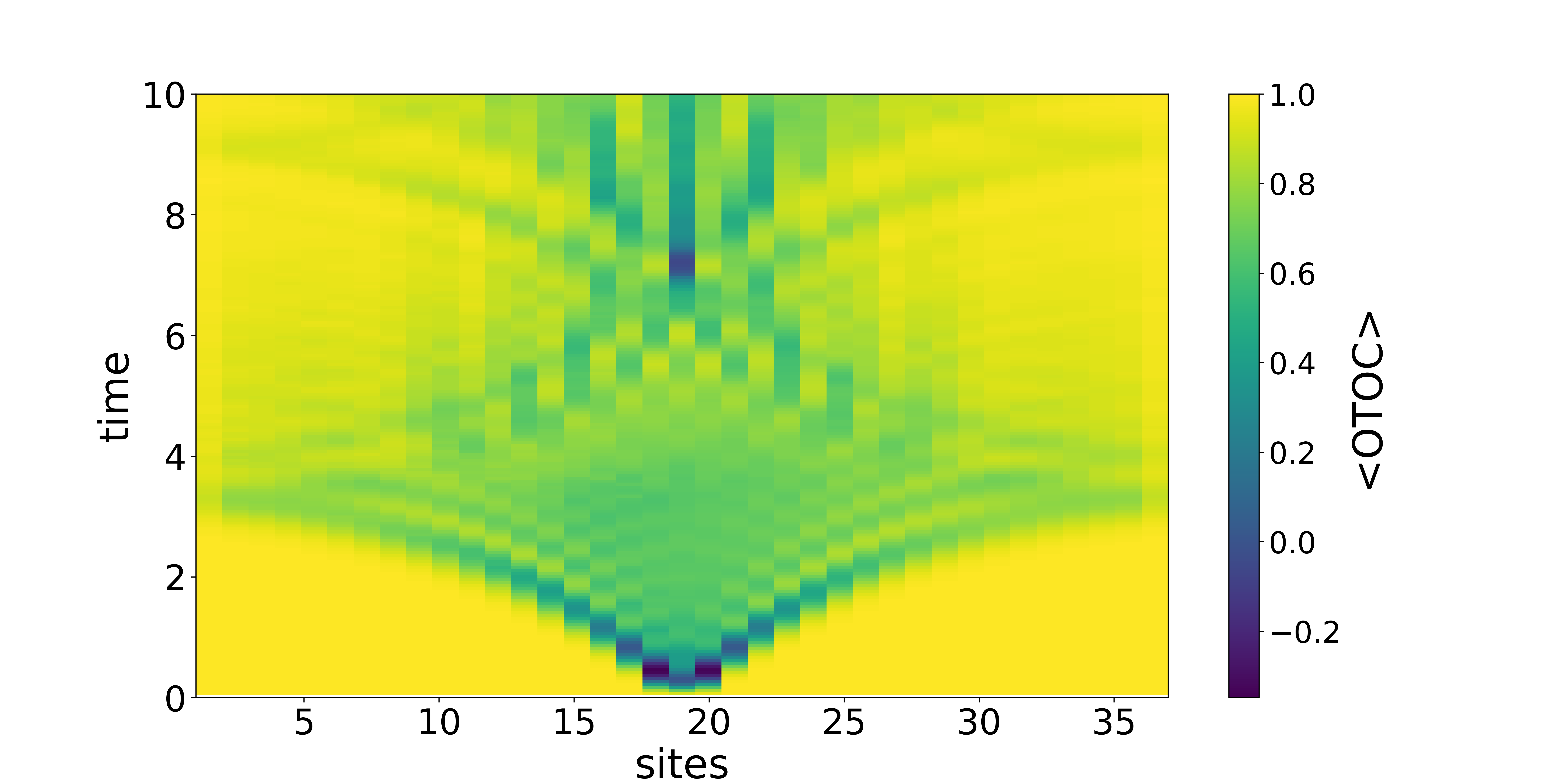}
	\caption{$XX$-OTOC for $N=37,J=6.0, h=3.05, m=0.25$}
	\label{fig:otoc_xx}
\end{figure}

Even though we focused solely on the  choice of $W(t)=\sigma^z(t)$ and $V=\sigma^z$ for the OTOC calculations it is possible to choose other combinations of operators. One such choice corresponds to taking $\sigma^x$ operators for both $W(t)$ and $V$ operators which results in the plot shown in Fig.\ref{fig:otoc_xx}. This behavior
of the $XX$-OTOC is consistent with previous computations of  OTOCs in the flat-space transverse Ising model by Lin and Motrunich in \cite{PhysRevB.97.144304} where they found a shell like structure for the $XX$-OTOC operator. As can be seen from the Fig.~\ref{fig:otoc_xx} the points inside the light cone are significantly less prominent as compared their $ZZ$ counterparts. 

\begin{figure}[!h]
	\centering
	\includegraphics[width=0.5\textwidth]{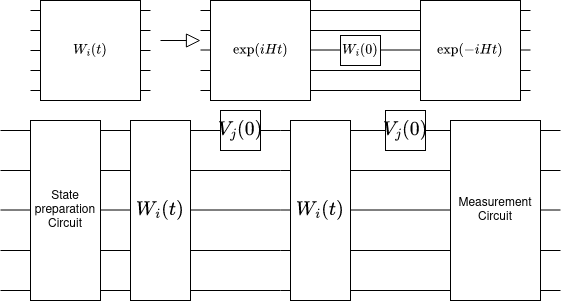}
	\caption{Schematic circuit diagram of OTOC using the definition at Eq.~(\ref{otoc_alt_def}).}
	\label{otoc_plain_ckt}
\end{figure}

\begin{figure*}[!htb]
	\subfloat[\label{local_protocol1}]{%
		\includegraphics[height=0.3\textwidth]{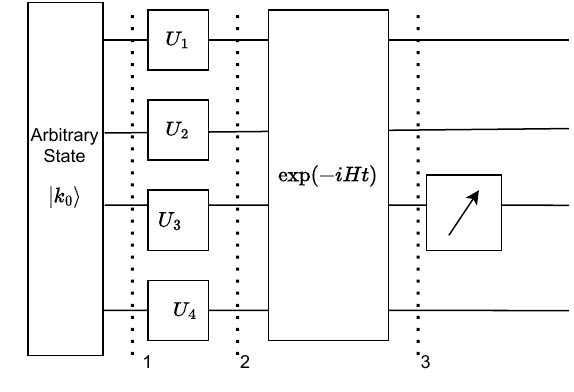}%
	}\hfill
	\subfloat[\label{local_protocol2}]{%
		\includegraphics[height=0.3\textwidth]{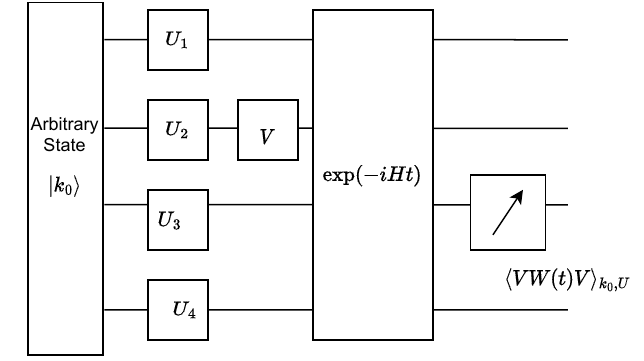}%
	}
	\caption{Modified OTOCs are computed from the correlation of the measurement of two different operators (a) $\langle W(t) \rangle $ and (b) $\langle V^\dagger W(t) V \rangle$. The same set of unitaries are required to find the correlation between the measurements. The procoess is repeated for many different sets of unitaries.}
	\label{fig_OTOC_local_protocol}
\end{figure*}

\begin{figure*}[!htb]
	\subfloat[\label{corr_t0}]{%
		\includegraphics[width=.3\textwidth]{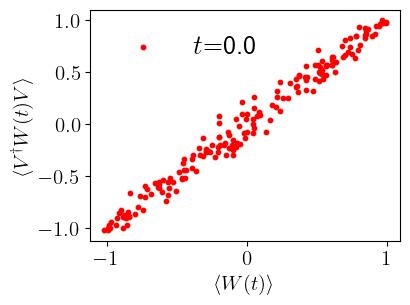}%
	}\hfill
	\subfloat[\label{corr_t1.5}]{%
		\includegraphics[width=.3\textwidth]{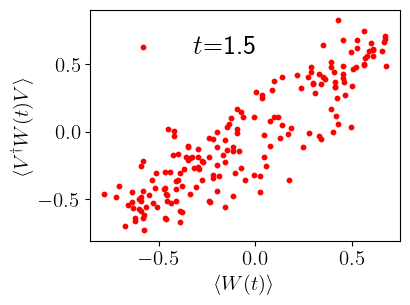}%
	}\hfill
	\subfloat[\label{corr_t3.5}]{%
		\includegraphics[width=.3\textwidth]{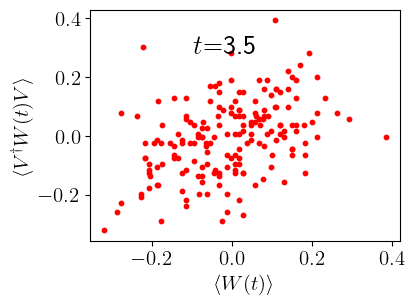}%
	}
	\caption{Change in correlation of the operators $\langle W(t) \rangle=\langle\sigma^z_3(t)\rangle$ and $\langle VW(t)V \rangle=\langle \sigma^z_2\sigma^z_3(t)\sigma^z_2 \rangle$ over time. Parameters: $l_{\rm max}=3.0,\,J=-0.5$, $h=-0.525$, $W(t)=\sigma^z_3(t)$, and $V=\sigma^z_2$.}
	\label{correlations}
\end{figure*}

\begin{figure}[!h]
	\centering
	\includegraphics[width=0.5\textwidth]{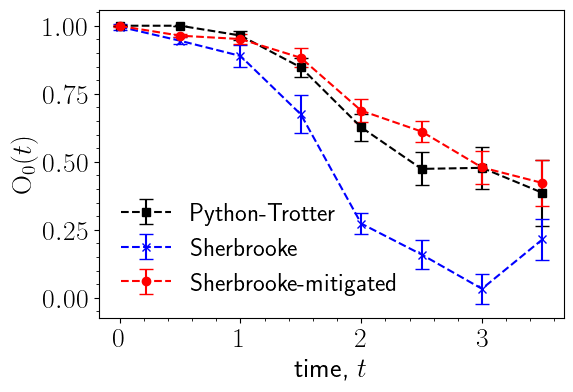}
	\caption{Modified OTOC of the zeroth order, $O_{0}(t)$ for $l_{\rm max}=3.0,\,J=-0.5$, $h=-0.525$, $W(t)=\sigma^z_3(t)$, and $V=\sigma^z_2$.}
	\label{mod_otoc_plot1}
\end{figure} 

\begin{figure}[!h]
	\centering
	\includegraphics[width=0.5\textwidth]{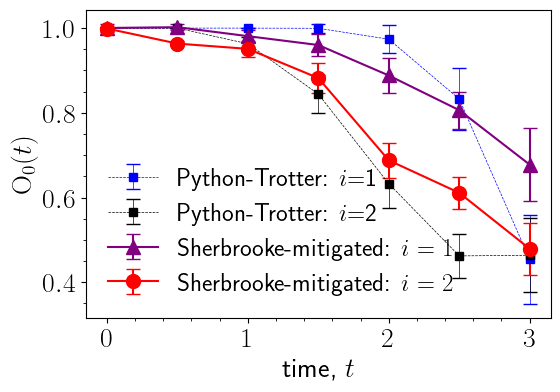}
	\caption{Modified OTOC as the position $i$ of the  $V$ operator varies. Parameters: $l_{\rm max}=3.0,\,J=-0.5$, $h=-0.525$, $W(t)=\sigma^z_3(t)$, and $V=\sigma^z_i$.}
	\label{mod_otoc_vary_Vqubit}
\end{figure} 

\subsection{\label{qm_OTOC}Quantum Simulation of OTOCs}

In this subsection, the computation of the OTOC with digital quantum computers is discussed. First, let us write down an alternative definition of the OTOC for a $N$-qubit system
\begin{equation}
	O_i^{\rm eig}(t) = \frac{\bra{\psi} (W(t)_{\frac{N+1}{2}} V_i^\dagger W(t)_{\frac{N+1}{2}} V_i)\ket{\psi}}{\bra{\psi}(W_{\frac{N+1}{2}}(t)^2 V_i^\dagger V_i)\ket{\psi}},\label{otoc_alt_def}
\end{equation}
where $\ket{\psi}$ represents an arbitrary state. The schematic circuit diagram to compute this quantity is shown in the Fig~\ref{otoc_plain_ckt}. From this schematic diagram and the discussion of the Trotter evolution in the previous section, it is evident that to compute the OTOC with Trotterized evolution operator requires $8n(N-1)$ CNOT gates for the n\textsubscript{th} Trotter step. In our work we considered a spin chain of length $N=7$, and used a trotter step $\delta t =0.5$ up to a maximum time $t_{\rm max}=3.5$. This indicates that a quantum computation of the OTOC with a quantum circuit like that of Fig.~\ref{otoc_plain_ckt} would require more that 200 CNOT gates in just four Trotter steps. Hence extracting any useful results would
become impossible at early times due to coherent and incoherent noise in the device. Using a weaved Trotterization technique, similar circuits were implemented to compute OTOCs for a small system of four qubits in \cite{geller2022quantum}.

Our goal in this section
is to investigate if we can extract the scrambling time at infinite temperature ($\rho \propto \mathbf{I}$) with the current IBM devices for a system with $7$ spins. As for
tensor network
simulations we position the $W$ operator at the center of the lattice chain 
and vary the position $i$ of the $V$ operator. Our choice for the $W$ and $V$ operators 
remain the same as that of the previous section. With quantum simulation, we also would like to see if the simulation can identify the difference in the scrambling time as we vary the position of the $V$ operator.  Many protocols for computing OTOCs have been proposed \cite{yao2016interferometric, vermersch_probing_2019, joshi2020quantum, swingle2016measuring,halpern2017jarzynski,PhysRevA.98.012132} and many authors have also suggested some modified quantities that also contain scrambling information  \cite{geller2022quantum,swingle2016measuring}. For example, to reduce the computational cost the magnitude-squared of OTOC ($|F|^2$, see Eq.~(\ref{F_def}) for definition of $F$) can be computed ignoring the phases \cite{swingle2016measuring}. In this paper, we have used the protocol 
proposed by Vermersch \textit{et. al.} to compute both the  OTOC and the  modified OTOC \cite{joshi_quantum_2020}.  The gate cost per circuit for computing the modified OTOC of zeroth order using this protocol is  $\sim 2n(N-1) $, which is significantly lower than the gate-count needed in the straightforward evaluation presented by Fig.~\ref{otoc_plain_ckt}. Also the protocol we have chosen does not require any ancilla qubits unlike some other OTOC computation protocols.\\

Vermersch \textit{et. al.}\cite{vermersch2019probing} discussed a `global protocol' to compute the OTOC and a `local protocol' for computing modified OTOCs. Both 
protocols require state preparation of random states created from random unitary operators. The idea is to sample enough random states to mimic a thermalized scenario for the computation of the OTOC. Mathematically, the global protocol relies upon the following equation
\begin{align}
	&\operatorname{Tr}\left[W(t) V^{\dagger} W(t) V\right] \nonumber\\
	&= \frac{1}{ \mathcal{D}(\mathcal{D}+1) } 
	\overline{\langle W(t)\rangle_{u, k_0}\left\langle V^{\dagger} W(t) V\right\rangle_{u, k_0}},  \label{eqn_trace_ensemble}
\end{align}
where, $\mathcal{D}$ is the dimension of the Hilbert space. On the right hand side, the overline denotes an ensemble average of measurements over a set $U=\{u_0,u_1,\cdots u_{N_U}\}$ of random unitary operators and $k_0$ is an abitrary initial state. Each unitary in the set $U$ is a $N$-qubit unitary.
Implementation of the global protocol requires creating a $N$-qubit random unitary operator that is applied to an input state of $N$ qubits. 
Decomposition of an
$N$-qubit unitary is costly in terms of the entangling gates. Moreover, for a specific precision, the local protocol needs a smaller number of measurements \cite{vermersch_probing_2019}. 
As a result, we have found it convenient to implement the local protocol in Fig.~\ref{fig_OTOC_local_protocol}  which requires just $N$ random unitaries per run.  Depending on the number of initial states $|k_i\rangle=\{k_0,k_1,\cdots\,k_{2^n}\}$ being used, 
the modified OTOCs of different orders $n$ can be computed.  The larger the order $n$ of the modified OTOC, the better it approximates the original OTOC while the specific $n$ needed is model dependent.
Indeed, there is evidence that the modified OTOCs contain the needed information on entanglement spreading \cite{vermersch_probing_2019}.
For numerical justification of the Eq.~(\ref{eqn_trace_ensemble}) and for the connection of the different OTOC definitions, readers are advised to consult Appendix~\ref{app_otoc_protocol}.
Here, for  completeness, we outline the steps to compute the modified OTOC of zeroth order:
\begin{itemize}
	\item We prepare an arbitrary initial state $|k_0\rangle$ (position 1 in Fig~\ref{local_protocol1}). The initial state preparation step can be avoided if the all-zero state $\ket{\mathbf{0}}=\ket{0000 000}$ is chosen as the starting quantum state. Then, a set of unitary gates $u^i=\{U^i_0,U^i_1,\cdots, U^i_N, \}$ are applied to each qubit, which results in a random state $|\psi_1\rangle = U^i_0 \otimes U^i_1 \otimes \cdots \otimes U^i_N \ket{\mathbf{0}} $ at position 2 in the Fig.~\ref{local_protocol1}.
	
	\item Next the time evolution of the random state is computed using the Trotterized evolution operator $U(n)=\Big[\exp(-i \hat{H} \delta t) \Big]^n$. This yields $|\psi_2\rangle= U(n) |\psi_1\rangle$ at position 3 in the Fig.~\ref{local_protocol1}.
	
	\item The necessary gates are then applied to compute the observable $W$ in the computational basis. In our case, since $W=\sigma^z_i$, projective measurements of qubit $i$ allows us to compute $\langle W(t)\rangle =p_0 -p_1$, where $p_{0(1)}$ is the probability of measuring the qubit in the zero (one) state. We use $N_{\rm shots}=200$ for computing the expectation value of the operator.
	
	\item In a similar fashion, if we include the $V$ operator after creating the random state $|\psi_1\rangle$ (Fig.~\ref{local_protocol2}), the previous two steps can be applied to compute $\langle V^\dagger W(t) V \rangle$.
	
	\item The process is repeated $N_R=180$ times. Thus, measuring $\langle W \rangle$ (or $\langle VWV \rangle$) requires generating a total of $N_U=N_R \times N$ unitary matrices of size $2\times 2$, with each unitary matrix drawn randomly from the Circular Unitary Ensemble (CUE) \cite{mezzadri2006generate}. CUE($n$) represents a uniform distribution over the unitary square matrices of dimension $n$ -- the Haar measure of the unitary group $U(n)$.
	
	\item Finally, an ensemble average of the quantity $\overline{\langle W(t)\rangle_{u, \mathbf{k}_0} \langle V W(t) V\rangle_{u, \mathbf{k}_0}}$ is computed which is a measure of the modified OTOC of the zeroth order.
\end{itemize}
With the proper normalization, the modified OTOC of the zeroth order $O_{0}(t)$, can be described by the following equation
\begin{equation}
	O_{0}(t)=\frac{\overline{\langle W(t)\rangle_{u, \mathbf{k}_0}\langle V W(t) V\rangle_{u, \mathbf{k}_0}}}{ \overline{\langle W(t)\rangle_{u, \mathbf{k}_0}\langle W(t)\rangle_{u, \mathbf{k}_0}}}.
\end{equation}

Using the steps described above, operator expectation values $\langle W(t)\rangle$ and $\langle VW(t)V \rangle$ are computed with the same set of unitaries. Fig.~\ref{correlations} shows
measurements of these operators. Initially the operators are correlated (Fig.~\ref{corr_t0}) while
over time due to operator spreading the operators become decorrelated (Fig.~\ref{corr_t3.5}) which signifies a loss of memory of the initial state.  
As the resources required for the computation of higher order OTOCs is large
we have only computed the zeroth order OTOC in this study corresponding to
the plot in Fig~\ref{mod_otoc_plot1}. $N_U=180\times N$ unitaries were used for this 
simulation and  each measurement required $N_M=200$ shots. 
These numbers were chosen carefully using a noise model simulation 
so as to minimize the overall cost for implementing the protocol with current quantum devices. 
From the figure, it is seen that mitigated results with the IBM Sherbrooke machine
compare well with results from exact diagonalization. 
Dynamical decoupling (DD) was used to compensate coherent noise and 
M3 was used for the readout error mitigation. Our studies show that applying noise mitigation techniques is important in recovering scrambling information with current NISQ-era devices.

The dependence of the speed of information spread on the position of the $V$ operator can be seen in Fig.~\ref{mod_otoc_vary_Vqubit} where it is compared with classical Python-Trotter simulations. The error bars in the simulation indicate 
the jackknife error due to the choice of different sets of random unitaries. 
For a fixed number of unitaries $N_U$, the error can be reduced at the expense of increased computational resources, that is, by increasing the number of shots $N_{\rm shots}$. On the other hand, increasing the number of unitaries $N_U$ also reduces the error, allowing us to better approximate the trace in Eq.~(\ref{eqn_trace_ensemble}) with the ensemble average on the right-hand side. Clearly, the measured values obtained with the IBM device without mitigation
deviate from the ideal Python Trotter results, indicating the presence of different sources of noise in the device. The mitigated results agree rather well and can depict the difference in speed due to the varied distance $d=|j-i|$ of the $W_i$ and $V_j$ operators.
It would be intriguing to see in the future
whether we can use more computational resources to compute higher-order modified OTOCs. 

Additionally, investigating the scrambling time and quantum Lyapunov exponents with quantum computers could be an exciting avenue for the future research.

\section{Conclusion}

In this paper, we have investigated 
the transverse quantum Ising model discretized on two dimensional anti-de Sitter space.
In practice this is implemented by using site dependent couplings which mock up the metric
factors corresponding to a one dimensional hyperbolic space.
We computed the time evolution and OTOCs of the
model using both tensor network methods and 
quantum simulations using both gate based quantum computers 
as well as analog quantum computers that use Rydberg arrays. 
We showed that the time evolution and OTOCs obtained from the quantum 
simulations agree well with the tensor network calculations. 

The use of new publicly available universal quantum computers and new mitigation techniques allowed reliable time-evolution calculations with up to 13 qubits. In previous work on related real time evolution of systems of comparable difficulty \cite{gustafson2021indexed,Gustafson:2021imb,Gustafson:2021mky,asaduzzaman2022quantum}, reliable 4 qubit calculations were reported but extensions to 8 qubits were unsuccessful. Additionally, to the authors' best knowledge, this is the first time a protocol to compute OTOCs has been
implemented for a seven qubit system using an IBM QPU, superseding a previous attempt with four qubits. From this perspective, the results presented here give a sense of progress in quantum hardware and software in the last few years. Nevertheless, this remains a relatively small number of qubits and the boundary effects are significant. 
These boundary effects are of potential interest \cite{Takayanagi:2011zk,Fujita:2011fp,Dey:2020jlc} and could be studied in more detail for their own sake.

We found that depending on the parameters of the model it's possible to have different profiles for the light cones that
describe the propagation of information in the system.
Perhaps most intriguingly we find a regime of the critical system
where the direction of the light cones in global coordinates displays a logarithmic 
dependence on bulk distance. This behavior implies that the scrambling
time characterizing thermalization in this system depends only logarithmically on
the number of degrees of freedom. Such a behavior 
is usually seen in models with 
long or even infinite range interactions 
while our model has only nearest neighbor interactions. We believe that this makes this 
model a very interesting candidate for future studies of scrambling in quantum spin models.

\section*{Acknowledgements}
We acknowledge useful discussions with the members of QuLat collaboration, Richard Brower and Evan Owen. We thank the IBM-Q hub at Brookhaven National
Laboratory for providing access to the IBMQ quantum
computers. S.C, M.A and G.T were supported under U.S.\ Department of Energy grants DE-SC0009998 and DE-SC0019139.
Y. M. is supported under U.S.\ Department of Energy grant DE-SC0019139.

\appendix

\section{Digital Quantum Simulation of the Magnetization} \label{quantum_ap}

\begin{figure*}[!htb]
	\subfloat[\label{t0}]{%
		\includegraphics[width=.45\textwidth]{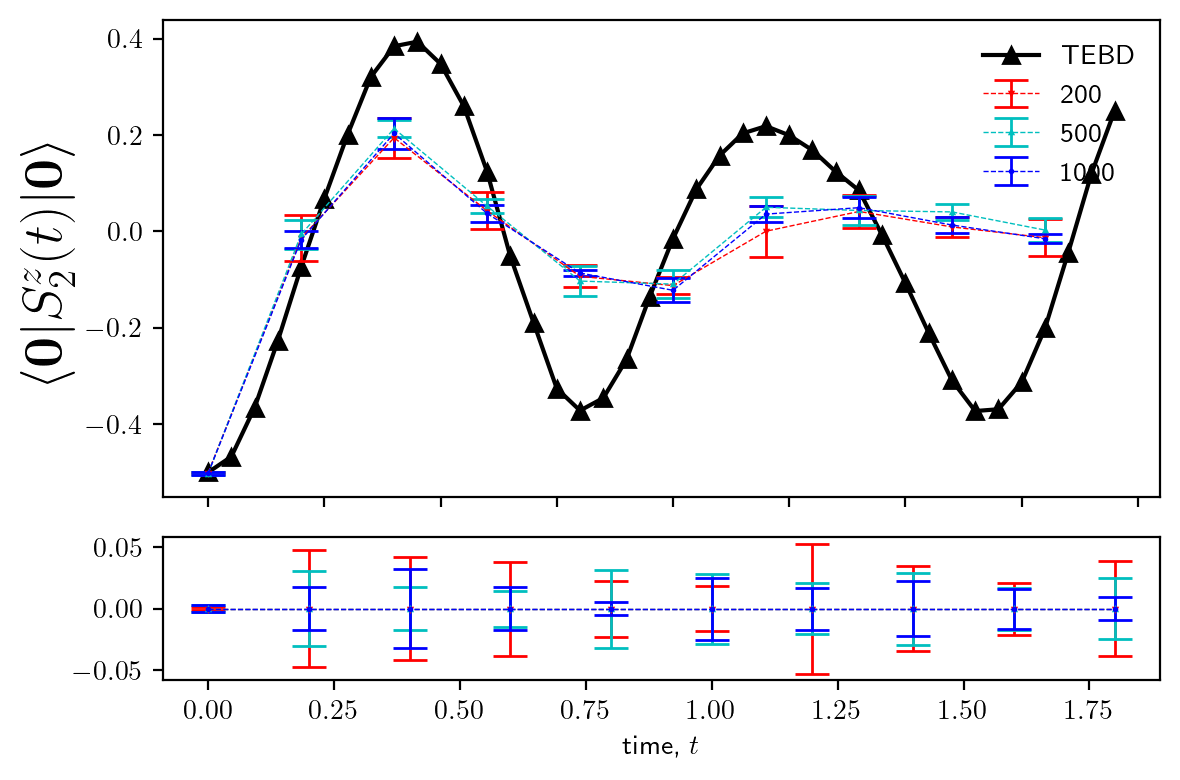}%
	}\hfill
	\subfloat[\label{t1}]{%
		\includegraphics[width=.45\textwidth]{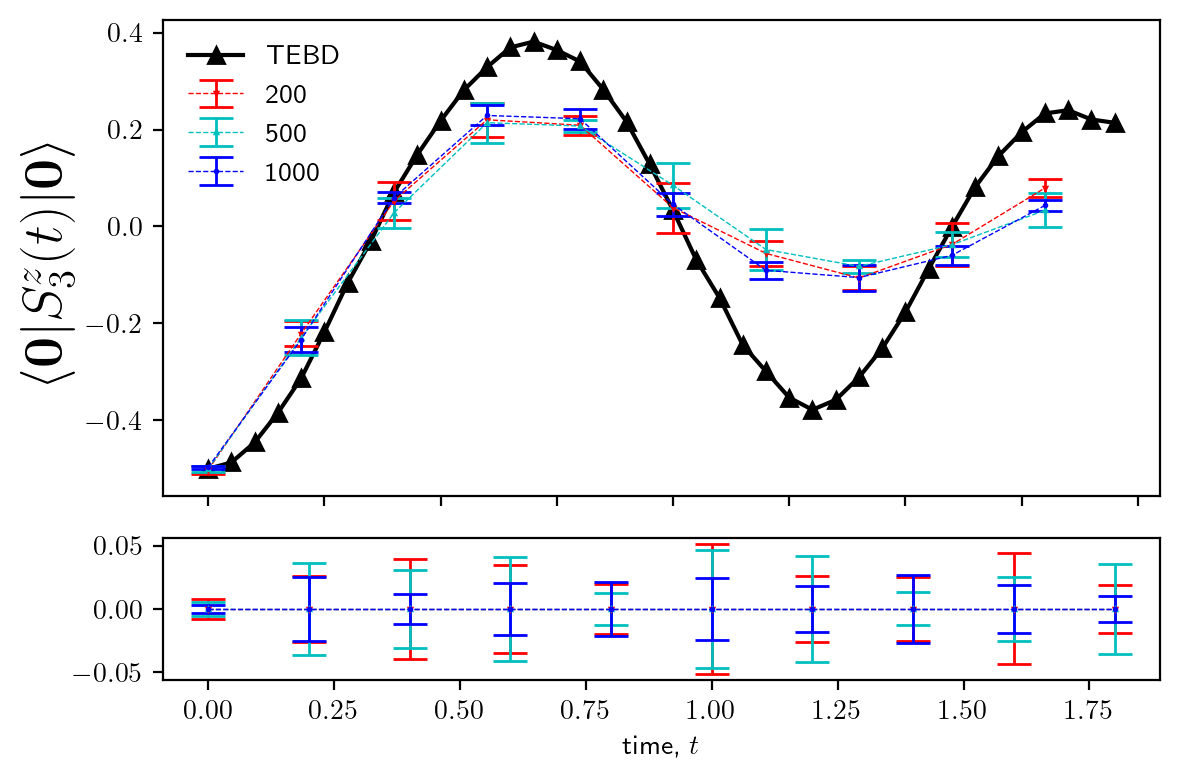}%
	}\vfill
	\subfloat[\label{t1}]{%
		\includegraphics[width=.45\textwidth]{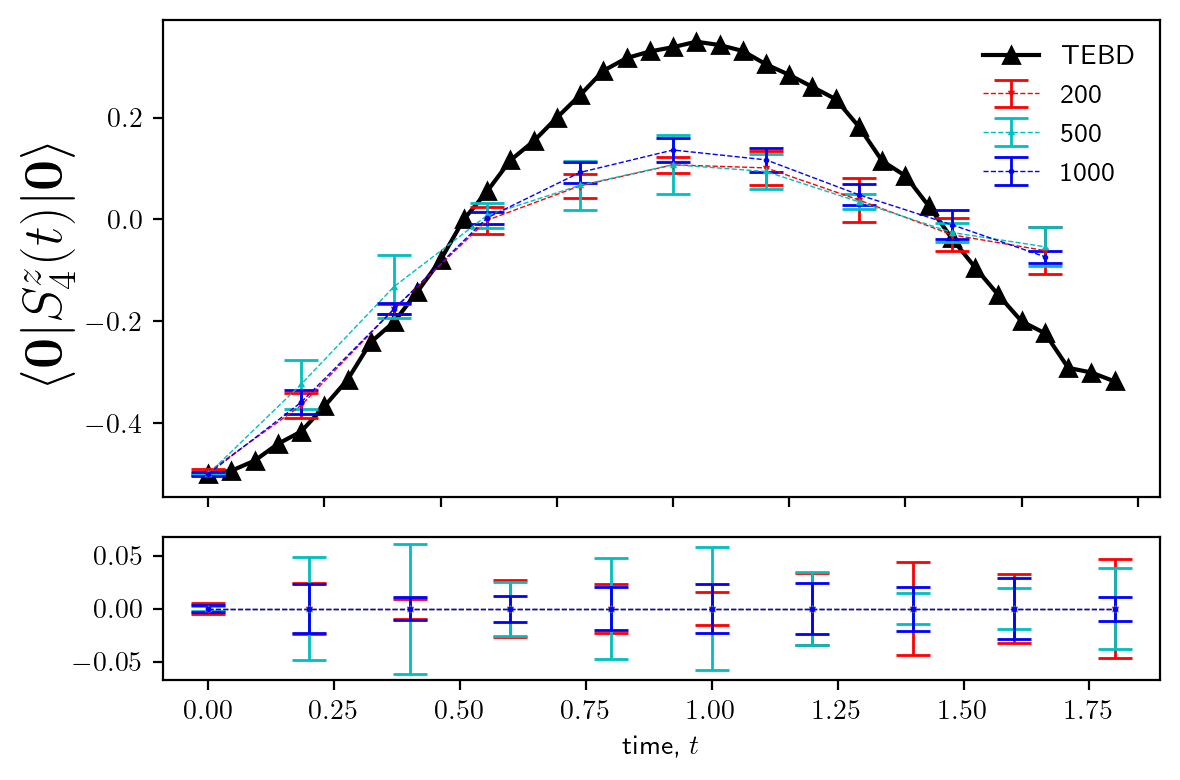}%
	}\hfill
	\subfloat[\label{t1}]{%
		\includegraphics[width=.45\textwidth]{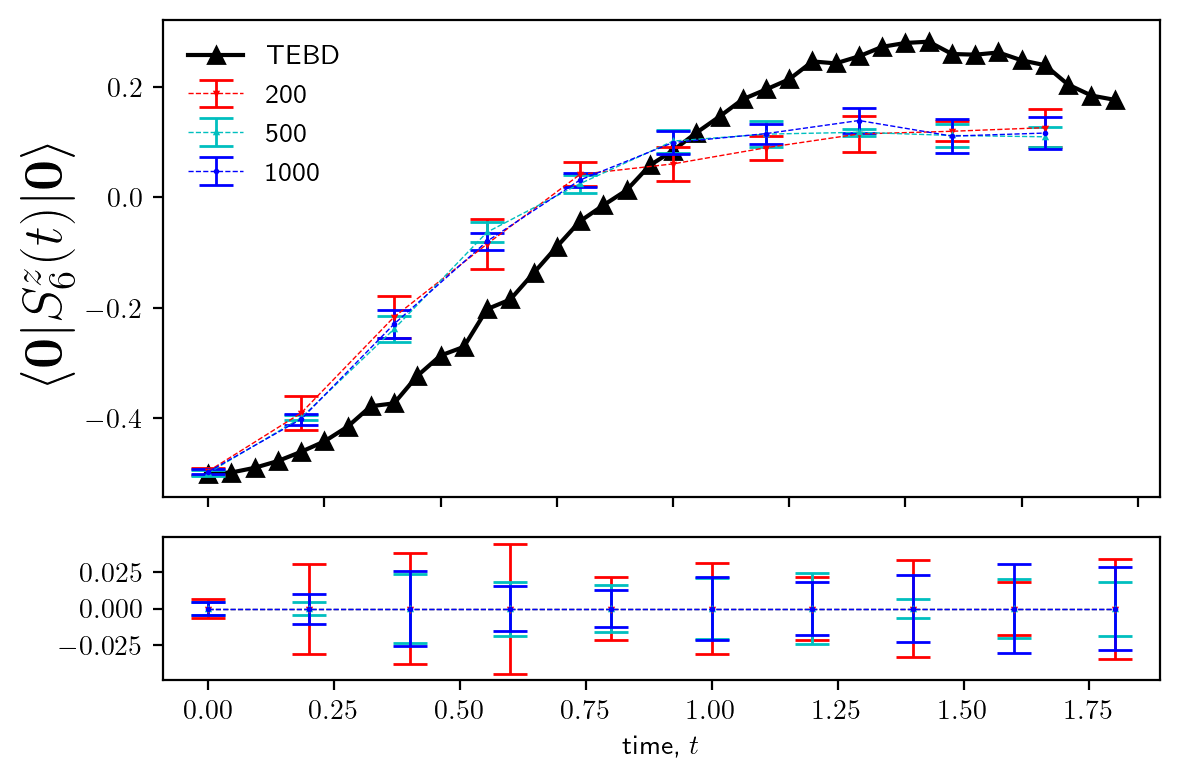}%
	}
	\caption{Shot noise analysis at Guadalupe machine is presented with local magnetization data. Shot noise associated for each trotter step is demonstrated in the bottom panel for the better visualization. The number in the labels denote the number of shots applied for measurements. Gap between the corresponding classical TEBD simulation results and QPU results indicate the presence of other coherent and incoherent sources of noise. Parameters: $J=2.0$, $h=1.05$, $l_{\rm max}=3.0$. }
	\label{comparison_statistical_error}
\end{figure*}

\begin{figure*}[!ht]
	\subfloat[\label{ordering1}]{%
		\includegraphics[width=.3\textwidth]{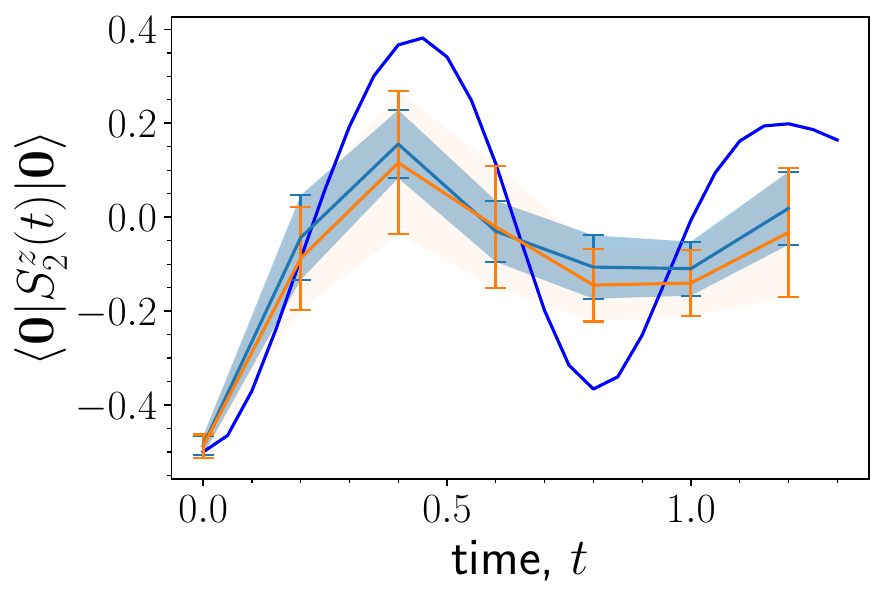}%
	}\hfill
	\subfloat[\label{ordering2}]{%
		\includegraphics[width=.3\textwidth]{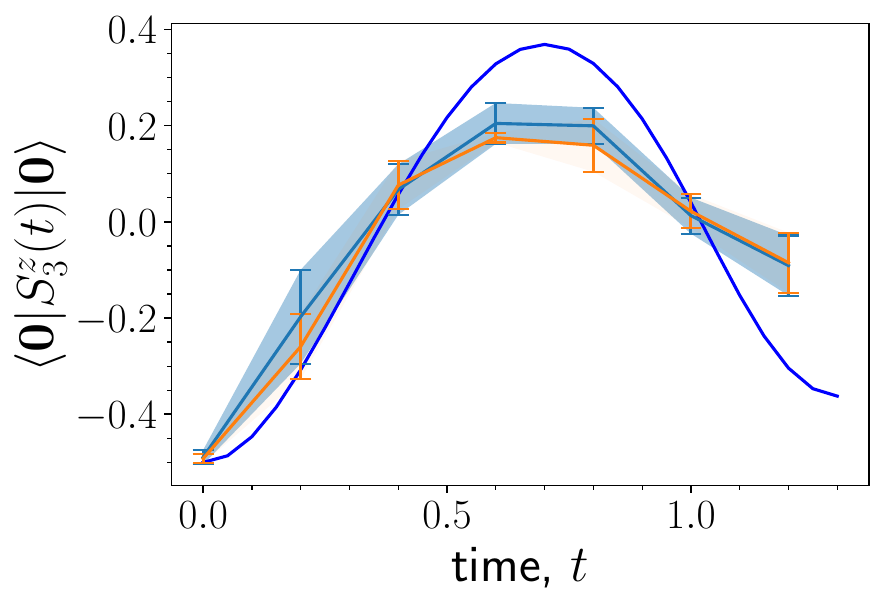}%
	}\hfill
	\subfloat[\label{ordering3}]{%
		\includegraphics[width=.3\textwidth]{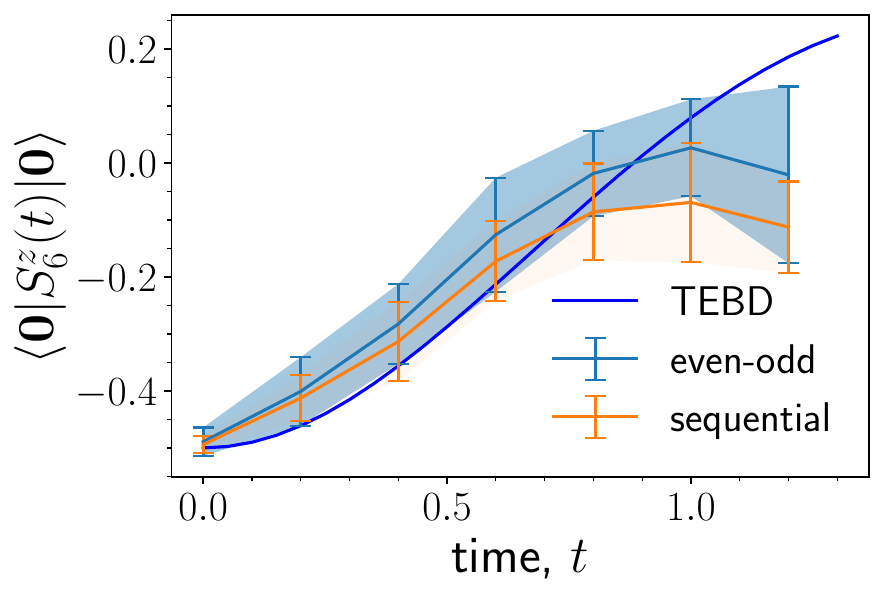}%
	}
	\caption{Comparison of Trotter evolution of magnetization results with different operator ordering. Parameters: $J=2.0$, $h=1.05$, $l_{\rm max}=3.0$.}
	\label{comparison_evenodd}
\end{figure*}

\begin{figure*}[!ht]
	\subfloat[\label{mitigation1}]{%
		\includegraphics[width=.3\textwidth]{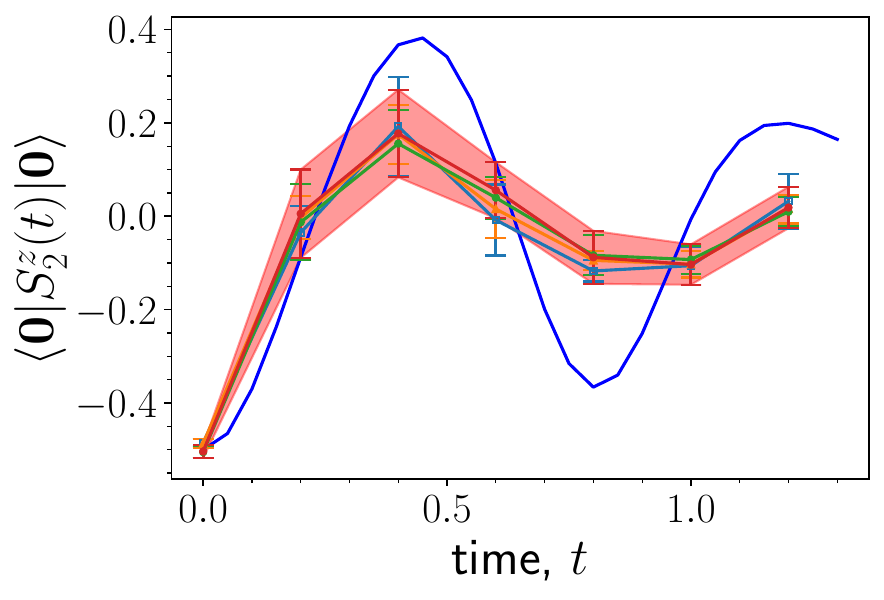}%
	}\hfill
	\subfloat[\label{mitigation2}]{%
		\includegraphics[width=.3\textwidth]{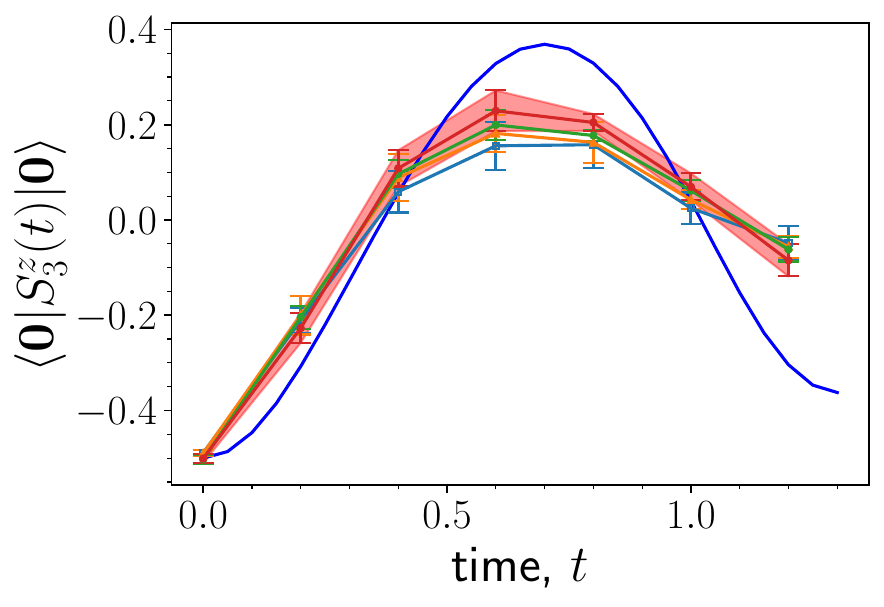}%
	}\hfill
	\subfloat[\label{mitigation3}]{%
		\includegraphics[width=.3\textwidth]{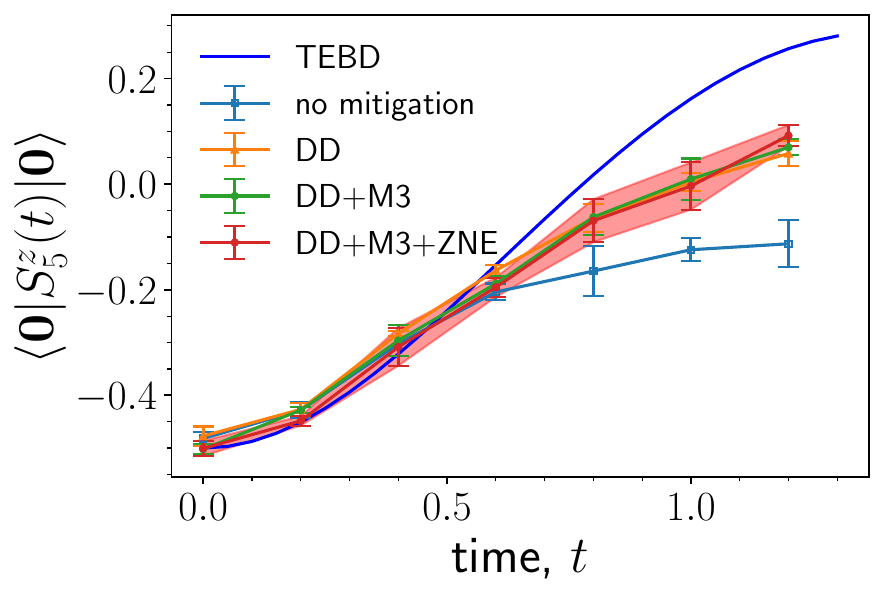}%
	}
	\caption{Comparison of magnetization results with different mitigation techniques and their combinations. Parameters: $J=2.0$, $h=1.05$, $l_{\rm max}=3.0$.}
	\label{comparison_different_mitigation}
\end{figure*}

\begin{figure*}[!ht]
	\subfloat[\label{noisescale1}]{%
		\includegraphics[width=.3\textwidth]{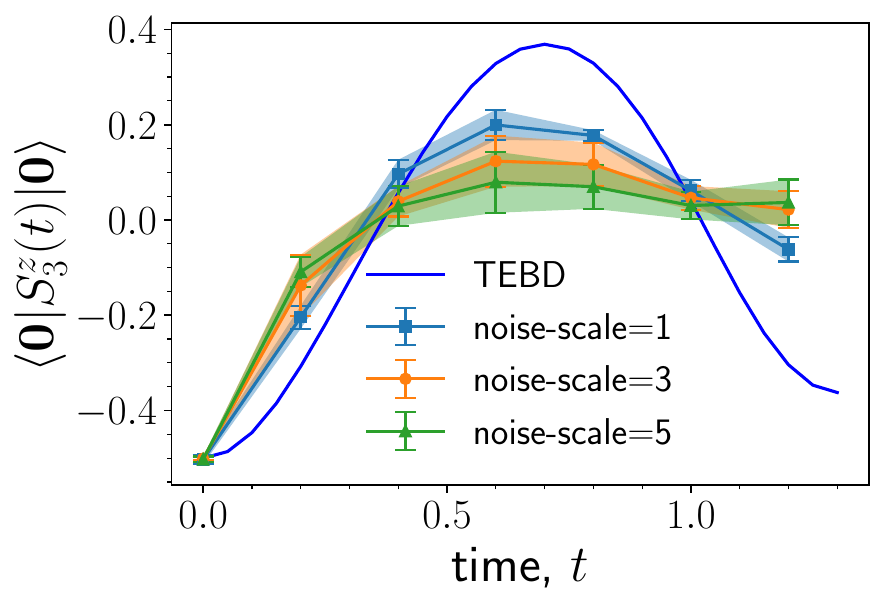}%
	}\hfill
	\subfloat[\label{noisescale2}]{%
		\includegraphics[width=.3\textwidth]{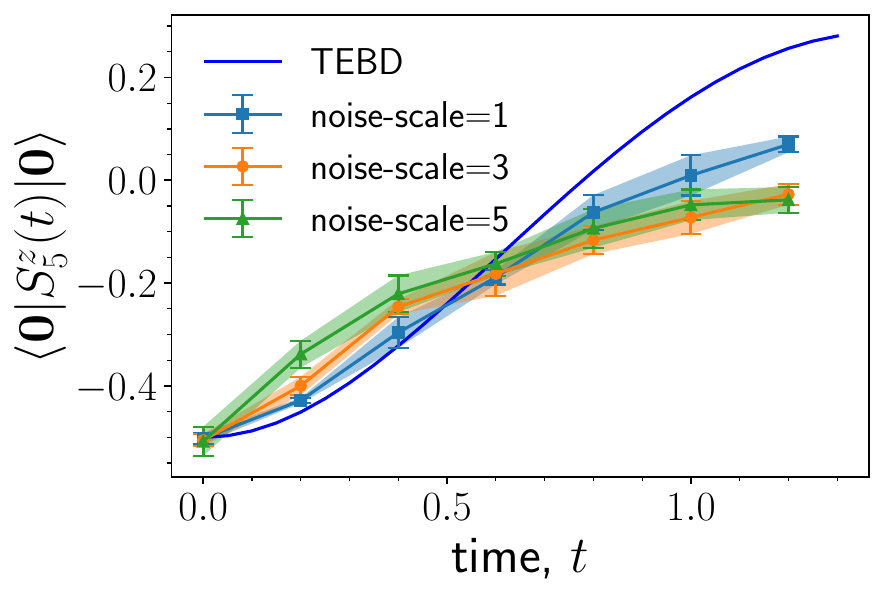}%
	}\hfill
	\subfloat[\label{noisescale3}]{%
		\includegraphics[width=.3\textwidth]{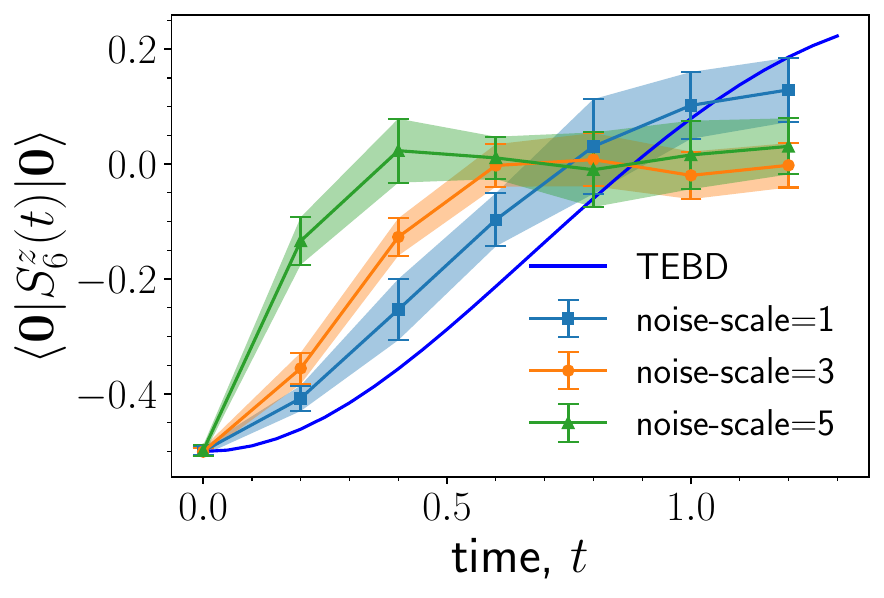}%
	}
	\caption{Comparison of Trotter evolution of magnetization results in different noise scaled circuits. Noise scale=$n$ indicates $n$-fold noise compared to the original circuit for the Trotter evolution of the local magnetization. Parameters: $J=2.0$, $h=1.05$, $l_{\rm max}=3.0$.}
	\label{noise_scaling_results}
\end{figure*}

\begin{figure*}[!ht]
	\subfloat[\label{znefit1}]{%
		\includegraphics[width=.3\textwidth]{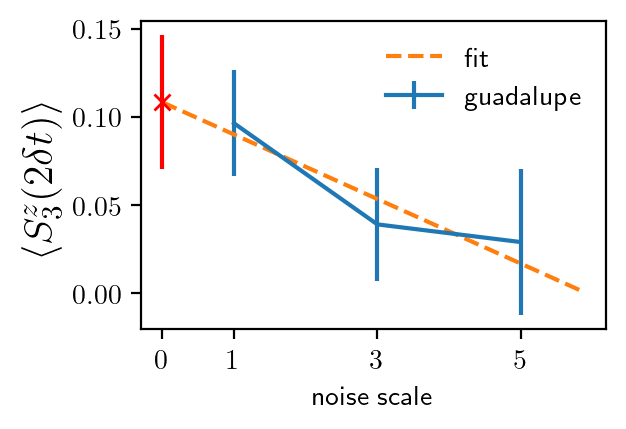}%
	}\hfill
	\subfloat[\label{znefit2}]{%
		\includegraphics[width=.3\textwidth]{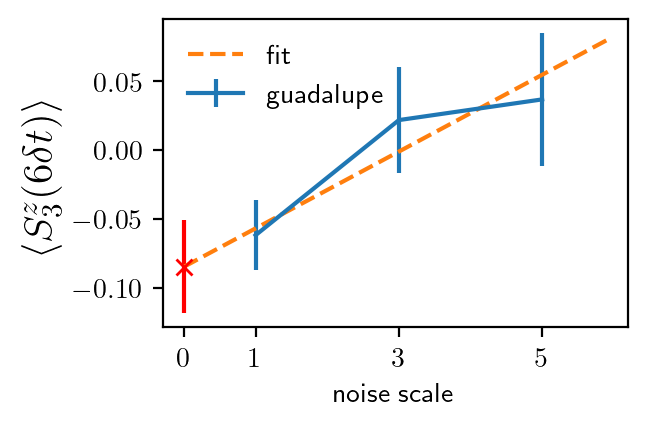}%
	}\hfill
	\subfloat[\label{znefit3}]{%
		\includegraphics[width=.3\textwidth]{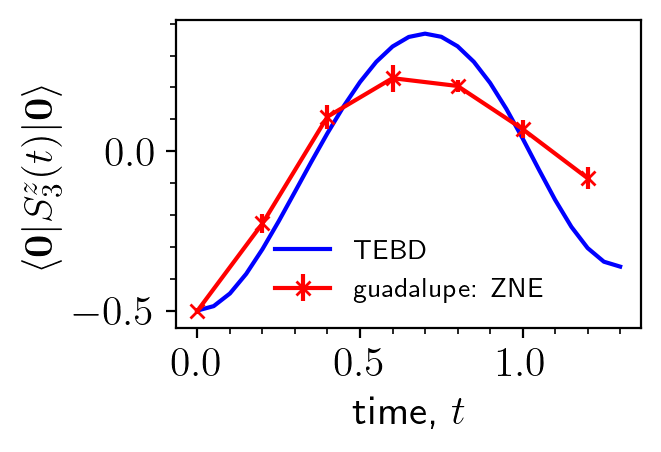}%
	}
	\caption{(a-b) Example of the extraction of the zero noise extrapolated data (red cross) at the second and the sixth trotter step, obtained from the measurements of the noise-scaled-circuits at guadalupe machine. (c) Extrapolated values are obtained for all trotter steps to plot ZNE data of the local magnetization. Parameters: $J=2.0$, $h=1.05$, $l_{\rm max}=3.0$, $\delta t=0.2$.}
	\label{zne_extrapolations}
\end{figure*}

In this appendix, we present some observations of the simulation with digital quantum computing processors which would be useful for investigation of quantum field theories with quantum computers for interested readers.

\subsection{\label{shot_error}Statistical error}
In this subsection, we discuss statistical errors associated with different number of shots. Fig.~\ref{comparison_statistical_error} shows magnetization results obtained from the Guadalupe quantum computer with 200, 500 and 1000 shots. With our choice of the parameters, we find that the information about the magnetization is completely lost after $\sim 7$ Trotter steps for some of the cases. As a result, for the followup discussion, we considered data up-to the sixth trotter step. Varying the number of shots ($N_{\rm shots}$) reduces the statistical error ($\epsilon_{\rm stat}$) and is roughly consistent with the relation $\epsilon_{\rm stat} \propto 1/\sqrt{N_{\rm shots}}$. Statistical errors were computed from data obtained from the six measurement sessions at different times. It's noteworthy that we do not see significant differences in the central value of the measurements. The central value stabilizes with the increase in the number of sessions. From our analysis, we find that the systematic error is much larger than the shot noise error. Hence, it is necessary to develop error-correction routines to recover correct results. With the NISQ-era devices, fault-tolerant computation is not feasible due to conflicting requirements of low fidelity of the qubits and the large qubit overhead for error-correction protocols. However, different error mitigation techniques can be applied to scale up the number of qubits for simulation in the current NISQ-devices.  In the following section, we discuss the application of the different error mitigation techniques to improve results obtained from the quantum processing units.

\subsection{\label{ordering}Operator ordering}
Fig.~\ref{comparison_evenodd} demonstrates how the local magnetization of $N=13$ qubit lattice chain obtained from the Guadalupe QPU compares with different operator-ordering. To address the question of the operator-ordering we exclude mitigation techniques and circuit optimization techniques. Each data point was obtained from the average of 
six experiments each with 200 shots.  Here, the label `sequential' implies that
the continuum evolution operator is approximated as 
\begin{equation}
	U_{\rm seq}=  \prod_{i } h_{x}  \prod_{\langle ij \rangle} h_{\rm int}^{ij},  
\end{equation}
whereas, the following ordering of operator denotes `odd-even' ordering of operators
\begin{equation}
	U_{\rm seq}= \prod_{i} h_{ x}     \prod_{\langle ij \rangle,i \, \mathrm{even}} h_{\rm int}^{ij} \prod_{\langle ij \rangle,i \, \mathrm{odd}} h_{\rm int}^{ij}. 
\end{equation}
Local operators are defined as $h_x=\exp(i\frac{h}{2}) \eta_{ij} \sigma_i^x$ and $h_{\rm int}=\exp(i\frac{T}{4}) \eta_i \sigma_i^z \sigma_{i+1}^z $. 

We did not find a particular choice of operator ordering to be an important factor in the noisy Guadalupe device. Indeed, it is likely that the systematic errors are much larger than the differences in measurements associated with different choices of operator ordering. 

\subsection{\label{mitigation}Error mitigation}
In this subsection, we discuss the importance of different error mitigation techniques in the context of computations of the real time evolution of the magnetization of our model. We first analyze results obtained with dynamical decoupling (DD), then with a combination of dynamical decoupling and M3 (DD+M3) mitigation techniques, and finally with a combination of dynamical decoupling, M3 and Zero Noise Extrapolation (DD+M3+ZNE) techniques.\\

Further observation on the combined cases of error mitigation revealed that for some cases like $\langle S^z_3(t) \rangle$ in Fig~\ref{mitigation1}, local magnetization data did not improve the results much. In contrast, for some cases like Fig~\ref{mitigation2}, the results were significantly improved and for the rest (Fig~\ref{mitigation3}), it is found that the results were improved only for large trotter steps.

On top of the dynamical decoupling and readout error correction technique, we applied Zero Noise Extrapolation (ZNE) to mitigate incoherent noise. The first step in the process is to scale up the noise systematically by generating unitary gate-folding or pulse-stretching \footnote{Pulse stretching needs pulse level access to device where the amount of noise introduced is controlled by the duration of the pulse applied to implement different gates.}. We used unitary folding by mapping a two-qubit operator $U \to U U^\dagger U $. For pair of CX gates that are added one increases the noise-level by a factor of three.  The second step is to perform measurements in the folded circuits and finally use these measurements  with different noise levels to extrapolate a zero noise limit of the observables. Fig. ~\ref{noise_scaling_results} clearly demonstrates that increasing the noise by adding more unitaries causes the experimental values to deviate further away from the classically computed results with TEBD. The noise scaled values that
are obtained for local magnetization $\langle S^z_i\rangle$ at a time $t_0$ are then used to extrapolate zero-noise value by linear extrapolations (Fig~\ref{znefit1},\ref{znefit2}). Extrapolated values obtained at different trotter step are then combined to produce the time dependent magnetization curve Fig~\ref{znefit3}.

\section{Analog Quantum simulation of Magnetization\label{ryd_app}}

\begin{figure}[!h]
	\centering
	\includegraphics[scale=.32]{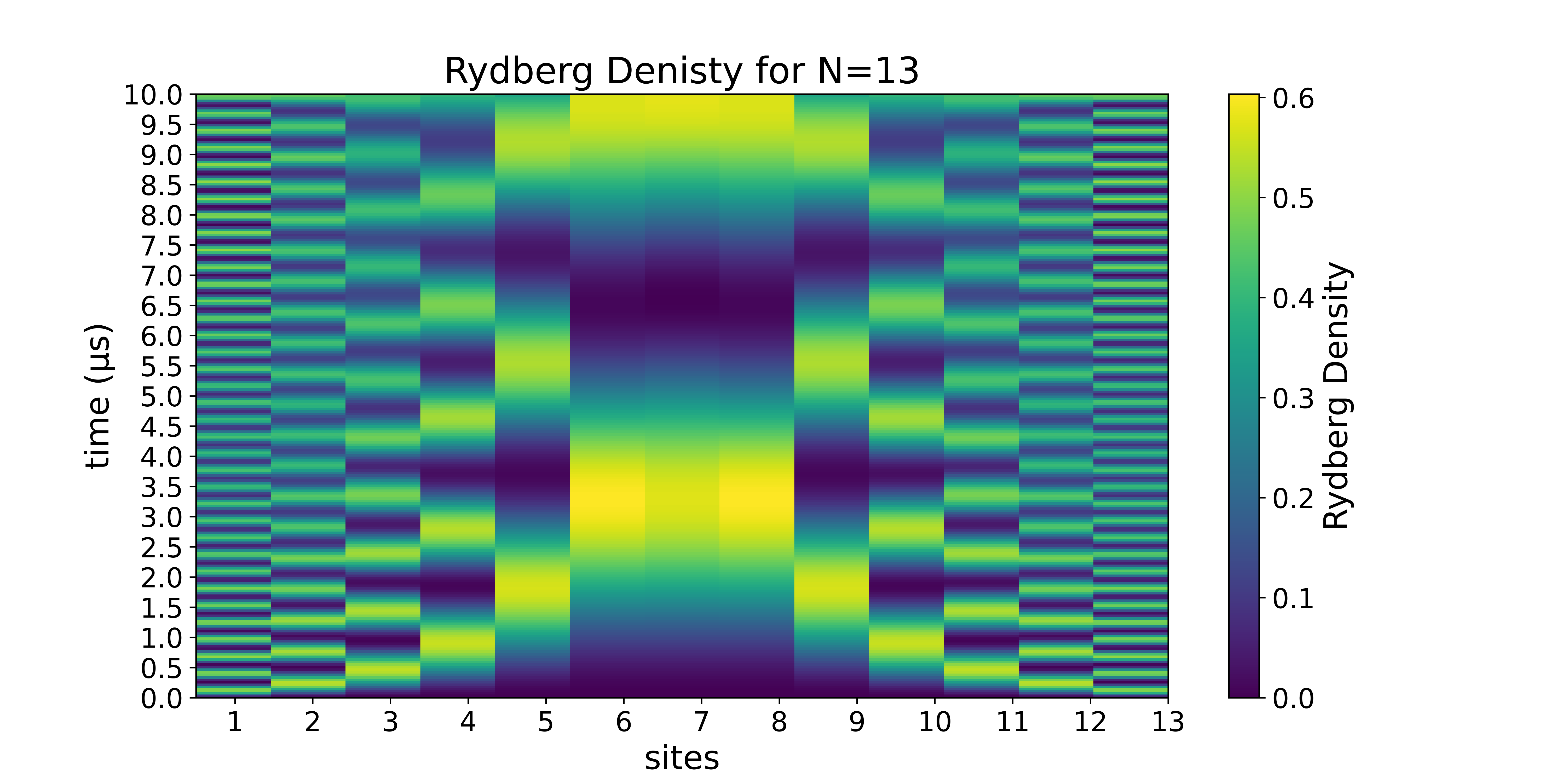}
	\caption{Time evolution of the Rydberg density}
	\label{fig:bloqade_sim}
\end{figure}

In this appendix, we report on quantum simulations of this model using Rydberg arrays.  The Hamiltonian that governs the Rydberg simulator can be written as,  
\begin{align}
	\hat{H}_R(t) &= \sum_j\frac{\Omega_j(t)}{2}(e^{i\phi_j(t)}\ket{g_j}\bra{r_j}+e^{-i\phi_j(t)}\ket{r_j}\bra{g_j}) \nonumber \\
	&-\sum_j\Delta_j(t)\hat{n}_j + \sum_{j<k}V_{jk}\hat{n}_j \hat{n}_k, 
\end{align}
where, $\Omega_j(t)$ is the Rabi frequency, $\phi_j(t)$ denotes the laser phase, $\Delta_j(t)$ the detuning
parameter at site $j$. Van der Walls interaction  $V_{jk}=C_6/\abs{r_j - r_k}^6$ is known as the Rydberg interaction term with $C_6=2\pi\times862690 \si{\MHz}\si{\um^6}$ \cite{Cong:2021nhm, Ebadi:2020ldi,Keesling:2018ish}.

Different operators in the hyperbolic Ising Hamiltonian can be mapped to different operators of the Rydberg Hamiltonian with the choice of zero laser phase $\phi_j(t)$ at all sites,
\begin{align}
	\hat{H}_R(t) &= \sum_j\frac{\Omega_j(t)}{2}\underbrace{\left(\ket{g_j}\bra{r_j}+\ket{r_j}\bra{g_j}\right)}_{\sigma^x_j} \nonumber \\
	&-\sum_j\Delta_j(t)\underbrace{\hat{n}_j}_{(1-\sigma^z_j)} + \sum_{j<k}V_{jk}\underbrace{\hat{n}_j \hat{n}_k}_{(1-\sigma^z_j)(1- \sigma^z_k)}.
\end{align}

The Rydberg interaction potential, $V_{jk}$ determines the position of the atoms to quantum simulate the hyperbolic Hamiltonian. Due to the hyperbolic deformation, it is expected that we need to position the atoms non-uniformly. This is achieved by placing the atoms starting at location $(0,0)$ and 
using Eq.~(\ref{distance_ryd}) to find the distances between successive spins: 
\begin{equation}\label{distance_ryd}
	\delta_{i+1}=(A /\eta_i)^{1/6} + r_{i}.   
\end{equation}

This equation is just the rearranged form of $\frac{A}{(r_{i+1}-r_i)^6}=\cosh{l_i}$ which is the form of the Rydberg potential. Here, $A=2\pi\times 512$ is a constant for adjusting the scale,  $\eta_i=J\cosh(l_i)$ is the hyperbolic deformation and $r_i$ is the location for the $i\textsubscript{th}$ site. We set $J=1$ for the rest of our discussion of Rydberg simulations.  

Using this procedure we get the following locations for the Rydberg atoms for $l_{\rm max}=3.0$ where the resulting distances between atoms range from $12.13\si{\um}$,  to $17.72 \si{\um}$ with the furthest atom located at $ 180.77 \si{\um}$ from the origin.

The form of $\Delta_j$ and $\Omega_j$ is then given by equating the coefficients to the form of the Rydberg potential between the atoms
\begin{equation}
	\Delta_j,\Omega_j= \frac{10\times C_6}{(r_{j+1}-r_j)^6}.
\end{equation}

However, currently commercially available Rydberg machines are constrained to have only global laser parameters. Hence we have turned to the Bloqade Simulator developed by QuEra to perform simulations \cite{Bloqade}. Fig. ~\ref{fig:bloqade_sim} shows a picture of the time evolution of the
Rydberg density (essentially $\langle S_z \rangle$). Notice that Fig. ~\ref{fig:bloqade_sim} exhibits similar warping effects to those seen in the TEBD simulations of the model. This shows us that our model can be simulated with Rydberg Arrays.  We hope that in the future with advancements in the Rydberg array technologies, we will be able to probe information propagation  in this model with Rydberg simulators. However even with a local detuning it might not be possible to probe the whole spectrum of the model due to limitations in chain length and largeness of the Rabi and detuning term.

\section{\label{app_otoc_protocol}Digital Quantum Simulation: OTOC}

\begin{figure}[!h]
	\centering
	\includegraphics[scale=.35]{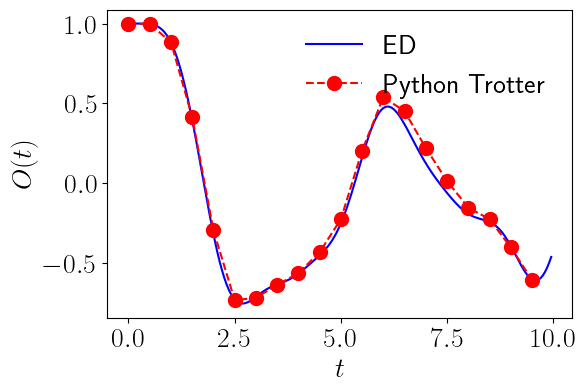}
	\caption{Choice of the Trotter step $\delta t \sim 0.5$ seems a good choice for the OTOC computation with our choice of parameters}
	\label{otoc_trotter_step}
\end{figure}

\begin{figure}[!h]
	\centering
	\includegraphics[scale=.35]{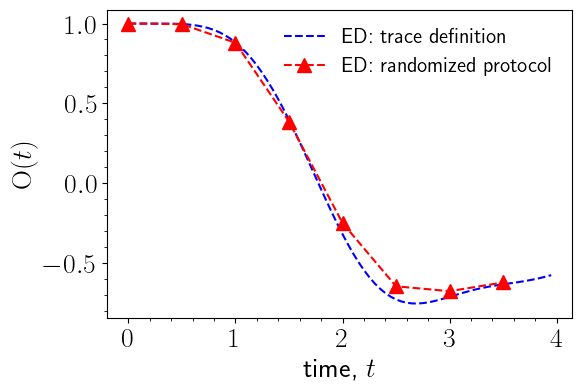}
	\caption{OTOC computed with the protocol with global unitaries match with traced data of products of operators.}
	\label{global_protocol_data_comparison}
\end{figure}

In this section of the appendix, we will discuss some of the details of the OTOC computation with quantum simulators and quantum processing units. 

Just like the magnetization, we need to pick a suitable Trotter step to observe physics with current NISQ era machines. Fig~\ref{otoc_trotter_step} demonstrates that $\delta t=0.5$ is a suitable choice. As the OTOC drops from one to zero in four Trotter steps, the entangling gate-cost for the measurements of $\langle W \rangle $ and $\langle VWV \rangle$ (see Fig~\ref{fig_OTOC_local_protocol} in the main text) is manageable with current NISQ devices and a comparison of the Trotterized version (without shot noise) of the results and the exact-diagonalized results reveal that the Trotter error associated with the trotter step is not large enough to obscure the physics we are interested in (Fig~\ref{otoc_trotter_step}). 

We conclude this section of the appendix by justifying Eq.~(\ref{eqn_trace_ensemble}) numerically. In Fig~\ref{global_protocol_data_comparison}, we compared the OTOC computed from the trace definition with the results obtained from the global protocol developed by Vermersch \textit{et. al.} \cite{vermersch_probing_2019} with numerics . For a mathematical proof of the identity, please see the appendix in \cite{chen2018measuring}. Higher order modified OTOCs computed from the local protocol yields the same result as that of global protocol \cite{vermersch2019probing}.

\bibliography{hyp_ising}

\begin{thebibliography}{92}
\expandafter\ifx\csname natexlab\endcsname\relax\def\natexlab#1{#1}\fi
\expandafter\ifx\csname bibnamefont\endcsname\relax
  \def\bibnamefont#1{#1}\fi
\expandafter\ifx\csname bibfnamefont\endcsname\relax
  \def\bibfnamefont#1{#1}\fi
\expandafter\ifx\csname citenamefont\endcsname\relax
  \def\citenamefont#1{#1}\fi
\expandafter\ifx\csname url\endcsname\relax
  \def\url#1{\texttt{#1}}\fi
\expandafter\ifx\csname urlprefix\endcsname\relax\def\urlprefix{URL }\fi
\providecommand{\bibinfo}[2]{#2}
\providecommand{\eprint}[2][]{\url{#2}}

\bibitem[{\citenamefont{Beisert et~al.}(2012)}]{Beisert:2010jr}
\bibinfo{author}{\bibfnamefont{N.}~\bibnamefont{Beisert}} \bibnamefont{et~al.},
  \bibinfo{journal}{Lett. Math. Phys.} \textbf{\bibinfo{volume}{99}},
  \bibinfo{pages}{3} (\bibinfo{year}{2012}), \eprint{1012.3982}.

\bibitem[{\citenamefont{Hubeny}(2015)}]{Hubeny:2014bla}
\bibinfo{author}{\bibfnamefont{V.~E.} \bibnamefont{Hubeny}},
  \bibinfo{journal}{Class. Quant. Grav.} \textbf{\bibinfo{volume}{32}},
  \bibinfo{pages}{124010} (\bibinfo{year}{2015}), \eprint{1501.00007}.

\bibitem[{\citenamefont{Asaduzzaman et~al.}(2020)\citenamefont{Asaduzzaman,
  Catterall, Hubisz, Nelson, and Unmuth-Yockey}}]{Asaduzzaman:2020hjl}
\bibinfo{author}{\bibfnamefont{M.}~\bibnamefont{Asaduzzaman}},
  \bibinfo{author}{\bibfnamefont{S.}~\bibnamefont{Catterall}},
  \bibinfo{author}{\bibfnamefont{J.}~\bibnamefont{Hubisz}},
  \bibinfo{author}{\bibfnamefont{R.}~\bibnamefont{Nelson}}, \bibnamefont{and}
  \bibinfo{author}{\bibfnamefont{J.}~\bibnamefont{Unmuth-Yockey}},
  \bibinfo{journal}{Phys. Rev. D} \textbf{\bibinfo{volume}{102}},
  \bibinfo{pages}{034511} (\bibinfo{year}{2020}), \eprint{2005.12726}.

\bibitem[{\citenamefont{Brower et~al.}(2021{\natexlab{a}})\citenamefont{Brower,
  Cogburn, Fitzpatrick, Howarth, and Tan}}]{Brower:2019kyh}
\bibinfo{author}{\bibfnamefont{R.~C.} \bibnamefont{Brower}},
  \bibinfo{author}{\bibfnamefont{C.~V.} \bibnamefont{Cogburn}},
  \bibinfo{author}{\bibfnamefont{A.~L.} \bibnamefont{Fitzpatrick}},
  \bibinfo{author}{\bibfnamefont{D.}~\bibnamefont{Howarth}}, \bibnamefont{and}
  \bibinfo{author}{\bibfnamefont{C.-I.} \bibnamefont{Tan}},
  \bibinfo{journal}{Phys. Rev. D} \textbf{\bibinfo{volume}{103}},
  \bibinfo{pages}{094507} (\bibinfo{year}{2021}{\natexlab{a}}),
  \eprint{1912.07606}.

\bibitem[{\citenamefont{Asaduzzaman
  et~al.}(2022{\natexlab{a}})\citenamefont{Asaduzzaman, Catterall, Hubisz,
  Nelson, and Unmuth-Yockey}}]{Asaduzzaman:2021bcw}
\bibinfo{author}{\bibfnamefont{M.}~\bibnamefont{Asaduzzaman}},
  \bibinfo{author}{\bibfnamefont{S.}~\bibnamefont{Catterall}},
  \bibinfo{author}{\bibfnamefont{J.}~\bibnamefont{Hubisz}},
  \bibinfo{author}{\bibfnamefont{R.}~\bibnamefont{Nelson}}, \bibnamefont{and}
  \bibinfo{author}{\bibfnamefont{J.}~\bibnamefont{Unmuth-Yockey}},
  \bibinfo{journal}{Phys. Rev. D} \textbf{\bibinfo{volume}{106}},
  \bibinfo{pages}{054506} (\bibinfo{year}{2022}{\natexlab{a}}),
  \eprint{2112.00184}.

\bibitem[{\citenamefont{Brower et~al.}(2022)\citenamefont{Brower, Cogburn, and
  Owen}}]{Brower:2022atv}
\bibinfo{author}{\bibfnamefont{R.~C.} \bibnamefont{Brower}},
  \bibinfo{author}{\bibfnamefont{C.~V.} \bibnamefont{Cogburn}},
  \bibnamefont{and} \bibinfo{author}{\bibfnamefont{E.}~\bibnamefont{Owen}},
  \bibinfo{journal}{Phys. Rev. D} \textbf{\bibinfo{volume}{105}},
  \bibinfo{pages}{114503} (\bibinfo{year}{2022}), \eprint{2202.03464}.

\bibitem[{\citenamefont{Swingle}(2012{\natexlab{a}})}]{swingle2012entanglement}
\bibinfo{author}{\bibfnamefont{B.}~\bibnamefont{Swingle}},
  \bibinfo{journal}{Physical Review D} \textbf{\bibinfo{volume}{86}},
  \bibinfo{pages}{065007} (\bibinfo{year}{2012}{\natexlab{a}}).

\bibitem[{\citenamefont{Swingle}(2012{\natexlab{b}})}]{swingle2012constructing}
\bibinfo{author}{\bibfnamefont{B.}~\bibnamefont{Swingle}},
  \bibinfo{journal}{arXiv preprint arXiv:1209.3304}
  (\bibinfo{year}{2012}{\natexlab{b}}).

\bibitem[{\citenamefont{Steinberg and Prior}(2022)}]{steinberg2022conformal}
\bibinfo{author}{\bibfnamefont{M.}~\bibnamefont{Steinberg}} \bibnamefont{and}
  \bibinfo{author}{\bibfnamefont{J.}~\bibnamefont{Prior}},
  \bibinfo{journal}{Scientific Reports} \textbf{\bibinfo{volume}{12}},
  \bibinfo{pages}{532} (\bibinfo{year}{2022}).

\bibitem[{\citenamefont{Jahn et~al.}(2022)\citenamefont{Jahn, Zimbor{\'a}s, and
  Eisert}}]{jahn2022tensor}
\bibinfo{author}{\bibfnamefont{A.}~\bibnamefont{Jahn}},
  \bibinfo{author}{\bibfnamefont{Z.}~\bibnamefont{Zimbor{\'a}s}},
  \bibnamefont{and} \bibinfo{author}{\bibfnamefont{J.}~\bibnamefont{Eisert}},
  \bibinfo{journal}{Quantum} \textbf{\bibinfo{volume}{6}}, \bibinfo{pages}{643}
  (\bibinfo{year}{2022}).

\bibitem[{\citenamefont{Basteiro et~al.}(2023)\citenamefont{Basteiro, Dusel,
  Erdmenger, Herdt, Hinrichsen, Meyer, and
  Schrauth}}]{basteiro2023breitenlohner}
\bibinfo{author}{\bibfnamefont{P.}~\bibnamefont{Basteiro}},
  \bibinfo{author}{\bibfnamefont{F.}~\bibnamefont{Dusel}},
  \bibinfo{author}{\bibfnamefont{J.}~\bibnamefont{Erdmenger}},
  \bibinfo{author}{\bibfnamefont{D.}~\bibnamefont{Herdt}},
  \bibinfo{author}{\bibfnamefont{H.}~\bibnamefont{Hinrichsen}},
  \bibinfo{author}{\bibfnamefont{R.}~\bibnamefont{Meyer}}, \bibnamefont{and}
  \bibinfo{author}{\bibfnamefont{M.}~\bibnamefont{Schrauth}},
  \bibinfo{journal}{Physical Review Letters} \textbf{\bibinfo{volume}{130}},
  \bibinfo{pages}{091604} (\bibinfo{year}{2023}).

\bibitem[{\citenamefont{Rigol et~al.}(2008)\citenamefont{Rigol, Dunjko, and
  Olshanii}}]{Rigol_2008}
\bibinfo{author}{\bibfnamefont{M.}~\bibnamefont{Rigol}},
  \bibinfo{author}{\bibfnamefont{V.}~\bibnamefont{Dunjko}}, \bibnamefont{and}
  \bibinfo{author}{\bibfnamefont{M.}~\bibnamefont{Olshanii}},
  \bibinfo{journal}{Nature} \textbf{\bibinfo{volume}{452}},
  \bibinfo{pages}{854} (\bibinfo{year}{2008}),
  \urlprefix\url{https://doi.org/10.1038%2Fnature06838}.

\bibitem[{\citenamefont{Srednicki}(1994)}]{PhysRevE.50.888}
\bibinfo{author}{\bibfnamefont{M.}~\bibnamefont{Srednicki}},
  \bibinfo{journal}{Phys. Rev. E} \textbf{\bibinfo{volume}{50}},
  \bibinfo{pages}{888} (\bibinfo{year}{1994}),
  \urlprefix\url{https://link.aps.org/doi/10.1103/PhysRevE.50.888}.

\bibitem[{\citenamefont{Kaufman et~al.}(2016)\citenamefont{Kaufman, Tai, Lukin,
  Rispoli, Schittko, Preiss, and Greiner}}]{Kaufman_2016}
\bibinfo{author}{\bibfnamefont{A.~M.} \bibnamefont{Kaufman}},
  \bibinfo{author}{\bibfnamefont{M.~E.} \bibnamefont{Tai}},
  \bibinfo{author}{\bibfnamefont{A.}~\bibnamefont{Lukin}},
  \bibinfo{author}{\bibfnamefont{M.}~\bibnamefont{Rispoli}},
  \bibinfo{author}{\bibfnamefont{R.}~\bibnamefont{Schittko}},
  \bibinfo{author}{\bibfnamefont{P.~M.} \bibnamefont{Preiss}},
  \bibnamefont{and} \bibinfo{author}{\bibfnamefont{M.}~\bibnamefont{Greiner}},
  \bibinfo{journal}{Science} \textbf{\bibinfo{volume}{353}},
  \bibinfo{pages}{794} (\bibinfo{year}{2016}),
  \urlprefix\url{https://doi.org/10.1126%2Fscience.aaf6725}.

\bibitem[{\citenamefont{Bekenstein}(1973)}]{Bekenstein:1973ur}
\bibinfo{author}{\bibfnamefont{J.~D.} \bibnamefont{Bekenstein}},
  \bibinfo{journal}{Phys. Rev. D} \textbf{\bibinfo{volume}{7}},
  \bibinfo{pages}{2333} (\bibinfo{year}{1973}).

\bibitem[{\citenamefont{Bekenstein}(2003)}]{Bekenstein:2003dt}
\bibinfo{author}{\bibfnamefont{J.~D.} \bibnamefont{Bekenstein}},
  \bibinfo{journal}{Contemp. Phys.} \textbf{\bibinfo{volume}{45}},
  \bibinfo{pages}{31} (\bibinfo{year}{2003}), \eprint{quant-ph/0311049}.

\bibitem[{\citenamefont{Eisert et~al.}(2010)\citenamefont{Eisert, Cramer, and
  Plenio}}]{Eisert:2008ur}
\bibinfo{author}{\bibfnamefont{J.}~\bibnamefont{Eisert}},
  \bibinfo{author}{\bibfnamefont{M.}~\bibnamefont{Cramer}}, \bibnamefont{and}
  \bibinfo{author}{\bibfnamefont{M.~B.} \bibnamefont{Plenio}},
  \bibinfo{journal}{Rev. Mod. Phys.} \textbf{\bibinfo{volume}{82}},
  \bibinfo{pages}{277} (\bibinfo{year}{2010}), \eprint{0808.3773}.

\bibitem[{\citenamefont{Aharony et~al.}(2000)\citenamefont{Aharony, Gubser,
  Maldacena, Ooguri, and Oz}}]{Aharony:1999ti}
\bibinfo{author}{\bibfnamefont{O.}~\bibnamefont{Aharony}},
  \bibinfo{author}{\bibfnamefont{S.~S.} \bibnamefont{Gubser}},
  \bibinfo{author}{\bibfnamefont{J.~M.} \bibnamefont{Maldacena}},
  \bibinfo{author}{\bibfnamefont{H.}~\bibnamefont{Ooguri}}, \bibnamefont{and}
  \bibinfo{author}{\bibfnamefont{Y.}~\bibnamefont{Oz}}, \bibinfo{journal}{Phys.
  Rept.} \textbf{\bibinfo{volume}{323}}, \bibinfo{pages}{183}
  (\bibinfo{year}{2000}), \eprint{hep-th/9905111}.

\bibitem[{\citenamefont{Xu and Swingle}(2020)}]{xu_accessing_2020}
\bibinfo{author}{\bibfnamefont{S.}~\bibnamefont{Xu}} \bibnamefont{and}
  \bibinfo{author}{\bibfnamefont{B.}~\bibnamefont{Swingle}},
  \bibinfo{journal}{Nature Physics} \textbf{\bibinfo{volume}{16}},
  \bibinfo{pages}{199} (\bibinfo{year}{2020}), ISSN \bibinfo{issn}{1745-2473,
  1745-2481}, \urlprefix\url{http://www.nature.com/articles/s41567-019-0712-4}.

\bibitem[{\citenamefont{Kusuki and Miyaji}(2019)}]{kusuki_entanglement_2019}
\bibinfo{author}{\bibfnamefont{Y.}~\bibnamefont{Kusuki}} \bibnamefont{and}
  \bibinfo{author}{\bibfnamefont{M.}~\bibnamefont{Miyaji}},
  \bibinfo{journal}{Journal of High Energy Physics}
  \textbf{\bibinfo{volume}{2019}}, \bibinfo{pages}{63} (\bibinfo{year}{2019}),
  ISSN \bibinfo{issn}{1029-8479},
  \urlprefix\url{https://link.springer.com/10.1007/JHEP08(2019)063}.

\bibitem[{\citenamefont{Yuan et~al.}(2022)\citenamefont{Yuan, Zhang, Wang,
  Duan, and Deng}}]{yuan_quantum_2022}
\bibinfo{author}{\bibfnamefont{D.}~\bibnamefont{Yuan}},
  \bibinfo{author}{\bibfnamefont{S.-Y.} \bibnamefont{Zhang}},
  \bibinfo{author}{\bibfnamefont{Y.}~\bibnamefont{Wang}},
  \bibinfo{author}{\bibfnamefont{L.-M.} \bibnamefont{Duan}}, \bibnamefont{and}
  \bibinfo{author}{\bibfnamefont{D.-L.} \bibnamefont{Deng}},
  \bibinfo{journal}{Physical Review Research} \textbf{\bibinfo{volume}{4}},
  \bibinfo{pages}{023095} (\bibinfo{year}{2022}), ISSN
  \bibinfo{issn}{2643-1564},
  \urlprefix\url{https://link.aps.org/doi/10.1103/PhysRevResearch.4.023095}.

\bibitem[{\citenamefont{Bhattacharyya et~al.}(2022)\citenamefont{Bhattacharyya,
  Joshi, and Sundar}}]{bhattacharyya_quantum_2022}
\bibinfo{author}{\bibfnamefont{A.}~\bibnamefont{Bhattacharyya}},
  \bibinfo{author}{\bibfnamefont{L.~K.} \bibnamefont{Joshi}}, \bibnamefont{and}
  \bibinfo{author}{\bibfnamefont{B.}~\bibnamefont{Sundar}},
  \bibinfo{journal}{The European Physical Journal C}
  \textbf{\bibinfo{volume}{82}}, \bibinfo{pages}{458} (\bibinfo{year}{2022}),
  ISSN \bibinfo{issn}{1434-6052}, \bibinfo{note}{arXiv:2111.11945 [hep-th,
  physics:quant-ph]}, \urlprefix\url{http://arxiv.org/abs/2111.11945}.

\bibitem[{\citenamefont{Xu and Swingle}(2022)}]{xu_scrambling_2022}
\bibinfo{author}{\bibfnamefont{S.}~\bibnamefont{Xu}} \bibnamefont{and}
  \bibinfo{author}{\bibfnamefont{B.}~\bibnamefont{Swingle}},
  \emph{\bibinfo{title}{Scrambling {Dynamics} and {Out}-of-{Time} {Ordered}
  {Correlators} in {Quantum} {Many}-{Body} {Systems}: a {Tutorial}}}
  (\bibinfo{year}{2022}), \bibinfo{note}{arXiv:2202.07060 [cond-mat,
  physics:hep-th, physics:quant-ph]},
  \urlprefix\url{http://arxiv.org/abs/2202.07060}.

\bibitem[{\citenamefont{Tsuji et~al.}(2018)\citenamefont{Tsuji, Shitara, and
  Ueda}}]{Tsuji:2017fxs}
\bibinfo{author}{\bibfnamefont{N.}~\bibnamefont{Tsuji}},
  \bibinfo{author}{\bibfnamefont{T.}~\bibnamefont{Shitara}}, \bibnamefont{and}
  \bibinfo{author}{\bibfnamefont{M.}~\bibnamefont{Ueda}},
  \bibinfo{journal}{Phys. Rev. E} \textbf{\bibinfo{volume}{98}},
  \bibinfo{pages}{012216} (\bibinfo{year}{2018}), \eprint{1706.09160}.

\bibitem[{\citenamefont{Bentsen et~al.}(2019)\citenamefont{Bentsen, Gu, and
  Lucas}}]{Bentsen:2018uph}
\bibinfo{author}{\bibfnamefont{G.}~\bibnamefont{Bentsen}},
  \bibinfo{author}{\bibfnamefont{Y.}~\bibnamefont{Gu}}, \bibnamefont{and}
  \bibinfo{author}{\bibfnamefont{A.}~\bibnamefont{Lucas}},
  \bibinfo{journal}{Proc. Nat. Acad. Sci.} \textbf{\bibinfo{volume}{116}},
  \bibinfo{pages}{6689} (\bibinfo{year}{2019}), \eprint{1805.08215}.

\bibitem[{\citenamefont{Campisi and Goold}(2017)}]{Campisi:2016qlj}
\bibinfo{author}{\bibfnamefont{M.}~\bibnamefont{Campisi}} \bibnamefont{and}
  \bibinfo{author}{\bibfnamefont{J.}~\bibnamefont{Goold}},
  \bibinfo{journal}{Phys. Rev. E} \textbf{\bibinfo{volume}{95}},
  \bibinfo{pages}{062127} (\bibinfo{year}{2017}), \eprint{1609.05848}.

\bibitem[{\citenamefont{Pappalardi et~al.}(2018)\citenamefont{Pappalardi,
  Russomanno, \ifmmode \check{Z}\else \v{Z}\fi{}unkovi\ifmmode~\check{c}\else
  \v{c}\fi{}, Iemini, Silva, and Fazio}}]{PhysRevB.98.134303}
\bibinfo{author}{\bibfnamefont{S.}~\bibnamefont{Pappalardi}},
  \bibinfo{author}{\bibfnamefont{A.}~\bibnamefont{Russomanno}},
  \bibinfo{author}{\bibfnamefont{B.}~\bibnamefont{\ifmmode \check{Z}\else
  \v{Z}\fi{}unkovi\ifmmode~\check{c}\else \v{c}\fi{}}},
  \bibinfo{author}{\bibfnamefont{F.}~\bibnamefont{Iemini}},
  \bibinfo{author}{\bibfnamefont{A.}~\bibnamefont{Silva}}, \bibnamefont{and}
  \bibinfo{author}{\bibfnamefont{R.}~\bibnamefont{Fazio}},
  \bibinfo{journal}{Phys. Rev. B} \textbf{\bibinfo{volume}{98}},
  \bibinfo{pages}{134303} (\bibinfo{year}{2018}),
  \urlprefix\url{https://link.aps.org/doi/10.1103/PhysRevB.98.134303}.

\bibitem[{\citenamefont{Bohrdt et~al.}(2017)\citenamefont{Bohrdt, Mendl,
  Endres, and Knap}}]{Bohrdt:2016vhv}
\bibinfo{author}{\bibfnamefont{A.}~\bibnamefont{Bohrdt}},
  \bibinfo{author}{\bibfnamefont{C.~B.} \bibnamefont{Mendl}},
  \bibinfo{author}{\bibfnamefont{M.}~\bibnamefont{Endres}}, \bibnamefont{and}
  \bibinfo{author}{\bibfnamefont{M.}~\bibnamefont{Knap}}, \bibinfo{journal}{New
  J. Phys.} \textbf{\bibinfo{volume}{19}}, \bibinfo{pages}{063001}
  (\bibinfo{year}{2017}), \eprint{1612.02434}.

\bibitem[{\citenamefont{Smith et~al.}(2019)\citenamefont{Smith, Knolle,
  Moessner, and Kovrizhin}}]{smith2019logarithmic}
\bibinfo{author}{\bibfnamefont{A.}~\bibnamefont{Smith}},
  \bibinfo{author}{\bibfnamefont{J.}~\bibnamefont{Knolle}},
  \bibinfo{author}{\bibfnamefont{R.}~\bibnamefont{Moessner}}, \bibnamefont{and}
  \bibinfo{author}{\bibfnamefont{D.~L.} \bibnamefont{Kovrizhin}},
  \bibinfo{journal}{Physical Review Letters} \textbf{\bibinfo{volume}{123}},
  \bibinfo{pages}{086602} (\bibinfo{year}{2019}).

\bibitem[{\citenamefont{Lieb and Robinson}(2004)}]{lieb2004finite}
\bibinfo{author}{\bibfnamefont{E.~H.} \bibnamefont{Lieb}} \bibnamefont{and}
  \bibinfo{author}{\bibfnamefont{D.~W.} \bibnamefont{Robinson}},
  \emph{\bibinfo{title}{The finite group velocity of quantum spin systems}}
  (\bibinfo{publisher}{Springer}, \bibinfo{year}{2004}).

\bibitem[{\citenamefont{Shenker and Stanford}(2014)}]{Shenker:2013pqa}
\bibinfo{author}{\bibfnamefont{S.~H.} \bibnamefont{Shenker}} \bibnamefont{and}
  \bibinfo{author}{\bibfnamefont{D.}~\bibnamefont{Stanford}},
  \bibinfo{journal}{JHEP} \textbf{\bibinfo{volume}{03}}, \bibinfo{pages}{067}
  (\bibinfo{year}{2014}), \eprint{1306.0622}.

\bibitem[{\citenamefont{Shenker and Stanford}(2015)}]{Shenker:2014cwa}
\bibinfo{author}{\bibfnamefont{S.~H.} \bibnamefont{Shenker}} \bibnamefont{and}
  \bibinfo{author}{\bibfnamefont{D.}~\bibnamefont{Stanford}},
  \bibinfo{journal}{JHEP} \textbf{\bibinfo{volume}{05}}, \bibinfo{pages}{132}
  (\bibinfo{year}{2015}), \eprint{1412.6087}.

\bibitem[{\citenamefont{Maldacena et~al.}(2016)\citenamefont{Maldacena,
  Shenker, and Stanford}}]{Maldacena:2015waa}
\bibinfo{author}{\bibfnamefont{J.}~\bibnamefont{Maldacena}},
  \bibinfo{author}{\bibfnamefont{S.~H.} \bibnamefont{Shenker}},
  \bibnamefont{and} \bibinfo{author}{\bibfnamefont{D.}~\bibnamefont{Stanford}},
  \bibinfo{journal}{JHEP} \textbf{\bibinfo{volume}{08}}, \bibinfo{pages}{106}
  (\bibinfo{year}{2016}), \eprint{1503.01409}.

\bibitem[{\citenamefont{Aleiner et~al.}(2016)\citenamefont{Aleiner, Faoro, and
  Ioffe}}]{Aleiner:2016eni}
\bibinfo{author}{\bibfnamefont{I.~L.} \bibnamefont{Aleiner}},
  \bibinfo{author}{\bibfnamefont{L.}~\bibnamefont{Faoro}}, \bibnamefont{and}
  \bibinfo{author}{\bibfnamefont{L.~B.} \bibnamefont{Ioffe}},
  \bibinfo{journal}{Annals Phys.} \textbf{\bibinfo{volume}{375}},
  \bibinfo{pages}{378} (\bibinfo{year}{2016}), \eprint{1609.01251}.

\bibitem[{\citenamefont{Sekino and Susskind}(2008)}]{sekino_fast_2008}
\bibinfo{author}{\bibfnamefont{Y.}~\bibnamefont{Sekino}} \bibnamefont{and}
  \bibinfo{author}{\bibfnamefont{L.}~\bibnamefont{Susskind}},
  \bibinfo{journal}{Journal of High Energy Physics}
  \textbf{\bibinfo{volume}{2008}}, \bibinfo{pages}{065} (\bibinfo{year}{2008}),
  ISSN \bibinfo{issn}{1029-8479}, \bibinfo{note}{arXiv:0808.2096 [hep-th,
  physics:quant-ph]}, \urlprefix\url{http://arxiv.org/abs/0808.2096}.

\bibitem[{\citenamefont{Lashkari et~al.}(2013)\citenamefont{Lashkari, Stanford,
  Hastings, Osborne, and Hayden}}]{lashkari_towards_2013}
\bibinfo{author}{\bibfnamefont{N.}~\bibnamefont{Lashkari}},
  \bibinfo{author}{\bibfnamefont{D.}~\bibnamefont{Stanford}},
  \bibinfo{author}{\bibfnamefont{M.}~\bibnamefont{Hastings}},
  \bibinfo{author}{\bibfnamefont{T.}~\bibnamefont{Osborne}}, \bibnamefont{and}
  \bibinfo{author}{\bibfnamefont{P.}~\bibnamefont{Hayden}},
  \bibinfo{journal}{Journal of High Energy Physics}
  \textbf{\bibinfo{volume}{2013}}, \bibinfo{pages}{22} (\bibinfo{year}{2013}),
  ISSN \bibinfo{issn}{1029-8479}, \bibinfo{note}{arXiv:1111.6580 [hep-th,
  physics:quant-ph]}, \urlprefix\url{http://arxiv.org/abs/1111.6580}.

\bibitem[{\citenamefont{Sachdev and Ye}(1993)}]{sachdev1993gapless}
\bibinfo{author}{\bibfnamefont{S.}~\bibnamefont{Sachdev}} \bibnamefont{and}
  \bibinfo{author}{\bibfnamefont{J.}~\bibnamefont{Ye}},
  \bibinfo{journal}{Physical review letters} \textbf{\bibinfo{volume}{70}},
  \bibinfo{pages}{3339} (\bibinfo{year}{1993}).

\bibitem[{noa({\natexlab{a}})}]{noauthor_alexei_nodate}
\emph{\bibinfo{title}{Alexei {Kitaev}, {Caltech} \& {KITP}, {A} simple model of
  quantum holography (part 1)}},
  \urlprefix\url{https://online.kitp.ucsb.edu/online/entangled15/kitaev/}.

\bibitem[{noa({\natexlab{b}})}]{noauthor_alexei_nodate2}
\emph{\bibinfo{title}{Alexei {Kitaev}, {Caltech}, {A} simple model of quantum
  holography (part 2)}},
  \urlprefix\url{https://online.kitp.ucsb.edu/online/entangled15/kitaev2/}.

\bibitem[{\citenamefont{Maldacena and Stanford}(2016)}]{PhysRevD.94.106002}
\bibinfo{author}{\bibfnamefont{J.}~\bibnamefont{Maldacena}} \bibnamefont{and}
  \bibinfo{author}{\bibfnamefont{D.}~\bibnamefont{Stanford}},
  \bibinfo{journal}{Phys. Rev. D} \textbf{\bibinfo{volume}{94}},
  \bibinfo{pages}{106002} (\bibinfo{year}{2016}),
  \urlprefix\url{https://link.aps.org/doi/10.1103/PhysRevD.94.106002}.

\bibitem[{\citenamefont{Polchinski and Rosenhaus}(2016)}]{Polchinski:2016xgd}
\bibinfo{author}{\bibfnamefont{J.}~\bibnamefont{Polchinski}} \bibnamefont{and}
  \bibinfo{author}{\bibfnamefont{V.}~\bibnamefont{Rosenhaus}},
  \bibinfo{journal}{JHEP} \textbf{\bibinfo{volume}{04}}, \bibinfo{pages}{001}
  (\bibinfo{year}{2016}), \eprint{1601.06768}.

\bibitem[{\citenamefont{White and Feiguin}(2004)}]{white2004real}
\bibinfo{author}{\bibfnamefont{S.~R.} \bibnamefont{White}} \bibnamefont{and}
  \bibinfo{author}{\bibfnamefont{A.~E.} \bibnamefont{Feiguin}},
  \bibinfo{journal}{Physical review letters} \textbf{\bibinfo{volume}{93}},
  \bibinfo{pages}{076401} (\bibinfo{year}{2004}).

\bibitem[{\citenamefont{White}(1993)}]{PhysRevB.48.10345}
\bibinfo{author}{\bibfnamefont{S.~R.} \bibnamefont{White}},
  \bibinfo{journal}{Phys. Rev. B} \textbf{\bibinfo{volume}{48}},
  \bibinfo{pages}{10345} (\bibinfo{year}{1993}),
  \urlprefix\url{https://link.aps.org/doi/10.1103/PhysRevB.48.10345}.

\bibitem[{\citenamefont{Schollw\"ock}(2005)}]{RevModPhys.77.259}
\bibinfo{author}{\bibfnamefont{U.}~\bibnamefont{Schollw\"ock}},
  \bibinfo{journal}{Rev. Mod. Phys.} \textbf{\bibinfo{volume}{77}},
  \bibinfo{pages}{259} (\bibinfo{year}{2005}),
  \urlprefix\url{https://link.aps.org/doi/10.1103/RevModPhys.77.259}.

\bibitem[{\citenamefont{Vidal}(2003{\natexlab{a}})}]{PhysRevLett.91.147902}
\bibinfo{author}{\bibfnamefont{G.}~\bibnamefont{Vidal}},
  \bibinfo{journal}{Phys. Rev. Lett.} \textbf{\bibinfo{volume}{91}},
  \bibinfo{pages}{147902} (\bibinfo{year}{2003}{\natexlab{a}}),
  \urlprefix\url{https://link.aps.org/doi/10.1103/PhysRevLett.91.147902}.

\bibitem[{\citenamefont{Verstraete et~al.}(2004)\citenamefont{Verstraete,
  Garcia-Ripoll, and Cirac}}]{verstraete2004matrix}
\bibinfo{author}{\bibfnamefont{F.}~\bibnamefont{Verstraete}},
  \bibinfo{author}{\bibfnamefont{J.~J.} \bibnamefont{Garcia-Ripoll}},
  \bibnamefont{and} \bibinfo{author}{\bibfnamefont{J.~I.} \bibnamefont{Cirac}},
  \bibinfo{journal}{Physical review letters} \textbf{\bibinfo{volume}{93}},
  \bibinfo{pages}{207204} (\bibinfo{year}{2004}).

\bibitem[{\citenamefont{Vidal}(2004)}]{PhysRevLett.93.040502}
\bibinfo{author}{\bibfnamefont{G.}~\bibnamefont{Vidal}},
  \bibinfo{journal}{Phys. Rev. Lett.} \textbf{\bibinfo{volume}{93}},
  \bibinfo{pages}{040502} (\bibinfo{year}{2004}),
  \urlprefix\url{https://link.aps.org/doi/10.1103/PhysRevLett.93.040502}.

\bibitem[{\citenamefont{Fishman et~al.}(2020)\citenamefont{Fishman, White, and
  Stoudenmire}}]{Fishman:2020gel}
\bibinfo{author}{\bibfnamefont{M.}~\bibnamefont{Fishman}},
  \bibinfo{author}{\bibfnamefont{S.~R.} \bibnamefont{White}}, \bibnamefont{and}
  \bibinfo{author}{\bibfnamefont{E.~M.} \bibnamefont{Stoudenmire}}
  (\bibinfo{year}{2020}), \eprint{2007.14822}.

\bibitem[{\citenamefont{Huang et~al.}(2017)\citenamefont{Huang, Zhang, and
  Chen}}]{Huang:2016knw}
\bibinfo{author}{\bibfnamefont{Y.}~\bibnamefont{Huang}},
  \bibinfo{author}{\bibfnamefont{Y.-L.} \bibnamefont{Zhang}}, \bibnamefont{and}
  \bibinfo{author}{\bibfnamefont{X.}~\bibnamefont{Chen}},
  \bibinfo{journal}{Annalen Phys.} \textbf{\bibinfo{volume}{529}},
  \bibinfo{pages}{1600318} (\bibinfo{year}{2017}), \eprint{1608.01091}.

\bibitem[{\citenamefont{Chen et~al.}(2017)\citenamefont{Chen, Zhou, Huse, and
  Fradkin}}]{Chen:2016qpx}
\bibinfo{author}{\bibfnamefont{X.}~\bibnamefont{Chen}},
  \bibinfo{author}{\bibfnamefont{T.}~\bibnamefont{Zhou}},
  \bibinfo{author}{\bibfnamefont{D.~A.} \bibnamefont{Huse}}, \bibnamefont{and}
  \bibinfo{author}{\bibfnamefont{E.}~\bibnamefont{Fradkin}},
  \bibinfo{journal}{Annalen Phys.} \textbf{\bibinfo{volume}{529}},
  \bibinfo{pages}{1600332} (\bibinfo{year}{2017}), \eprint{1610.00220}.

\bibitem[{\citenamefont{Vermersch
  et~al.}(2019{\natexlab{a}})\citenamefont{Vermersch, Elben, Sieberer, Yao, and
  Zoller}}]{vermersch_probing_2019}
\bibinfo{author}{\bibfnamefont{B.}~\bibnamefont{Vermersch}},
  \bibinfo{author}{\bibfnamefont{A.}~\bibnamefont{Elben}},
  \bibinfo{author}{\bibfnamefont{L.~M.} \bibnamefont{Sieberer}},
  \bibinfo{author}{\bibfnamefont{N.~Y.} \bibnamefont{Yao}}, \bibnamefont{and}
  \bibinfo{author}{\bibfnamefont{P.}~\bibnamefont{Zoller}},
  \bibinfo{journal}{Physical Review X} \textbf{\bibinfo{volume}{9}},
  \bibinfo{pages}{021061} (\bibinfo{year}{2019}{\natexlab{a}}), ISSN
  \bibinfo{issn}{2160-3308}, \bibinfo{note}{arXiv:1807.09087 [cond-mat,
  physics:hep-th, physics:physics, physics:quant-ph]},
  \urlprefix\url{http://arxiv.org/abs/1807.09087}.

\bibitem[{\citenamefont{Asaduzzaman
  et~al.}(2022{\natexlab{b}})\citenamefont{Asaduzzaman, Catterall, Hubisz,
  Nelson, and Unmuth-Yockey}}]{asaduzzaman2022holography}
\bibinfo{author}{\bibfnamefont{M.}~\bibnamefont{Asaduzzaman}},
  \bibinfo{author}{\bibfnamefont{S.}~\bibnamefont{Catterall}},
  \bibinfo{author}{\bibfnamefont{J.}~\bibnamefont{Hubisz}},
  \bibinfo{author}{\bibfnamefont{R.}~\bibnamefont{Nelson}}, \bibnamefont{and}
  \bibinfo{author}{\bibfnamefont{J.}~\bibnamefont{Unmuth-Yockey}},
  \bibinfo{journal}{Physical Review D} \textbf{\bibinfo{volume}{106}},
  \bibinfo{pages}{054506} (\bibinfo{year}{2022}{\natexlab{b}}).

\bibitem[{\citenamefont{Brower et~al.}(2021{\natexlab{b}})\citenamefont{Brower,
  Cogburn, Fitzpatrick, Howarth, and Tan}}]{brower2021lattice}
\bibinfo{author}{\bibfnamefont{R.~C.} \bibnamefont{Brower}},
  \bibinfo{author}{\bibfnamefont{C.~V.} \bibnamefont{Cogburn}},
  \bibinfo{author}{\bibfnamefont{A.~L.} \bibnamefont{Fitzpatrick}},
  \bibinfo{author}{\bibfnamefont{D.}~\bibnamefont{Howarth}}, \bibnamefont{and}
  \bibinfo{author}{\bibfnamefont{C.-I.} \bibnamefont{Tan}},
  \bibinfo{journal}{Physical Review D} \textbf{\bibinfo{volume}{103}},
  \bibinfo{pages}{094507} (\bibinfo{year}{2021}{\natexlab{b}}).

\bibitem[{\citenamefont{Ueda et~al.}(2011)\citenamefont{Ueda, Nakano, Kusakabe,
  and Nishino}}]{ueda_hyperbolic_2011}
\bibinfo{author}{\bibfnamefont{H.}~\bibnamefont{Ueda}},
  \bibinfo{author}{\bibfnamefont{H.}~\bibnamefont{Nakano}},
  \bibinfo{author}{\bibfnamefont{K.}~\bibnamefont{Kusakabe}}, \bibnamefont{and}
  \bibinfo{author}{\bibfnamefont{T.}~\bibnamefont{Nishino}},
  \bibinfo{journal}{Journal of the Physical Society of Japan}
  \textbf{\bibinfo{volume}{80}}, \bibinfo{pages}{094001}
  (\bibinfo{year}{2011}), ISSN \bibinfo{issn}{0031-9015, 1347-4073},
  \bibinfo{note}{arXiv:1102.0845 [cond-mat]},
  \urlprefix\url{http://arxiv.org/abs/1102.0845}.

\bibitem[{\citenamefont{Ueda et~al.}(2010{\natexlab{a}})\citenamefont{Ueda,
  Gendiar, Zauner, Iharagi, and Nishino}}]{ueda_transverse_2010}
\bibinfo{author}{\bibfnamefont{H.}~\bibnamefont{Ueda}},
  \bibinfo{author}{\bibfnamefont{A.}~\bibnamefont{Gendiar}},
  \bibinfo{author}{\bibfnamefont{V.}~\bibnamefont{Zauner}},
  \bibinfo{author}{\bibfnamefont{T.}~\bibnamefont{Iharagi}}, \bibnamefont{and}
  \bibinfo{author}{\bibfnamefont{T.}~\bibnamefont{Nishino}},
  \emph{\bibinfo{title}{Transverse {Field} {Ising} {Model} {Under} {Hyperbolic}
  {Deformation}}} (\bibinfo{year}{2010}{\natexlab{a}}),
  \bibinfo{note}{arXiv:1008.3458 [cond-mat, physics:quant-ph]},
  \urlprefix\url{http://arxiv.org/abs/1008.3458}.

\bibitem[{\citenamefont{Ueda et~al.}(2010{\natexlab{b}})\citenamefont{Ueda,
  Nakano, Kusakabe, and Nishino}}]{ueda_scaling_2010}
\bibinfo{author}{\bibfnamefont{H.}~\bibnamefont{Ueda}},
  \bibinfo{author}{\bibfnamefont{H.}~\bibnamefont{Nakano}},
  \bibinfo{author}{\bibfnamefont{K.}~\bibnamefont{Kusakabe}}, \bibnamefont{and}
  \bibinfo{author}{\bibfnamefont{T.}~\bibnamefont{Nishino}},
  \bibinfo{journal}{Progress of Theoretical Physics}
  \textbf{\bibinfo{volume}{124}}, \bibinfo{pages}{389}
  (\bibinfo{year}{2010}{\natexlab{b}}), ISSN \bibinfo{issn}{0033-068X,
  1347-4081}, \bibinfo{note}{arXiv:1006.2652 [cond-mat, physics:quant-ph]},
  \urlprefix\url{http://arxiv.org/abs/1006.2652}.

\bibitem[{\citenamefont{Suzuki}(1976)}]{suzuki1976generalized}
\bibinfo{author}{\bibfnamefont{M.}~\bibnamefont{Suzuki}},
  \bibinfo{journal}{Communications in Mathematical Physics}
  \textbf{\bibinfo{volume}{51}}, \bibinfo{pages}{183} (\bibinfo{year}{1976}).

\bibitem[{\citenamefont{Vidal}(2003{\natexlab{b}})}]{vidal2003efficient}
\bibinfo{author}{\bibfnamefont{G.}~\bibnamefont{Vidal}},
  \bibinfo{journal}{Physical review letters} \textbf{\bibinfo{volume}{91}},
  \bibinfo{pages}{147902} (\bibinfo{year}{2003}{\natexlab{b}}).

\bibitem[{\citenamefont{Gustafson
  et~al.}(2021{\natexlab{a}})\citenamefont{Gustafson, Dreher, Hang, and
  Meurice}}]{gustafson2021indexed}
\bibinfo{author}{\bibfnamefont{E.}~\bibnamefont{Gustafson}},
  \bibinfo{author}{\bibfnamefont{P.}~\bibnamefont{Dreher}},
  \bibinfo{author}{\bibfnamefont{Z.}~\bibnamefont{Hang}}, \bibnamefont{and}
  \bibinfo{author}{\bibfnamefont{Y.}~\bibnamefont{Meurice}},
  \bibinfo{journal}{Quantum Science and Technology}
  \textbf{\bibinfo{volume}{6}}, \bibinfo{pages}{045020}
  (\bibinfo{year}{2021}{\natexlab{a}}).

\bibitem[{\citenamefont{Meurice}(2021)}]{meurice2021quantum}
\bibinfo{author}{\bibfnamefont{Y.}~\bibnamefont{Meurice}},
  \emph{\bibinfo{title}{Quantum Field Theory}} (\bibinfo{publisher}{IOP
  Publishing}, \bibinfo{year}{2021}).

\bibitem[{\citenamefont{Asaduzzaman
  et~al.}(2022{\natexlab{c}})\citenamefont{Asaduzzaman, Toga, Catterall,
  Meurice, and Sakai}}]{asaduzzaman2022quantum}
\bibinfo{author}{\bibfnamefont{M.}~\bibnamefont{Asaduzzaman}},
  \bibinfo{author}{\bibfnamefont{G.~C.} \bibnamefont{Toga}},
  \bibinfo{author}{\bibfnamefont{S.}~\bibnamefont{Catterall}},
  \bibinfo{author}{\bibfnamefont{Y.}~\bibnamefont{Meurice}}, \bibnamefont{and}
  \bibinfo{author}{\bibfnamefont{R.}~\bibnamefont{Sakai}},
  \bibinfo{journal}{Physical Review D} \textbf{\bibinfo{volume}{106}},
  \bibinfo{pages}{114515} (\bibinfo{year}{2022}{\natexlab{c}}).

\bibitem[{\citenamefont{Viola and Lloyd}(1998)}]{viola1998dynamical}
\bibinfo{author}{\bibfnamefont{L.}~\bibnamefont{Viola}} \bibnamefont{and}
  \bibinfo{author}{\bibfnamefont{S.}~\bibnamefont{Lloyd}},
  \bibinfo{journal}{Physical Review A} \textbf{\bibinfo{volume}{58}},
  \bibinfo{pages}{2733} (\bibinfo{year}{1998}).

\bibitem[{\citenamefont{Charles et~al.}(2023)\citenamefont{Charles, Gustafson,
  Hardt, Herren, Hogan, Lamm, Starecheski, Van~de Water, and
  Wagman}}]{charles2023simulating}
\bibinfo{author}{\bibfnamefont{C.}~\bibnamefont{Charles}},
  \bibinfo{author}{\bibfnamefont{E.~J.} \bibnamefont{Gustafson}},
  \bibinfo{author}{\bibfnamefont{E.}~\bibnamefont{Hardt}},
  \bibinfo{author}{\bibfnamefont{F.}~\bibnamefont{Herren}},
  \bibinfo{author}{\bibfnamefont{N.}~\bibnamefont{Hogan}},
  \bibinfo{author}{\bibfnamefont{H.}~\bibnamefont{Lamm}},
  \bibinfo{author}{\bibfnamefont{S.}~\bibnamefont{Starecheski}},
  \bibinfo{author}{\bibfnamefont{R.~S.} \bibnamefont{Van~de Water}},
  \bibnamefont{and} \bibinfo{author}{\bibfnamefont{M.~L.}
  \bibnamefont{Wagman}}, \bibinfo{journal}{arXiv preprint arXiv:2305.02361}
  (\bibinfo{year}{2023}).

\bibitem[{\citenamefont{Nation et~al.}(2021)\citenamefont{Nation, Kang,
  Sundaresan, and Gambetta}}]{nation2021scalable}
\bibinfo{author}{\bibfnamefont{P.~D.} \bibnamefont{Nation}},
  \bibinfo{author}{\bibfnamefont{H.}~\bibnamefont{Kang}},
  \bibinfo{author}{\bibfnamefont{N.}~\bibnamefont{Sundaresan}},
  \bibnamefont{and} \bibinfo{author}{\bibfnamefont{J.~M.}
  \bibnamefont{Gambetta}}, \bibinfo{journal}{PRX Quantum}
  \textbf{\bibinfo{volume}{2}}, \bibinfo{pages}{040326} (\bibinfo{year}{2021}).

\bibitem[{\citenamefont{Temme et~al.}(2017)\citenamefont{Temme, Bravyi, and
  Gambetta}}]{temme2017error}
\bibinfo{author}{\bibfnamefont{K.}~\bibnamefont{Temme}},
  \bibinfo{author}{\bibfnamefont{S.}~\bibnamefont{Bravyi}}, \bibnamefont{and}
  \bibinfo{author}{\bibfnamefont{J.~M.} \bibnamefont{Gambetta}},
  \bibinfo{journal}{Physical review letters} \textbf{\bibinfo{volume}{119}},
  \bibinfo{pages}{180509} (\bibinfo{year}{2017}).

\bibitem[{\citenamefont{Li and Benjamin}(2017)}]{PhysRevX.7.021050}
\bibinfo{author}{\bibfnamefont{Y.}~\bibnamefont{Li}} \bibnamefont{and}
  \bibinfo{author}{\bibfnamefont{S.~C.} \bibnamefont{Benjamin}},
  \bibinfo{journal}{Phys. Rev. X} \textbf{\bibinfo{volume}{7}},
  \bibinfo{pages}{021050} (\bibinfo{year}{2017}),
  \urlprefix\url{https://link.aps.org/doi/10.1103/PhysRevX.7.021050}.

\bibitem[{\citenamefont{Swingle}(2018)}]{Swingle:2018ekw}
\bibinfo{author}{\bibfnamefont{B.}~\bibnamefont{Swingle}},
  \bibinfo{journal}{Nature Phys.} \textbf{\bibinfo{volume}{14}},
  \bibinfo{pages}{988} (\bibinfo{year}{2018}).

\bibitem[{\citenamefont{Lin and Motrunich}(2018)}]{PhysRevB.97.144304}
\bibinfo{author}{\bibfnamefont{C.-J.} \bibnamefont{Lin}} \bibnamefont{and}
  \bibinfo{author}{\bibfnamefont{O.~I.} \bibnamefont{Motrunich}},
  \bibinfo{journal}{Phys. Rev. B} \textbf{\bibinfo{volume}{97}},
  \bibinfo{pages}{144304} (\bibinfo{year}{2018}),
  \urlprefix\url{https://link.aps.org/doi/10.1103/PhysRevB.97.144304}.

\bibitem[{\citenamefont{Garc\'\i{}a-Mata
  et~al.}(2023)\citenamefont{Garc\'\i{}a-Mata, Jalabert, and
  Wisniacki}}]{Garcia-Mata:2022voo}
\bibinfo{author}{\bibfnamefont{I.}~\bibnamefont{Garc\'\i{}a-Mata}},
  \bibinfo{author}{\bibfnamefont{R.~A.} \bibnamefont{Jalabert}},
  \bibnamefont{and} \bibinfo{author}{\bibfnamefont{D.~A.}
  \bibnamefont{Wisniacki}}, \bibinfo{journal}{Scholarpedia}
  \textbf{\bibinfo{volume}{18}}, \bibinfo{pages}{55237} (\bibinfo{year}{2023}),
  \eprint{2209.07965}.

\bibitem[{\citenamefont{Goussev et~al.}(2012)\citenamefont{Goussev, Jalabert,
  Pastawski, and Wisniacki}}]{goussev2012loschmidt}
\bibinfo{author}{\bibfnamefont{A.}~\bibnamefont{Goussev}},
  \bibinfo{author}{\bibfnamefont{R.~A.} \bibnamefont{Jalabert}},
  \bibinfo{author}{\bibfnamefont{H.~M.} \bibnamefont{Pastawski}},
  \bibnamefont{and}
  \bibinfo{author}{\bibfnamefont{D.}~\bibnamefont{Wisniacki}},
  \bibinfo{journal}{arXiv preprint arXiv:1206.6348}  (\bibinfo{year}{2012}).

\bibitem[{\citenamefont{Joshi et~al.}(2022)\citenamefont{Joshi, Elben, Vikram,
  Vermersch, Galitski, and Zoller}}]{joshi_probing_2022}
\bibinfo{author}{\bibfnamefont{L.~K.} \bibnamefont{Joshi}},
  \bibinfo{author}{\bibfnamefont{A.}~\bibnamefont{Elben}},
  \bibinfo{author}{\bibfnamefont{A.}~\bibnamefont{Vikram}},
  \bibinfo{author}{\bibfnamefont{B.}~\bibnamefont{Vermersch}},
  \bibinfo{author}{\bibfnamefont{V.}~\bibnamefont{Galitski}}, \bibnamefont{and}
  \bibinfo{author}{\bibfnamefont{P.}~\bibnamefont{Zoller}},
  \bibinfo{journal}{Physical Review X} \textbf{\bibinfo{volume}{12}},
  \bibinfo{pages}{011018} (\bibinfo{year}{2022}), ISSN
  \bibinfo{issn}{2160-3308}, \bibinfo{note}{arXiv:2106.15530 [cond-mat,
  physics:hep-th, physics:nlin, physics:quant-ph]},
  \urlprefix\url{http://arxiv.org/abs/2106.15530}.

\bibitem[{\citenamefont{Fan et~al.}(2017)\citenamefont{Fan, Zhang, Shen, and
  Zhai}}]{fan2017out}
\bibinfo{author}{\bibfnamefont{R.}~\bibnamefont{Fan}},
  \bibinfo{author}{\bibfnamefont{P.}~\bibnamefont{Zhang}},
  \bibinfo{author}{\bibfnamefont{H.}~\bibnamefont{Shen}}, \bibnamefont{and}
  \bibinfo{author}{\bibfnamefont{H.}~\bibnamefont{Zhai}},
  \bibinfo{journal}{Science bulletin} \textbf{\bibinfo{volume}{62}},
  \bibinfo{pages}{707} (\bibinfo{year}{2017}).

\bibitem[{\citenamefont{Sundar et~al.}(2022)\citenamefont{Sundar, Elben, Joshi,
  and Zache}}]{sundar2022proposal}
\bibinfo{author}{\bibfnamefont{B.}~\bibnamefont{Sundar}},
  \bibinfo{author}{\bibfnamefont{A.}~\bibnamefont{Elben}},
  \bibinfo{author}{\bibfnamefont{L.~K.} \bibnamefont{Joshi}}, \bibnamefont{and}
  \bibinfo{author}{\bibfnamefont{T.~V.} \bibnamefont{Zache}},
  \bibinfo{journal}{New Journal of Physics} \textbf{\bibinfo{volume}{24}},
  \bibinfo{pages}{023037} (\bibinfo{year}{2022}).

\bibitem[{\citenamefont{Geller et~al.}(2022)\citenamefont{Geller, Arrasmith,
  Holmes, Yan, Coles, and Sornborger}}]{geller2022quantum}
\bibinfo{author}{\bibfnamefont{M.~R.} \bibnamefont{Geller}},
  \bibinfo{author}{\bibfnamefont{A.}~\bibnamefont{Arrasmith}},
  \bibinfo{author}{\bibfnamefont{Z.}~\bibnamefont{Holmes}},
  \bibinfo{author}{\bibfnamefont{B.}~\bibnamefont{Yan}},
  \bibinfo{author}{\bibfnamefont{P.~J.} \bibnamefont{Coles}}, \bibnamefont{and}
  \bibinfo{author}{\bibfnamefont{A.}~\bibnamefont{Sornborger}},
  \bibinfo{journal}{Physical Review E} \textbf{\bibinfo{volume}{105}},
  \bibinfo{pages}{035302} (\bibinfo{year}{2022}).

\bibitem[{\citenamefont{Yao et~al.}(2016)\citenamefont{Yao, Grusdt, Swingle,
  Lukin, Stamper-Kurn, Moore, and Demler}}]{yao2016interferometric}
\bibinfo{author}{\bibfnamefont{N.~Y.} \bibnamefont{Yao}},
  \bibinfo{author}{\bibfnamefont{F.}~\bibnamefont{Grusdt}},
  \bibinfo{author}{\bibfnamefont{B.}~\bibnamefont{Swingle}},
  \bibinfo{author}{\bibfnamefont{M.~D.} \bibnamefont{Lukin}},
  \bibinfo{author}{\bibfnamefont{D.~M.} \bibnamefont{Stamper-Kurn}},
  \bibinfo{author}{\bibfnamefont{J.~E.} \bibnamefont{Moore}}, \bibnamefont{and}
  \bibinfo{author}{\bibfnamefont{E.~A.} \bibnamefont{Demler}},
  \bibinfo{journal}{arXiv preprint arXiv:1607.01801}  (\bibinfo{year}{2016}).

\bibitem[{\citenamefont{Joshi et~al.}(2020{\natexlab{a}})\citenamefont{Joshi,
  Elben, Vermersch, Brydges, Maier, Zoller, Blatt, and
  Roos}}]{joshi2020quantum}
\bibinfo{author}{\bibfnamefont{M.~K.} \bibnamefont{Joshi}},
  \bibinfo{author}{\bibfnamefont{A.}~\bibnamefont{Elben}},
  \bibinfo{author}{\bibfnamefont{B.}~\bibnamefont{Vermersch}},
  \bibinfo{author}{\bibfnamefont{T.}~\bibnamefont{Brydges}},
  \bibinfo{author}{\bibfnamefont{C.}~\bibnamefont{Maier}},
  \bibinfo{author}{\bibfnamefont{P.}~\bibnamefont{Zoller}},
  \bibinfo{author}{\bibfnamefont{R.}~\bibnamefont{Blatt}}, \bibnamefont{and}
  \bibinfo{author}{\bibfnamefont{C.~F.} \bibnamefont{Roos}},
  \bibinfo{journal}{Physical Review Letters} \textbf{\bibinfo{volume}{124}},
  \bibinfo{pages}{240505} (\bibinfo{year}{2020}{\natexlab{a}}).

\bibitem[{\citenamefont{Swingle et~al.}(2016)\citenamefont{Swingle, Bentsen,
  Schleier-Smith, and Hayden}}]{swingle2016measuring}
\bibinfo{author}{\bibfnamefont{B.}~\bibnamefont{Swingle}},
  \bibinfo{author}{\bibfnamefont{G.}~\bibnamefont{Bentsen}},
  \bibinfo{author}{\bibfnamefont{M.}~\bibnamefont{Schleier-Smith}},
  \bibnamefont{and} \bibinfo{author}{\bibfnamefont{P.}~\bibnamefont{Hayden}},
  \bibinfo{journal}{Physical Review A} \textbf{\bibinfo{volume}{94}},
  \bibinfo{pages}{040302} (\bibinfo{year}{2016}).

\bibitem[{\citenamefont{Halpern}(2017)}]{halpern2017jarzynski}
\bibinfo{author}{\bibfnamefont{N.~Y.} \bibnamefont{Halpern}},
  \bibinfo{journal}{Physical Review A} \textbf{\bibinfo{volume}{95}},
  \bibinfo{pages}{012120} (\bibinfo{year}{2017}).

\bibitem[{\citenamefont{Dressel et~al.}(2018)\citenamefont{Dressel,
  Gonz\'alez~Alonso, Waegell, and Yunger~Halpern}}]{PhysRevA.98.012132}
\bibinfo{author}{\bibfnamefont{J.}~\bibnamefont{Dressel}},
  \bibinfo{author}{\bibfnamefont{J.~R.} \bibnamefont{Gonz\'alez~Alonso}},
  \bibinfo{author}{\bibfnamefont{M.}~\bibnamefont{Waegell}}, \bibnamefont{and}
  \bibinfo{author}{\bibfnamefont{N.}~\bibnamefont{Yunger~Halpern}},
  \bibinfo{journal}{Phys. Rev. A} \textbf{\bibinfo{volume}{98}},
  \bibinfo{pages}{012132} (\bibinfo{year}{2018}),
  \urlprefix\url{https://link.aps.org/doi/10.1103/PhysRevA.98.012132}.

\bibitem[{\citenamefont{Joshi et~al.}(2020{\natexlab{b}})\citenamefont{Joshi,
  Elben, Vermersch, Brydges, Maier, Zoller, Blatt, and
  Roos}}]{joshi_quantum_2020}
\bibinfo{author}{\bibfnamefont{M.~K.} \bibnamefont{Joshi}},
  \bibinfo{author}{\bibfnamefont{A.}~\bibnamefont{Elben}},
  \bibinfo{author}{\bibfnamefont{B.}~\bibnamefont{Vermersch}},
  \bibinfo{author}{\bibfnamefont{T.}~\bibnamefont{Brydges}},
  \bibinfo{author}{\bibfnamefont{C.}~\bibnamefont{Maier}},
  \bibinfo{author}{\bibfnamefont{P.}~\bibnamefont{Zoller}},
  \bibinfo{author}{\bibfnamefont{R.}~\bibnamefont{Blatt}}, \bibnamefont{and}
  \bibinfo{author}{\bibfnamefont{C.~F.} \bibnamefont{Roos}},
  \bibinfo{journal}{Physical Review Letters} \textbf{\bibinfo{volume}{124}},
  \bibinfo{pages}{240505} (\bibinfo{year}{2020}{\natexlab{b}}), ISSN
  \bibinfo{issn}{0031-9007, 1079-7114},
  \urlprefix\url{https://link.aps.org/doi/10.1103/PhysRevLett.124.240505}.

\bibitem[{\citenamefont{Vermersch
  et~al.}(2019{\natexlab{b}})\citenamefont{Vermersch, Elben, Sieberer, Yao, and
  Zoller}}]{vermersch2019probing}
\bibinfo{author}{\bibfnamefont{B.}~\bibnamefont{Vermersch}},
  \bibinfo{author}{\bibfnamefont{A.}~\bibnamefont{Elben}},
  \bibinfo{author}{\bibfnamefont{L.~M.} \bibnamefont{Sieberer}},
  \bibinfo{author}{\bibfnamefont{N.~Y.} \bibnamefont{Yao}}, \bibnamefont{and}
  \bibinfo{author}{\bibfnamefont{P.}~\bibnamefont{Zoller}},
  \bibinfo{journal}{Physical Review X} \textbf{\bibinfo{volume}{9}},
  \bibinfo{pages}{021061} (\bibinfo{year}{2019}{\natexlab{b}}).

\bibitem[{\citenamefont{Mezzadri}(2006)}]{mezzadri2006generate}
\bibinfo{author}{\bibfnamefont{F.}~\bibnamefont{Mezzadri}},
  \bibinfo{journal}{arXiv preprint math-ph/0609050}  (\bibinfo{year}{2006}).

\bibitem[{\citenamefont{Gustafson
  et~al.}(2021{\natexlab{b}})\citenamefont{Gustafson, Zhu, Dreher, Linke, and
  Meurice}}]{Gustafson:2021imb}
\bibinfo{author}{\bibfnamefont{E.}~\bibnamefont{Gustafson}},
  \bibinfo{author}{\bibfnamefont{Y.}~\bibnamefont{Zhu}},
  \bibinfo{author}{\bibfnamefont{P.}~\bibnamefont{Dreher}},
  \bibinfo{author}{\bibfnamefont{N.~M.} \bibnamefont{Linke}}, \bibnamefont{and}
  \bibinfo{author}{\bibfnamefont{Y.}~\bibnamefont{Meurice}},
  \bibinfo{journal}{Phys. Rev. D} \textbf{\bibinfo{volume}{104}},
  \bibinfo{pages}{054507} (\bibinfo{year}{2021}{\natexlab{b}}),
  \eprint{2103.06848}.

\bibitem[{\citenamefont{Gustafson
  et~al.}(2021{\natexlab{c}})\citenamefont{Gustafson, Dreher, Hang, and
  Meurice}}]{Gustafson:2021mky}
\bibinfo{author}{\bibfnamefont{E.}~\bibnamefont{Gustafson}},
  \bibinfo{author}{\bibfnamefont{P.}~\bibnamefont{Dreher}},
  \bibinfo{author}{\bibfnamefont{Z.}~\bibnamefont{Hang}}, \bibnamefont{and}
  \bibinfo{author}{\bibfnamefont{Y.}~\bibnamefont{Meurice}},
  \bibinfo{journal}{Quantum Sci. Technol.} \textbf{\bibinfo{volume}{6}},
  \bibinfo{pages}{045020} (\bibinfo{year}{2021}{\natexlab{c}}).

\bibitem[{\citenamefont{Takayanagi}(2011)}]{Takayanagi:2011zk}
\bibinfo{author}{\bibfnamefont{T.}~\bibnamefont{Takayanagi}},
  \bibinfo{journal}{Phys. Rev. Lett.} \textbf{\bibinfo{volume}{107}},
  \bibinfo{pages}{101602} (\bibinfo{year}{2011}), \eprint{1105.5165}.

\bibitem[{\citenamefont{Fujita et~al.}(2011)\citenamefont{Fujita, Takayanagi,
  and Tonni}}]{Fujita:2011fp}
\bibinfo{author}{\bibfnamefont{M.}~\bibnamefont{Fujita}},
  \bibinfo{author}{\bibfnamefont{T.}~\bibnamefont{Takayanagi}},
  \bibnamefont{and} \bibinfo{author}{\bibfnamefont{E.}~\bibnamefont{Tonni}},
  \bibinfo{journal}{JHEP} \textbf{\bibinfo{volume}{11}}, \bibinfo{pages}{043}
  (\bibinfo{year}{2011}), \eprint{1108.5152}.

\bibitem[{\citenamefont{Dey and S\"oderberg}(2021)}]{Dey:2020jlc}
\bibinfo{author}{\bibfnamefont{P.}~\bibnamefont{Dey}} \bibnamefont{and}
  \bibinfo{author}{\bibfnamefont{A.}~\bibnamefont{S\"oderberg}},
  \bibinfo{journal}{JHEP} \textbf{\bibinfo{volume}{07}}, \bibinfo{pages}{013}
  (\bibinfo{year}{2021}), \eprint{2012.11344}.

\bibitem[{\citenamefont{Cong et~al.}(2022)\citenamefont{Cong, Levine, Keesling,
  Bluvstein, Wang, and Lukin}}]{Cong:2021nhm}
\bibinfo{author}{\bibfnamefont{I.}~\bibnamefont{Cong}},
  \bibinfo{author}{\bibfnamefont{H.}~\bibnamefont{Levine}},
  \bibinfo{author}{\bibfnamefont{A.}~\bibnamefont{Keesling}},
  \bibinfo{author}{\bibfnamefont{D.}~\bibnamefont{Bluvstein}},
  \bibinfo{author}{\bibfnamefont{S.-T.} \bibnamefont{Wang}}, \bibnamefont{and}
  \bibinfo{author}{\bibfnamefont{M.~D.} \bibnamefont{Lukin}},
  \bibinfo{journal}{Phys. Rev. X} \textbf{\bibinfo{volume}{12}},
  \bibinfo{pages}{021049} (\bibinfo{year}{2022}), \eprint{2105.13501}.

\bibitem[{\citenamefont{Ebadi et~al.}(2021)}]{Ebadi:2020ldi}
\bibinfo{author}{\bibfnamefont{S.}~\bibnamefont{Ebadi}} \bibnamefont{et~al.},
  \bibinfo{journal}{Nature} \textbf{\bibinfo{volume}{595}},
  \bibinfo{pages}{227} (\bibinfo{year}{2021}), \eprint{2012.12281}.

\bibitem[{\citenamefont{Keesling et~al.}(2019)}]{Keesling:2018ish}
\bibinfo{author}{\bibfnamefont{A.}~\bibnamefont{Keesling}}
  \bibnamefont{et~al.}, \bibinfo{journal}{Nature}
  \textbf{\bibinfo{volume}{568}}, \bibinfo{pages}{207} (\bibinfo{year}{2019}),
  \eprint{1809.05540}.

\bibitem[{Blo(2023)}]{Bloqade}
\emph{\bibinfo{title}{Bloqade.jl: Package for the quantum computation and
  quantum simulation based on the neutral-atom architecture.}}
  (\bibinfo{year}{2023}),
  \urlprefix\url{https://github.com/QuEraComputing/Bloqade.jl}.

\bibitem[{\citenamefont{Chen et~al.}(2018)\citenamefont{Chen, Zhou, and
  Xu}}]{chen2018measuring}
\bibinfo{author}{\bibfnamefont{X.}~\bibnamefont{Chen}},
  \bibinfo{author}{\bibfnamefont{T.}~\bibnamefont{Zhou}}, \bibnamefont{and}
  \bibinfo{author}{\bibfnamefont{C.}~\bibnamefont{Xu}},
  \bibinfo{journal}{Journal of Statistical Mechanics: Theory and Experiment}
  \textbf{\bibinfo{volume}{2018}}, \bibinfo{pages}{073101}
  (\bibinfo{year}{2018}).

\end{thebibliography}

\end{document}